\documentclass[a4paper,11pt]{article}
\usepackage{amsmath,amssymb,amsfonts}
\usepackage{graphicx,epsfig, subfigure}
\usepackage{hyperref}
\usepackage{color}
\usepackage{amsfonts}
\usepackage{slashed}
\usepackage{epigraph}
\usepackage{pifont}
\usepackage{latexsym, multirow,comment,appendix} 
\usepackage[T1]{fontenc}
\usepackage[utf8]{inputenc}
\usepackage{authblk}
\pdfoutput=1
\usepackage[utopia]{mathdesign}
\usepackage[font=small,labelfont=bf]{caption}
\usepackage{longtable}
\usepackage{verbatim}
\usepackage{cite}
\usepackage{slashed} 
\usepackage{float}

\numberwithin{equation}{section}
\newcommand{\bear}{\begin{array}}
\newcommand{\ear}{\end{array}}

\newcommand{\no}{\nonumber}

\def\OMIT#1{{}}
\newcommand{\lsim}{\mathrel{\rlap{\lower4pt\hbox{\hskip1pt$\sim$}}
    \raise1pt\hbox{$<$}}}         
\newcommand{\gsim}{\mathrel{\rlap{\lower4pt\hbox{\hskip1pt$\sim$}}
    \raise1pt\hbox{$>$}}}         

\newcommand{\be}{\begin{eqnarray}}
\newcommand{\ee}{\end{eqnarray}}
\newcommand{\ba}{\begin{eqnarray}}
\newcommand{\ea}{\end{eqnarray}}

\title{Vacuum stability bounds on Higgs coupling deviations}
\author[1]{Kfir Blum}
\author[1]{Raffaele Tito D'Agnolo}
\author[2]{JiJi Fan}
\affil[1]{Institute for Advanced Study, Princeton, NJ 08540, USA}
\affil[2]{Department of Physics, Syracuse University, Syracuse, NY 13244, USA}
\date{}
\begin{document}
\maketitle

\begin{abstract}

We analyze the constraints imposed by Higgs vacuum stability on models with new fermions beyond the Standard Model.  
We focus on the phenomenology of Higgs couplings accessible at the Large Hadron Collider. New fermions that affect Higgs couplings lead to vacuum instability of the Higgs potential. Above the scale of vacuum instability, bosonic states must stabilize the potential, implying a cut-off to the pure fermion model. 
Conservatively tuning the models to produce the maximal cut-off for a given Higgs coupling effect, we show that observing a deviation in the $Htt$, $H$-diphoton, or $H$-digluon coupling, larger than 20\%, would require that new bosons exist in order to stabilize the Higgs potential below about 100~TeV. For generic parameter configurations, and unless the new fermions are made as light as they can possibly be given current experimental constraints, observing a 10\% deviation in any of these couplings would suggest an instability cut-off below 10-100~TeV. Similarly, if new bosons are absent up to a high scale, then a deviation in the $Hbb$ or $H\tau\tau$ coupling, larger than about 20\%, should be accompanied by a sizable deviation in the $Zbb$ or $Z\tau\tau$ couplings that can be conclusively tested with electroweak precision measurements at planned lepton colliders.  
\end{abstract}

\section{Introduction}
The large hadron collider (LHC) Run-I gave us the Higgs boson, but the weak scale hierarchy problem does not seem closer to a solution than it did decades ago. This may change with new experimental information in Run-II, of which improved Higgs coupling measurements~\cite{Chatrchyan:2014vua, Khachatryan:2014ira, CMScom, Aad:2014xzb, Aad:2014eha, Aad:2014eva} are a guaranteed outcome. A natural question to ask, is whether this Higgs particle that was found is the only one of its kind, namely, the only scalar particle up to very high energies.

Indeed, proposed solutions to the hierarchy problem include new bosonic states beyond the Standard Model (SM). Examples are the scalar super-partners of the SM fields in supersymmetry~\cite{Dimopoulos:1981zb} and the bosonic resonances in composite Higgs models~\cite{Kaplan:1983fs, Kaplan:1983sm}. In these examples, the scale at which the new bosonic states become dynamical marks the cut-off of the quadratic divergence in the quantum corrections to the Higgs mass. 

In this paper we show that measuring deviations in Higgs couplings at the LHC can establish the presence of new bosonic states, even if these bosons do not directly affect any Higgs coupling and are beyond reach of direct production. 
To show this, we proceed by elimination: we analyze the possibility that Higgs coupling modifications arise due to new fermionic states, without any new bosons up to a scale $\Lambda_{UV}$. Our task is then to derive an upper bound on $\Lambda_{UV}$. As we show, this upper bound is found from vacuum stability. 

To explain the logic, note that the only way to couple new fermions to the Higgs is through Yukawa couplings. New Yukawa interactions can certainly affect Higgs couplings to SM states, e.g. through new fermions running in the $H\gamma\gamma$ or $HGG$ loop amplitudes, or mixing with the SM leptons or quarks at tree level. However, as we shall show, the new Yukawa couplings must be sizable to generate a measurable deviation. Large Yukawa couplings have a definite effect in the renormalization group evolution (RGE) of the Higgs-self quartic coupling, driving the quartic negative and leading to an instability in the effective potential~\cite{Zhang:2000zy}. To fix this instability, at least within the domain of validity of a perturbative analysis, new bosonic states are needed~\footnote{An alternative logical possibility is that the Higgs scalar itself ceases to exist as a fundamental state above a cut-off scale $\Lambda_{UV}$. This alternative is even more exciting but we are not aware of a framework in which this happens without new bosonic degrees of freedom (fundamental or composite) becoming dynamical close to the same scale.}.

Vacuum stability has been invoked as a constraint on the SM effective theory in the past (for an early review see, e.g.~\cite{Sher:1988mj}). A point that drew much attention in the days prior to the Higgs discovery and the precise measurement of the top quark mass, was the fact that measuring a heavy top quark or a light Higgs would have indirectly but robustly excluded the SM and required new physics to enter around a cut-off scale $\Lambda$ that, for certain (then plausible) values of $m_t$ and $m_h$, could have been within reach of collider experiments such as the (then futuristic) LHC~\cite{Altarelli:1994rb, Casas:1994qy, Casas:1994us, Espinosa:1995se, Hung:1995in,Casas:1996aq}. Our paper can be thought of as an update circa-2015 of this logic. Today, having measured both the top and the Higgs masses to impressive accuracy (establishing that the SM Higgs potential is  consistent with vacuum stability up to very high scales~\cite{Degrassi:2012ry, Andreassen:2014gha, Branchina:2014rva}), the missing crucial experimental information is the precise values of the Higgs couplings. 

By the end of the LHC~14~TeV program we expect uncertainties in the ballpark of 5-10\% on $Hbb$, $H\tau\tau$, $H\gamma\gamma$, $HGG$, $Htt$, $HZZ$, $HWW$ with $300\;{\rm fb}^{-1}$ of data~\cite{Peskin:2012we}. Postponing more details to the body of the paper, our generic quantitative statement here is that a deviation at the ${\cal O}$(10\%) level in any of the $H\gamma\gamma$, $HGG$, or $Htt$ couplings would imply that new bosonic states must stabilize the Higgs potential below a scale $\Lambda_{UV}\sim$~100~TeV. An ${\cal O}$(10\%) deviation in the $Hbb$ or $H\tau\tau$ coupling should be accompanied by a corresponding deviation in the $Zbb$ or $Z\tau\tau$ couplings at the permille level, well within the expected resolution of planned GigaZ machines such as the international linear collider (ILC) or circular electron-positron colliders in China or at CERN; ruling out this corresponding $Z$-pole deviation would rule out pure fermion models. In addition, while we do not report a detailed analysis of this point here, applying our results to the $HWW$ and $HZZ$ couplings suggests strong vacuum stability constraints, strong enough to imply that observing a deviation in one of these channels at the LHC would rule out pure fermion models.

Our results can be turned around to serve as generic prediction for the Higgs couplings in theories that do not contain new bosonic states up to very high scales, such as split supersymmetry~\cite{ArkaniHamed:2004fb,Giudice:2004tc,ArkaniHamed:2004yi,Wells:2004di} and its variants. According to our analysis, this general class of models predicts that Higgs coupling modifications will not be discoverable (or just very barely) at the LHC. This is not a trivial point because low-lying fermions, protected by chiral symmetries, could in principle be accommodated in these theories and couple to SM fields. 

Our results are relevant to the high-luminosity LHC as well as to a future lepton collider such as the ILC, the electron-positron mode of the future circular collider  FCC-ee (formerly known as TLEP) and the circular electron positron collider (CEPC), that promise percent and even sub-percent accuracy on Higgs coupling measurements~\cite{Peskin:2012we, Asner:2013psa, Gomez-Ceballos:2013zzn, Dawson:2013bba, ZhuTalk, Fan:2014vta}.

The body of this paper deals with the calculation of the cut-off scale $\Lambda_{UV}$ that could be inferred from measuring a deviation in the Higgs couplings to SM states. Since our derivation requires that we add only new fermions but no scalars or vector bosons, and since exotic chiral fermions are either ruled out already or, where they are not, require very large Yukawa couplings and so make our instability analysis trivial, we will only be dealing with new vector-like fermion representations. 
The fact that a sizable deviation in Higgs couplings in a pure-fermion theory renders the Higgs potential unstable was noted in several works~\cite{Joglekar:2012vc,Kearney:2012zi, Batell:2012ca, Altmannshofer:2013zba, Ellis:2014dza} (see also more general discussions in~\cite{Fairbairn:2013xaa, Maru:2013ooa, Feng:2013mea, Huo:2012tw, Carena:2012mw, Davoudiasl:2012tu, Batell:2012zw, Lee:2012wz, Chun:2012jw, Kobakhidze:2012wb, Kitahara:2012pb, Giudice:2012pf}). None of these works, however, have made the instability issue their primary quantitative focus. The closest in spirit to our current analysis are Refs.~\cite{ArkaniHamed:2012kq} and~\cite{Reece:2012gi}, where the instability calculation was done for the specific channels of $H\gamma\gamma$ and $HGG$. Here we expand on these works significantly by refining the instability calculation (using the one-loop, two-loop RGE-improved Coleman-Weinberg potential) and by generalizing to additional Higgs couplings.

The rest of this paper is organized as follows. In Sec.~\ref{sec:Hff} we consider Higgs couplings to SM fermions $Hff$. We begin by analyzing the $Hbb$ coupling in Sec.~\ref{ssec:Hqq}, and use this first example to illustrate our calculation of the Higgs effective potential and our definition of the vacuum instability scale $\Lambda_{\rm UV}$. We then proceed with the $H\tau\tau$ and $Htt$ analyses in Secs.~\ref{ssec:Hll} and~\ref{ssec:Htt}. We survey all of the new physics vector-like fermion representations that could mix with SM fermions at tree level, inducing Higgs-SM fermion coupling modifications. Analyzing together the Higgs and $Z$-boson couplings, we show that the $Hff$ and $Zff$ coupling deviations are correlated. This leads to strong constraints on the $Hbb$ and $H\tau\tau$ couplings. In Sec.~\ref{sssec:num} we devote special attention to validating our analytical results with numerical calculations for the case of light vector-like fermions. 
In Sec.~\ref{sec:HAA} we analyze the $H\gamma\gamma$ and $HGG$ couplings. We conclude in Sec.~\ref{sec:sum}. App.~\ref{app:cw} provides details of our calculation of the Higgs effective potential. App.~\ref{app:Hqq} contains details of the Higgs-fermion coupling analysis, surveying all of the relevant vector-like fermion representations that can mix with the SM states and presenting some useful  formulae derived using effective field theory. In App.~\ref{app:collider} we give a rough, but rather inclusive, assessment of the collider constraints on new physics fermions, based on the results of Run-I of the LHC. This survey of the current experimental constraints is important in particular for the $Htt$, $HGG$, and $H\gamma\gamma$ analyses.

\section{Higgs couplings to fermions}
\label{sec:Hff}

Our task in this section is to consider the possible ways in which new vector-like fermions could mix with the SM fermions $f$, inducing $\delta r_f\neq0$ where 
\be\delta r_f\equiv \left(g_{hff}-g_{hff}^{\rm SM}\right)/g_{hff}^{\rm SM}.\ee 

Only a few vector fermion representations can mix with the SM fermions. In App.~\ref{app:Hqq} we survey these representations, as well as define our notation for the SM and new physics fields. Higgs couplings and electroweak observables in some of these models have been studied extensively in the literature~\cite{Choudhury:2001hs, Morrissey:2003sc, Dawson:2012di, Kearney:2012zi, Batell:2012ca}. LEP and LHC constraints imply that new charged fermions cannot be too light. In App.~\ref{app:collider} we summarize the existing collider constraints, finding that $M>600$~GeV is a conservative lower limit on the mass scale of $b$ and $t$ quark partners in all cases of interest to us. This relatively heavy mass scale justifies the use of effective field theory (EFT) in analyzing the induced modifications to Higgs-quark couplings. For non-colored fermions the collider bounds are weaker, and we will devote some effort to extend the EFT analysis when dealing with Higgs-lepton couplings. 

Integration-out of heavy vector fermions up to non-renormalizable operators of dimension six is done in App.~\ref{app:Hqq}. From this exercise one finds that the effective Higgs and $Z$-boson couplings to the SM fermions exhibit correlated modifications: large deviations in the effective $Hff$ couplings imply sizable deviations in the $Zff$ couplings. This leads to vacuum stability constraints that can be very relevant for the interpretation of upcoming $Hbb$ and $H\tau\tau$ data, as we show in detail below. In addition, for the top quark, the fact that the SM top Yukawa coupling is $\mathcal{O}(1)$ implies that large new physics effects are required to deform the $Htt$ vertex appreciably, leading again to a strong vacuum stability constraint in the case of pure fermion models.

\subsection{$Hbb$}
\label{ssec:Hqq}

To analyze the vacuum stability constraints on the effective $Hbb$ coupling, and the interplay with precision $Zbb$ data, we begin with a concrete model example. 
Consider the vector-like fermion representation  
\be\label{eq:exdqs} Q(3,2)_{\frac{1}{6}},\;\;Q^c(\bar{3},2)_{-\frac{1}{6}},\;\;D(3,1)_{-\frac{1}{3}},\;\;D^c(\bar{3},1)_{\frac{1}{3}}\ee 
with the potential 
\be\label{eq:QDSexample}\mathcal{V}_{NP}=Y_{Qd^c}H^\dag Qd^c+Y_{qD^c}H^\dag qD^c+Y_{QD^c}H^\dag QD^c+Y_{Q^cD}H^T\epsilon Q^cD+M_QQ^T\epsilon Q^c+M_DDD^c+cc.\ee  This representation is denoted as rep' $DI$ in App.~\ref{app:Hqq}. 

\begin{figure}[htbp]
\begin{center}
\includegraphics[width=0.65\textwidth]{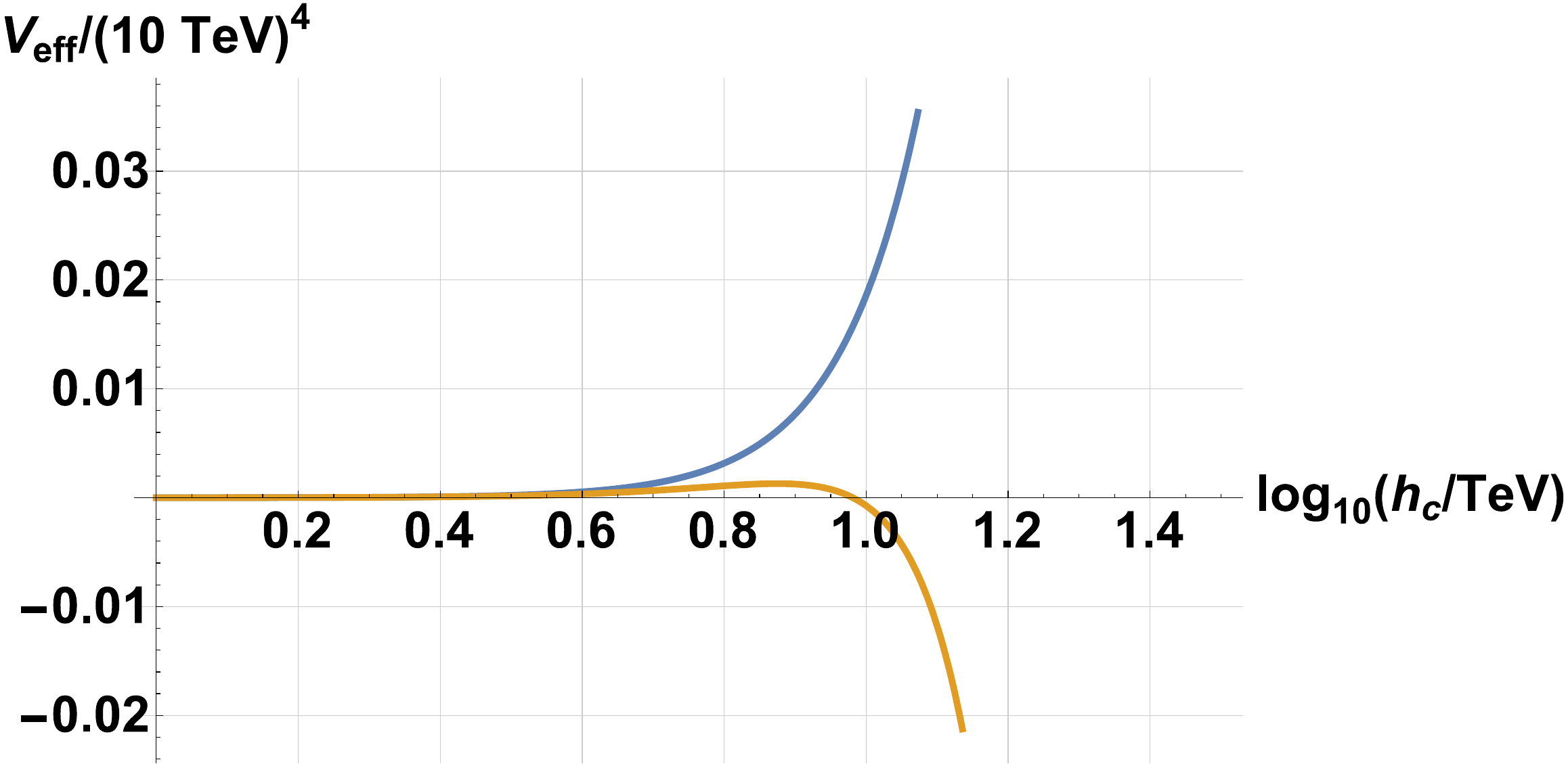}\quad
\caption{The one-loop Higgs effective potential, shown for the SM (blue) and in the presence of the vector-like fermions of Eqs.~(\ref{eq:exdqs}-\ref{eq:QDSexample}) (orange). We set $M_Q=M_D=1$~TeV, $Y_{Q^cD}=1.25$, and $Y_{QD^c}=Y_{qD^c}=Y_{Qd^c}=0$, defined at the scale $\mu=1$~TeV.}
\label{fig:VeffRGEhbb}
\end{center}
\end{figure}

Given a concrete example we can study the Higgs effective potential $V_{\rm eff}$ in the presence of the new fermions. In Fig.~\ref{fig:VeffRGEhbb} we plot $V_{\rm eff}$ as a function of the classical field $h_c$. We set $M_Q=M_D=1$~TeV, $Y_{Q^cD}=1.25$, and $Y_{QD^c}=Y_{qD^c}=Y_{Qd^c}=0$, defined at the scale $\mu=1$~TeV. The details of the calculation of $V_{\rm eff}$ are given in App.~\ref{app:cw}. The vacuum instability corresponds to the negative runaway of the effective potential, seen in Fig.~\ref{fig:VeffRGEhbb} (orange curve) to occur about an order of magnitude above the vector-like fermion mass scale.  

\begin{figure}[htbp]
\begin{center}
\includegraphics[width=0.65\textwidth]{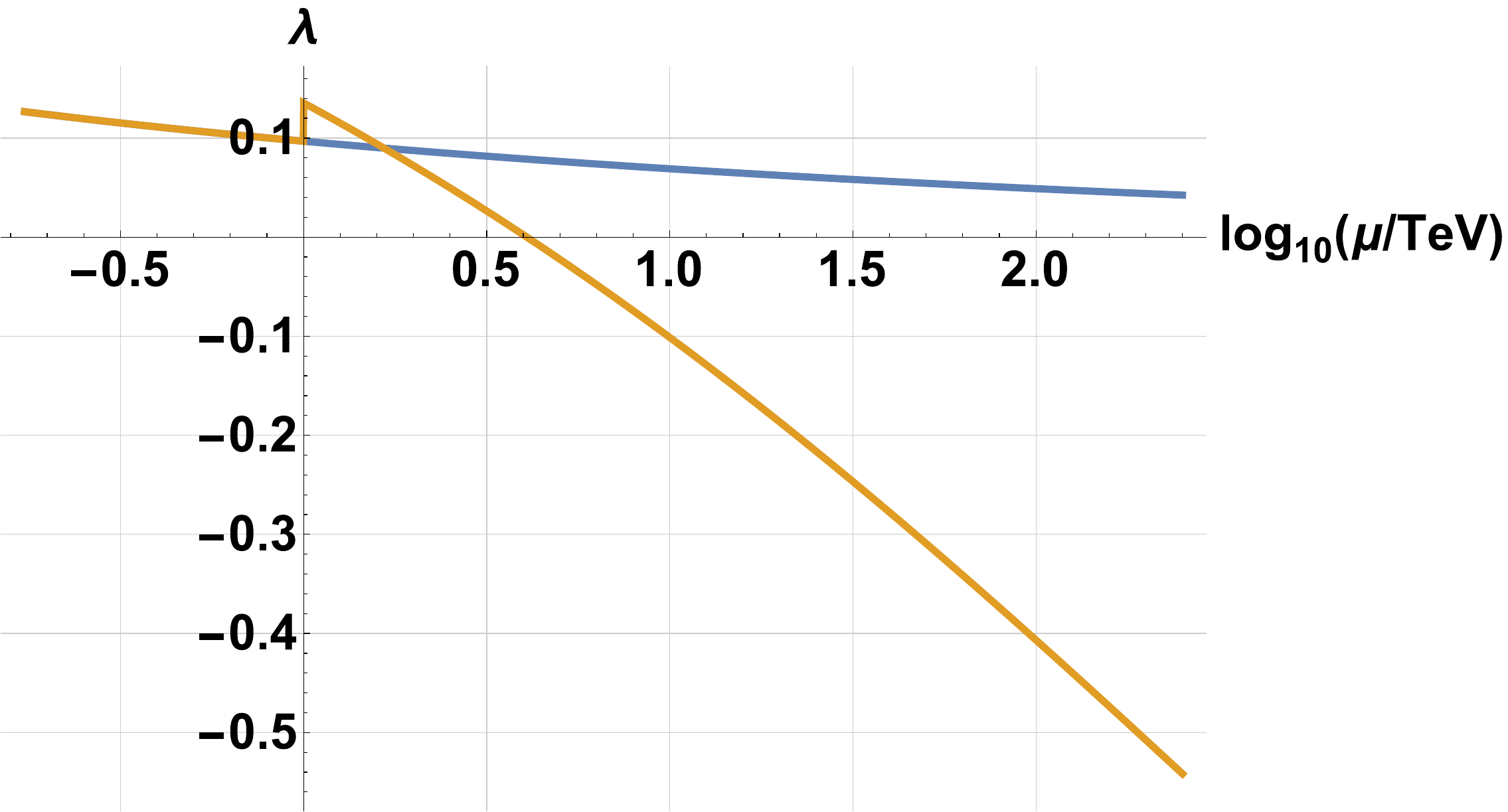}\quad
\caption{The Higgs quartic coupling in the SM (blue) and with the model example of Fig.~\ref{fig:VeffRGEhbb} (orange).}
\label{fig:lamRGEhbb2}
\end{center}
\end{figure}

The vacuum instability depicted in Fig.~\ref{fig:VeffRGEhbb} can be understood as due to the negative RGE running of the Higgs quartic coupling in the presence of extra fermions. In Fig.~\ref{fig:lamRGEhbb2} we plot the Higgs quartic coupling $\lambda$ for the same model example of Fig.~\ref{fig:VeffRGEhbb}. In the plot, the blue line shows the SM running of $\lambda$ and the orange line gives $\lambda$ with the new fermions included. The jump at $\mu=1$~TeV is due to the threshold correction in matching the SM EFT below the vector mass scale to the full theory above it. 

Figs.~\ref{fig:VeffRGEhbb}-\ref{fig:lamRGEhbb2} demonstrate that the field content of Eq.~(\ref{eq:exdqs}) cannot be considered as a complete theory, if implemented with a large Yukawa coupling at the vector-like fermion scale. Other fields, coupled to the Higgs, must be added in order to eliminate the runaway of the effective potential. Adding more fermions would make the instability worse, so the new fields need to be bosons. 

\begin{figure}[htbp]
\begin{center}
\includegraphics[width=0.6\textwidth]{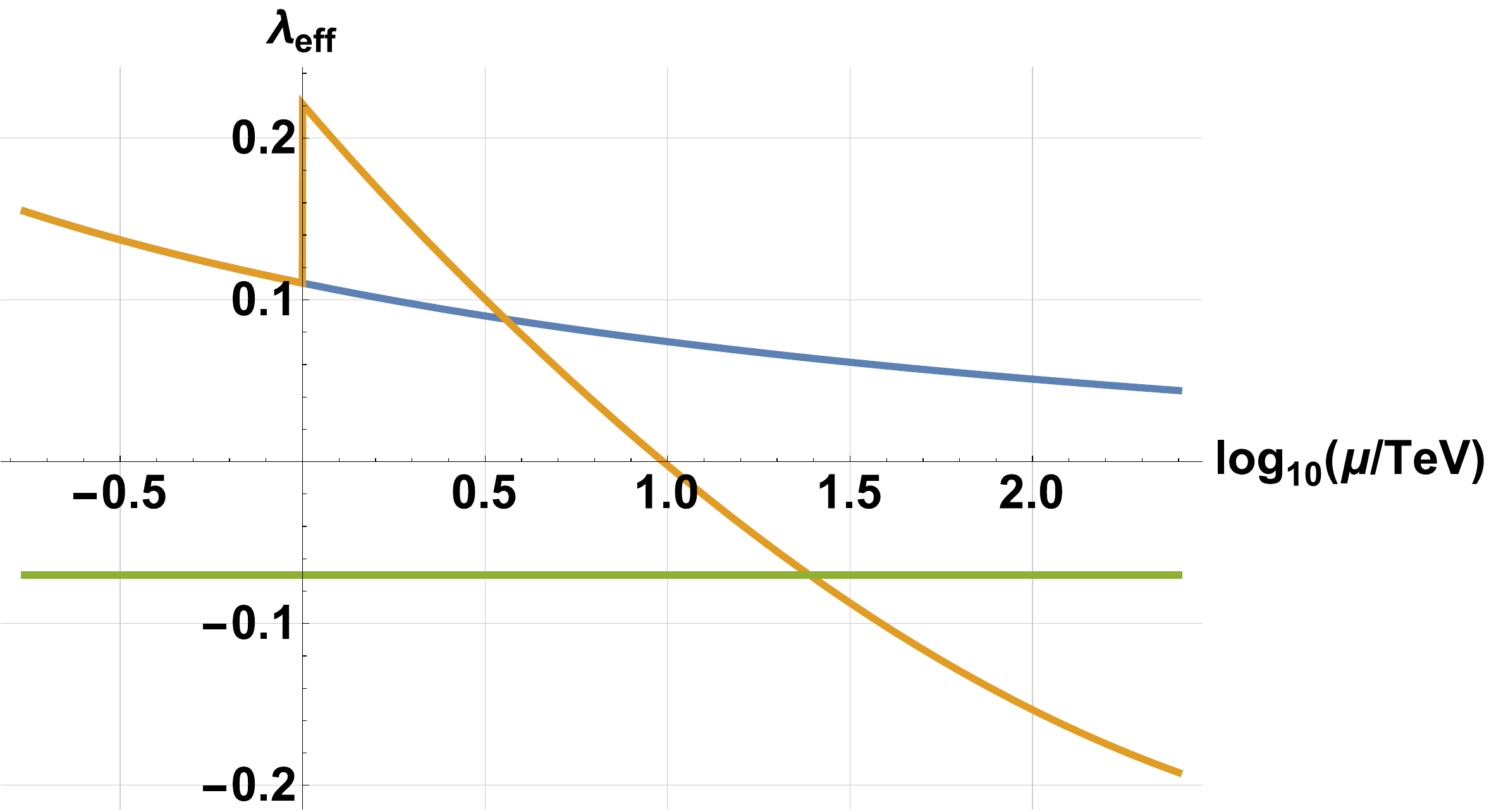}\quad
\caption{The effective Higgs quartic coupling in the SM (blue) and with the model example of Fig.~\ref{fig:VeffRGEhbb} (orange). The horizontal green line marks $\lambda_{\rm eff} = - 0.07$ indicating a rough criterion for vacuum instability.}
\label{fig:lamRGEhbb}
\end{center}
\end{figure}

A precise determination of the mass scale at which the new bosons need to be introduced, given an apparent runaway in the effective potential as in Fig.~\ref{fig:VeffRGEhbb}, is beyond the scope of this paper. Instead, we define a rough criterion for the scale $\Lambda_{\rm UV}$ at which the  effective potential should be stabilized, through the condition $\lambda_{\rm eff}(\Lambda_{\rm UV})=-0.07$ where $\lambda_{\rm eff}$ parametrizes the effective potential at large $h_c$ values, where we can set all dimensional parameters to zero besides from the Higgs field itself: $V_{\rm eff}\to\frac{\lambda_{\rm eff}}{4}h_c^4$. The value $\lambda_{\rm eff}=-0.07$ indicates roughly the onset of vacuum instability; for more negative $\lambda_{\rm eff}$, the time scale for tunneling from $h_c=246$~GeV to a remote vacuum with $h_c\gtrsim\Lambda_{\rm UV}$ becomes exponentially shorter than the age of the Universe~\cite{Isidori:2001bm}\footnote{A rough estimate of the tunneling probability through true vacuum bubbles of nucleation size $1/\Lambda$ is $p\sim(\Lambda/H_0)^{4}e^{-S(\Lambda)}$, where $S(\Lambda)\approx\frac{8\pi^2}{3|\lambda(\Lambda)|}$ and $H_0\sim10^{-42}$~GeV is the Hubble constant. Setting $p=1$ gives $\lambda(\Lambda)\approx-0.065\left(1+0.02\log_{10}\left(\Lambda/1~{\rm TeV}\right)\right)^{-1}$.}. 
In Fig.~\ref{fig:lamRGEhbb} we plot $\lambda_{\rm eff}$ for the same example of Figs.~\ref{fig:VeffRGEhbb}-\ref{fig:lamRGEhbb2}.  

\begin{figure}[htbp]
\begin{center}
\includegraphics[width=0.6\textwidth]{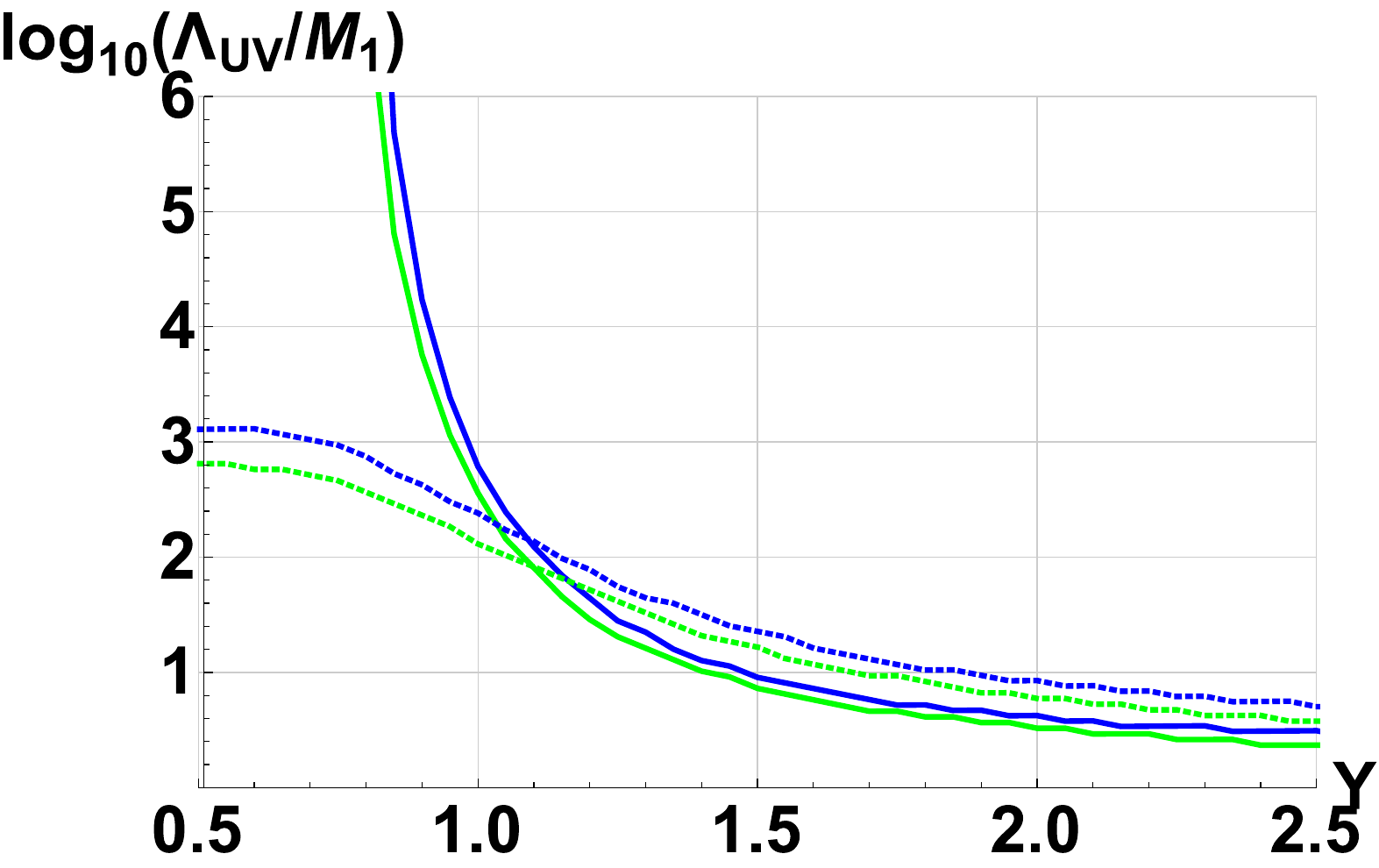}\quad
\caption{$\Lambda_{\rm UV}/M_1$ vs. the Yukawa coupling $Y=Y_{Q^cD}$. The model is the  vector-like quarks of Eqs.~(\ref{eq:exdqs}-\ref{eq:QDSexample}). Smooth (dotted) lines show the results fixing $Y_{QD^c}$, an additional Yukawa coupling in the model, to $Y_{QD^c}=0$ ($Y_{QD^c}=1$). Blue (green) lines are for $M_Q=M_D=1$~TeV (5~TeV).}
\label{fig:QDSLam2M1}
\end{center}
\end{figure}

The scale $\Lambda_{\rm UV}$, or any similar characteristic scale parametrizing the potential runaway, depends on the size of the Yukawa couplings. For $Y_{Q^cD}$ slightly larger than unity, as in the case depicted in Figs.~\ref{fig:VeffRGEhbb}-\ref{fig:lamRGEhbb}, we find $\Lambda_{\rm UV}$ about an order of magnitude above the vector-like fermion mass scale. Larger values of $Y_{Q^cD}$ within the perturbative window ($|Y|\lesssim4\pi/\sqrt{\mathcal{N}}$, with $\mathcal{N}=4N_c=12$ in the current case) change $\Lambda_{\rm UV}$ by an order one amount. Similarly, turning on, in addition to $Y_{Q^cD}$, some of the other Yukawa couplings in the model, also has a modest order one effect. 
To substantiate this point, in Fig.~\ref{fig:QDSLam2M1} we plot the ratio $\Lambda_{\rm UV}/M_1$, where $M_1$ is the lightest vector-like fermion mass eigenstate, vs. the Yukawa coupling $Y_{Q^cD}$. In obtaining the green (blue) smooth lines, we set $Y=Y_{Q^cD}$, all other Yukawa couplings to zero, and $M_Q=M_D=1$~TeV (5~TeV). In the dotted lines we repeat the same values for $M_Q,M_D$, but turn on the additional Yukawa coupling $Y_{QD^c}=1$. We have verified that varying $M_Q\neq M_D$ does not change the results. 

The bottom line is that allowing any of the Yukawa couplings in Eq.~(\ref{eq:QDSexample}) to become larger than $\sim1.5$ would render the vector-like fermion model inconsistent right above the new fermion mass scale. However, it is also apparent in Fig.~\ref{fig:QDSLam2M1} that if the Yukawa coupling is made sufficiently small, then the instability can be pushed up to high energies. This decoupling of $\Lambda_{\rm UV}$, that occurs as the Yukawa couplings are made small, is important because it means that for certain parameter configurations the vector-like fermion model can form a consistent EFT up to a high scale, in which case the vacuum stability argument looses phenomenological relevance. 

The next step is to exhibit the relation with  $Hbb$ and $Zbb$ measurements. 
Integrating out the heavy fields we find (see~\cite{Kearney:2012zi,Batell:2012ca} for related analysis)
\be
\delta r_b&\approx&-2\delta g_{Ab}+\frac{2|Y_{Q^cD}|v}{\sqrt{2}m_b}\sqrt{\left|\delta g_{Vb}^2-\delta g_{Ab}^2\right|}e^{i\phi},\label{eq:drbex1}
\ee
where $\phi$ is a complex phase (to be explained shortly) and $\delta g_{Vb}$ and $\delta g_{Ab}$ are the modification to the vector and axial $Zbb$ couplings, respectively, defined in the usual way (see, e.g.~\cite{Ciuchini:2013pca}) and constraint by LEP data to be in the ballpark of a few percent~\cite{Agashe:2014kda}. 
In App.~\ref{app:Hqq} (see Tab.~\ref{tab:qHZ}) we give a derivation of Eq.~(\ref{eq:drbex1}) by integrating out the heavy states in an expansion to leading order in $Y^2v^2/M^2$, where $Y$ is any of the Yukawa couplings in the problem and $M=M_{Q,D}$. At leading order in $Y^2v^2/M^2$, as done in App.~\ref{app:Hqq}, the phase $\phi$ is given by $\phi={\rm arg}\left(Y_{Q^cD}Y_{qD^c}Y_{Qd^c}y_b^*M_Q^*M_D^*\right)$. 
However, and usefully for our purpose, Eq.~(\ref{eq:drbex1}) is actually valid to good accuracy even when the vector-like states are light and allowing the Yukawa couplings in the vector-like sector to be large, namely $Y_{QD^c}^2v^2/M^2,\,Y_{Q^cD}^2v^2/M^2=\mathcal{O}(1)$. To see this, note that one could also derive Eq.~(\ref{eq:drbex1}) expanding in the mixing couplings $Y_{qD^c}$, $Y_{Qd^c}$, and in the mostly-SM bottom Yukawa coupling $y_b$, but keeping all orders in $Y_{Q^cD}$ and $Y_{QD^c}$. The couplings $y_b,Y_{Qd^c},Y_{qD^c}$ are guaranteed to be good expansion parameters because of the smallness of the bottom quark mass (combined with the fact that $\delta r_b$ is constrained to be smaller than unity by existing LHC data) and by the experimental constraints on $\delta g_{Ab,Vb}$. Following this route and allowing $Y_{QD^c}^2v^2/M^2,\,Y_{Q^cD}^2v^2/M^2=\mathcal{O}(1)$, we find that Eq.~(\ref{eq:drbex1}) remains valid as is, and the only modification that needs to be done is a generalization of the phase $\phi={\rm arg}\left(\frac{Y_{Q^cD}Y_{qD^c}Y_{Qd^c}M_QM_D}{|\mathcal{M}||\mathcal{M}_{2\times2}|}\right)$, where $|\mathcal{M}|$ is the determinant of the $3\times3$ mass matrix given by Eq.~(\ref{eq:QDSexample}) and $|\mathcal{M}_{2\times2}|=M_QM_D-Y_{Q^cD}Y_{QD^c}v^2/2$ is the determinant of the vector-like $2\times2$ sub-matrix. 
In Sec.~\ref{sssec:num}, devoted to vector-like leptons but dealing essentially with the same formula, we demonstrate the validity of Eq.~(\ref{eq:drbex1}) for very light vector-like states with a numerical analysis. 

Eq.~(\ref{eq:drbex1}) shows that a large deviation in $Hbb$ requires some corresponding deviation in the $Zbb$ couplings. This correlation can be relaxed, but only at the cost of a sizable Yukawa coupling, $Y_{Q^cD}$ in our example.  
To make a quantitative estimate, note that for $|\delta r_b|\gtrsim0.2$ or so, we can neglect the $-2\delta g_{Ad}$ contribution on the RHS of Eq.~(\ref{eq:drbex1}). Setting $|\delta g_{Vd},\,\delta g_{Ad}|\lesssim10^{-3}$, within the expected resolution of future lepton collider experiments, we have $|\delta r_b|\approx0.1|Y_{Q^cD}|\left(\sqrt{\left|\delta g_{Vb}^2-\delta g_{Ab}^2\right|}/10^{-3}\right)$. Thus, given our analysis above summarized in Fig.~\ref{fig:QDSLam2M1}, vacuum stability can make a powerful discriminator for a consistent explanation of Higgs couplings deviations, once the $Zbb$ couplings have been determined to about a permille accuracy.  

\begin{figure}[htbp]
\begin{center}
\includegraphics[width=0.45\textwidth]{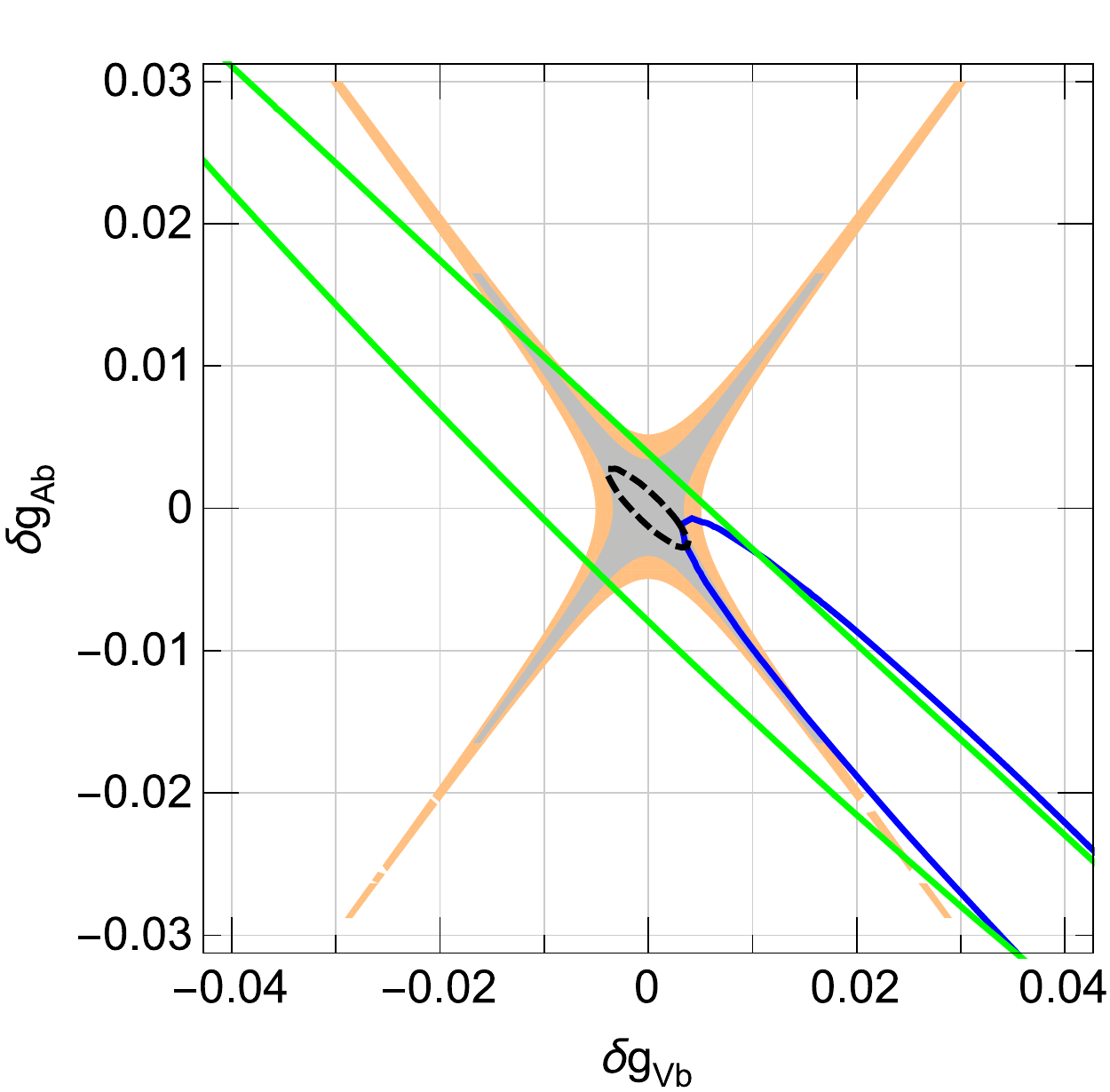}\quad
\includegraphics[width=0.45\textwidth]{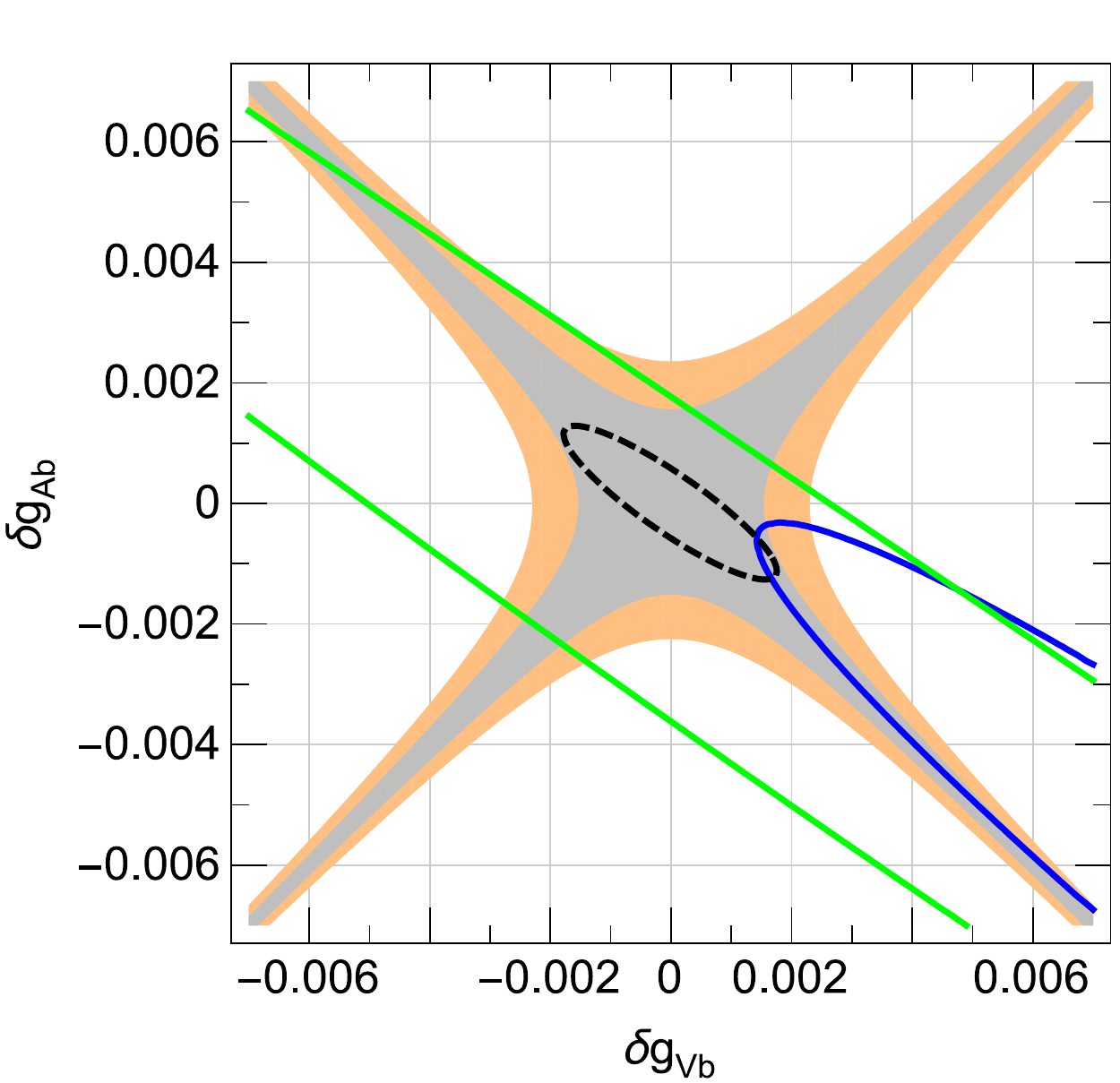}
\caption{Left panel: blue solid, green solid and black dashed contours mark the 95\%CL allowed region in the $\delta g_{Vb},\delta g_{Ab}$ plane, considering current LEP $Zbb$ data including $A_b^{FB}$, LEP data omitting $A_b^{FB}$ and future precision measurements from an experiment such as the ILC, respectively. Inside the orange-shaded (gray-shaded) region, the coupling $|Y_{Q^cD}|$ needs to be larger than 1 (1.5) in order to induce an $Hbb$ deviation of $\delta r_{b}\geq0.2$. Right panel: close-up of the same results.}
\label{fig:ZZHHb}
\end{center}
\end{figure}

In Fig.~\ref{fig:ZZHHb} we demonstrate our results quantitatively. On the $x$- and $y$-axes we plot $\delta g_{Vb}$ and $\delta g_{Ab}$, respectively. Inside the orange-shaded region, the coupling $|Y_{Q^cD}|$ needs to be larger than 1 in order to induce an $Hbb$ deviation of $\delta r_{b}\geq0.2$. Inside the gray-shaded region, $|Y_{Q^cD}|$ needs to be larger than 1.5 to achieve the same $\delta r_b$. To compare with current $Zbb$ data, the blue contour shows the 95\%CL allowed range for $\delta g_{Vb,Ab}$, using the reduced covariance matrix from Ref.~\cite{Ciuchini:2013pca} that includes the anomalous LEP result for $A_b^{FB}$. The green contour shows the 95\%CL allowed region, obtained, ignoring correlations, using the measurements for $R_b$ and $A_b$ as taken from~\cite{Agashe:2014kda}, and omitting the measurement of $A_b^{FB}$. 

We learn that the current $Zbb$ coupling measurements still allow a significant $Hbb$ deviation consistent with vacuum stability. Indeed, in vector-like quark interpretations of the $A_b^{FB}$ anomaly~\cite{Choudhury:2001hs}, an $Hbb$ deviation would be generically expected~\footnote{Note, however, that for $m_h\approx125$~GeV vacuum stability analysis indicates that the models of~\cite{Choudhury:2001hs,Morrissey:2003sc} exhibit instability on scales much lower than the scale of gauge coupling unification.}. Improved experimental determination of the $Zbb$ couplings, in the ballpark of planned future experiments, can sharpen these results and test conclusively a fermionic interpretation. To compare with future $Zbb$ data, the dashed black contour shows an estimate of the 95\%CL allowed region following future measurements at the ILC or another experiment of comparable precision, for which we adopt the SM central values for $R_b$ and $A_b$ and assume $\sigma_{R_b}=0.00014$, $\sigma_{A_b}=0.001$~\cite{Baer:2013cma} (to be compared with the current $\sigma_{R_b}\approx0.0007$ and $\sigma_{A_b}\approx0.02$ from LEP data~\cite{Agashe:2014kda}). The right panel shows an expanded version of the left. The fact that the dashed black error ellipse is completely contained in the orange shaded region, signaling low scale vacuum instability, shows the potential power of the vacuum stability argument to constrain Higgs coupling deviations in conjunction with improved $Zbb$ data. 

Before we conclude this section, we pause to generalize the results to other vector-like fermion representations that can mix with the SM $b$ quark. These representations are collected in App.~\ref{app:Hqq}, Tab.~\ref{tab:qrkL}. For all of these cases, the vacuum stability analysis proceeds in a similar manner, leading to results similar to what we have shown in Fig.~\ref{fig:QDSLam2M1}. 
Considering the $Zbb$ couplings, models $DII-DIV$ in Tab.~\ref{tab:qrkL} possess the same basic structure as in Eqs.~(\ref{eq:exdqs}-\ref{eq:QDSexample}) and produce relations between the $Hbb$ and $Zbb$ effective couplings, that are analogous to Eq.~(\ref{eq:drbex1}) and that were illustrated in Fig.~\ref{fig:ZZHHb}. For completeness, we summarize the $Hbb$ and $Zbb$ relations for all of these models in Tab.~\ref{tab:qHZ}.  
In contrast, model $DV$ provides only one vector-like fermion state that mixes with the SM right-handed $b$ quark, with no corresponding partner for the left-handed SM quark. As a result, the $Hbb$ coupling and the $Zbb$ coupling are directly correlated in model $DV$, predicting $\delta r_b\approx-2\delta g_{Vb}$. Using the current LEP constraints, model $DV$ is limited to induce at most $|\delta r_b|\lesssim0.1$ at 95\%CL. This discussion makes clear why we do not devote attention, in the context of the $Hbb$ coupling, to vector-like fermion representations in which only either the left-handed SM quark or the right-handed SM quark mixes with new fermions, but not both. Examples for such models are given by simply deleting, e.g., either the pair $Q,Q^c$ or the pair $D,D^c$ in Eq.~(\ref{eq:exdqs}), and similarly for the other cases $DII-DIV$. For all such vector-like fermion models, the $Hbb$ deviation is directly tied to a corresponding $Zbb$ deviation, implying that $|\delta r_b|>0.1$ is ruled out by existing precision $Z$-pole data.

Finally we comment on loop corrections to Eq.~(\ref{eq:drbex1}), that was derived at tree level. The dominant loop effect is due to the  renormalization of the bottom quark Yukawa coupling from the scale $m_b$ to the scale $m_h$, relevant for $h\to bb$ decay. This effect is captured by using the running $\overline{\rm MS}$ bottom quark mass $m_b(m_h)\approx3$~GeV in Eq.~(\ref{eq:drbex1}). Another effect is the RGE evolution, within the SM EFT, of the nonrenormalizable operators produced by integrating out the vector-like quarks at their mass threshold down to the scale $m_h$~\cite{Elias-Miro:2013mua,Jenkins:2013zja,Jenkins:2013wua,Alonso:2013hga}. This running corrects Eq.~(\ref{eq:drbex1}) at the 10\% level, in the direction of suppressing the right-hand side of Eq.~(\ref{eq:drbex1}) compared to its value when using $|Y_{Q^cD}|$ as given at the vector-like quark mass threshold.

To summarize: if a pure-fermion model produces a deviation $\gtrsim 20\%$ in $Hbb$, it should also produce an observable deviation in $Zbb$. If the former if found without the latter, then bosonic states should exist in the spectrum not far above the fermion mass scale.

\subsection{$H\tau\tau$}
\label{ssec:Hll}

Our analysis of $H\tau\tau$ follows closely the $Hbb$ discussion of the previous section. 
Again, we work out the details for one representative model example, assuming the field content 
\be \label{eq:exlds} L(1,2)_{-\frac{1}{2}},\;L^c(1,2)_{\frac{1}{2}},\;E(1,1)_{-1},\;E^c(1,1)_{1}\ee 
and the potential
\be\label{eq:LDS}\mathcal{V}_{NP}=Y_{Le^c}H^\dag Le^c+Y_{lE^c}H^\dag lE^c+Y_{LE^c}H^\dag LE^c+Y_{L^cE}H^T\epsilon L^cE+M_LL^T\epsilon L^c+M_EEE^c+cc.\ee

\begin{figure}[htbp]
\begin{center}
\includegraphics[width=0.65\textwidth]{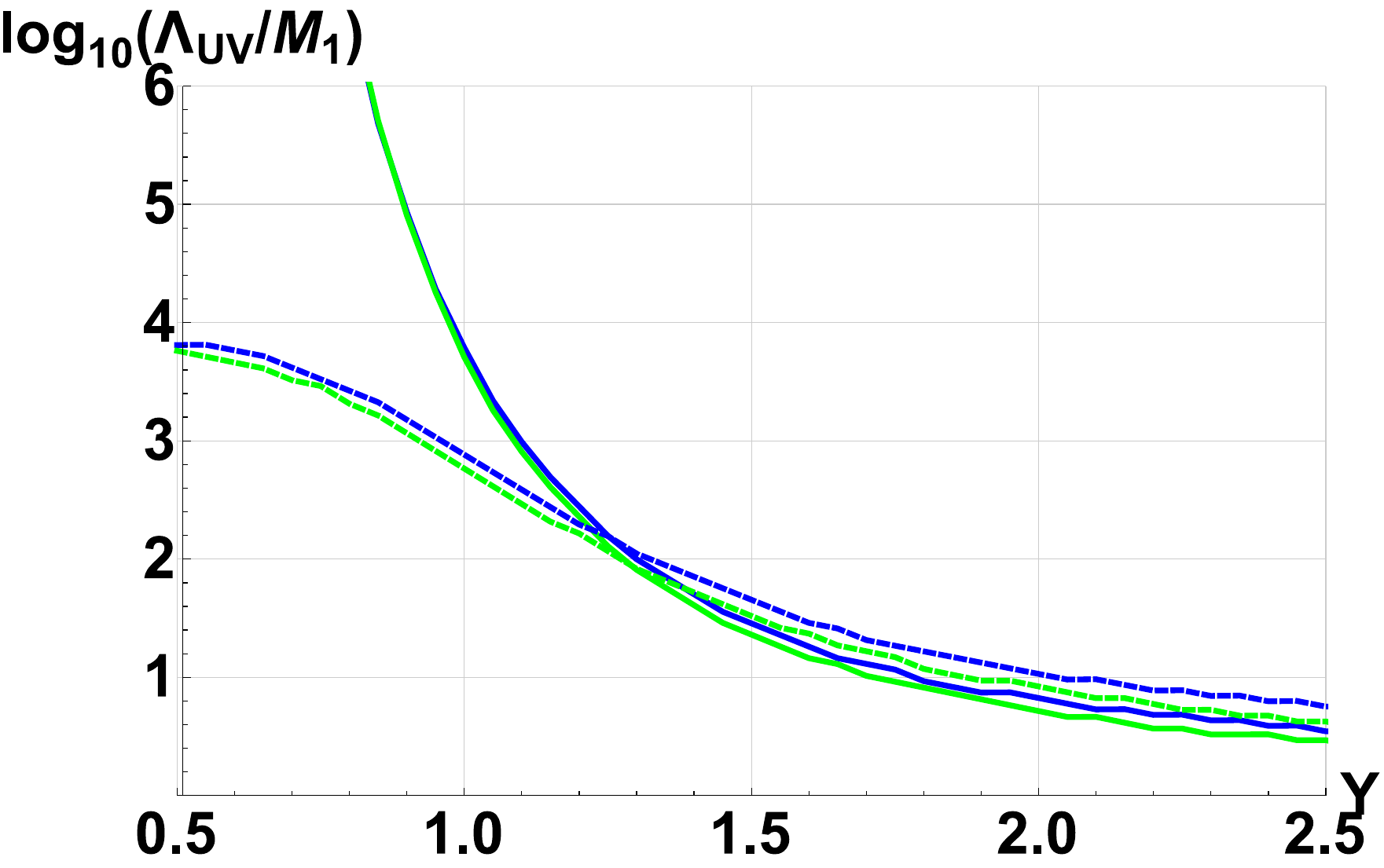}\quad
\caption{$\Lambda_{\rm UV}/M_1$ vs. the Yukawa coupling $Y=Y_{L^cE}$. The model is the  vector-like leptons of Eqs.~(\ref{eq:exlds}-\ref{eq:LDS}). Smooth (dotted) lines show the results fixing $Y_{LE^c}$, an additional Yukawa coupling in the model, to $Y_{LE^c}=0$ ($Y_{LE^c}=1$). Blue (green) lines are for $M_L=M_E=1$~TeV (5~TeV).}
\label{fig:LDSLam2M1}
\end{center}
\end{figure}

We consider the Higgs vacuum stability first. In Fig.~\ref{fig:LDSLam2M1} we plot the ratio $\Lambda_{\rm UV}/M_1$, where $M_1$ is the lightest vector-like fermion mass eigenstate, vs. the Yukawa coupling $Y_{L^cE}$. In obtaining the green (blue) smooth lines, we set $Y=Y_{L^cE}$, all other Yukawa couplings zero, and $M_L=M_E=1$~TeV (5~TeV). In the dotted lines we repeat the same values for $M_L,M_E$, but turn on $Y_{LE^c}=1$. Varying $M_L\neq M_E$ does not change the results. We see that allowing any of the Yukawa couplings in Eq.~(\ref{eq:LDS}) to become larger than $\sim2$ would render the vector-like fermion model inconsistent right above the new fermion mass scale. 

We next explore the relation with precision $Z\tau\tau$ measurements. Integrating out the heavy fields, we find (see App.~\ref{app:Hqq}, Tab.~\ref{tab:lHZ})
\be
\delta r_\tau&\approx&-2\delta g_{A\tau}+\frac{2|Y_{L^cE}|v}{\sqrt{2}m_\tau}\sqrt{\left|\delta g_{V\tau}^2-\delta g_{A\tau}^2\right|}e^{i\phi}.\label{eq:drtauex1}
\ee
The $Z\tau\tau$ coupling deviations $\delta g_{V\tau}$ and $\delta g_{A\tau}$ are constrained by LEP data to be in the ballpark of a few permille~\cite{Agashe:2014kda}.  
In analogy with our results in Sec.~\ref{ssec:Hqq}, Eq.~(\ref{eq:drtauex1}) is valid to good accuracy even when the vector-like states are light and the Yukawa couplings $Y_{L^cE}$ and $Y_{LE^c}$ are $\mathcal{O}(1)$, the only subtlety being the proper definition of the phase $\phi$. When the vector-like states are heavy we have $\phi={\rm arg}\left(Y_{L^cE}Y_{lE^c}Y_{Le^c}y_\tau^*M_L^*M_E^*\right)$, while if one allows light vector-like states then $\phi$ needs to be generalized in exact analogy with the discussion in Sec.~\ref{ssec:Hqq}. In the rest of this section we first use the analytical result in Eq.~(\ref{eq:drtauex1}), and subsequently turn to a numerical verification for very light vector-like states.

\begin{figure}[h]
\begin{center}
\includegraphics[width=0.5\textwidth]{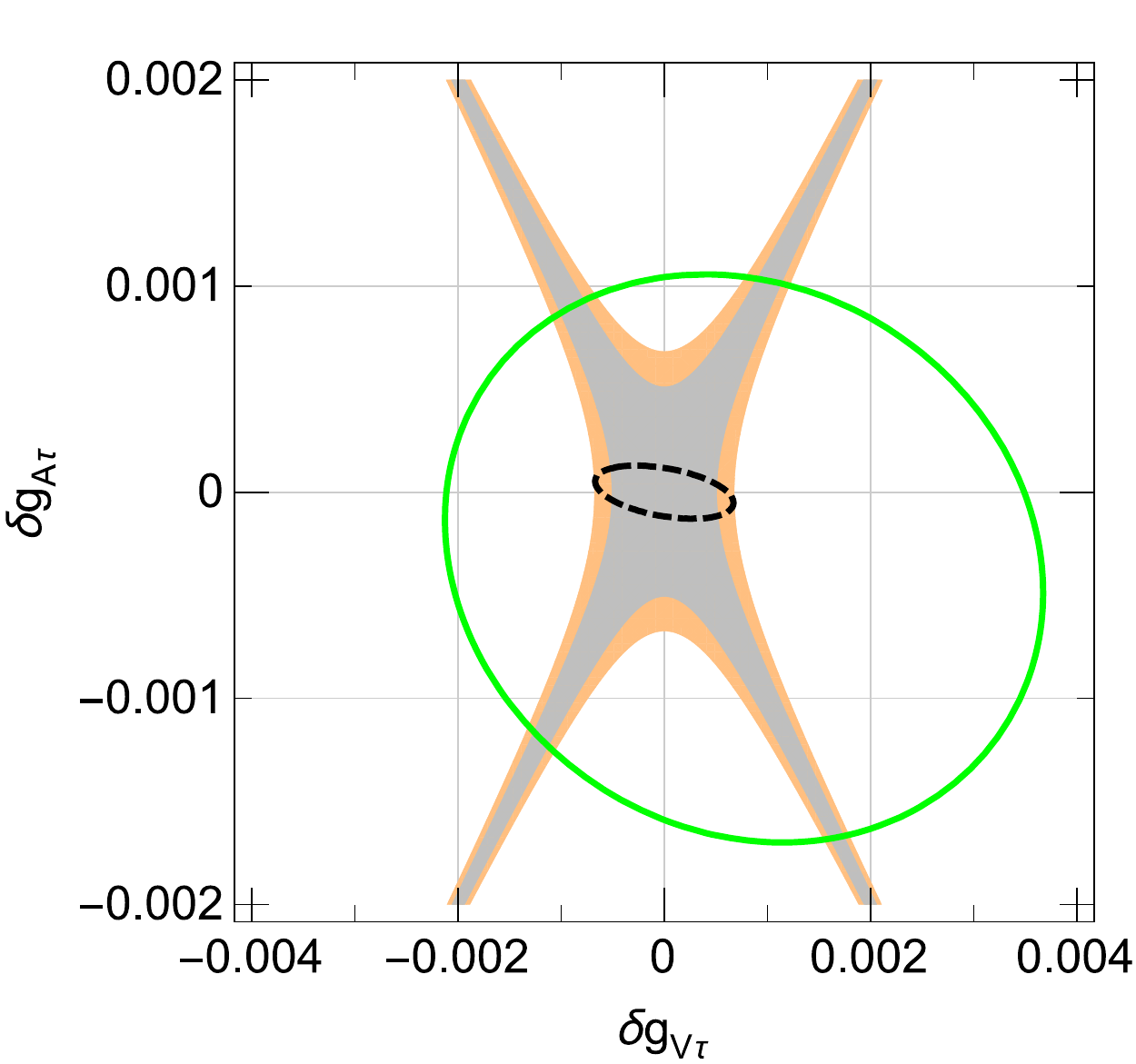}\quad
\caption{Green and black contours mark the 95\%CL allowed region in the $\delta g_{V\tau},\delta g_{A\tau}$ plane, considering current LEP $Z\tau\tau$ data, and future precision measurements from an experiment such as the ILC, respectively. Inside the orange-shaded (gray-shaded) region, the coupling $|Y_{L^cE}|$ needs to be larger than 1.5 (2) in order to induce an $H\tau\tau$ deviation of $\delta r_{\tau}\geq0.2$, indicating vacuum instability. }
\label{fig:ZZHHtau}
\end{center}
\end{figure}

In Fig.~\ref{fig:ZZHHtau} we illustrate the vacuum stability constraint vs. $Z\tau\tau$ data, computed using Eq.~(\ref{eq:drtauex1}), in the $\delta g_{V\tau},\delta g_{A\tau}$ plane. Inside the orange-shaded (gray-shaded) region, the coupling $|Y_{L^cE}|$ needs to be larger than 1.5 (2) in order to induce an $H\tau\tau$ deviation of $\delta r_{\tau}\geq0.2$.  
The green solid contour shows the 95\%CL allowed range for $\delta g_{V\tau,A\tau}$,  using the measurements for $R_\tau$ and $A_\tau$ at LEP~\cite{Agashe:2014kda}. To compare the result with future $Z\tau\tau$ data, the black dashed contour shows an estimate of the 95\%CL allowed region from future measurements at the ILC/GigaZ or another experiment of comparable precision, for which we adopt the SM central values for $R_\tau$ and $A_\tau$ and assume $\sigma_{R_\tau}=0.004$, $\sigma_{A_\tau}=0.001$ (to be compared with the current $\sigma_{R_\tau}\approx0.045$ and $\sigma_{A_\tau}\approx0.004$ from LEP data~\cite{Agashe:2014kda}). The estimate of future $\sigma_{R_\tau}$ at the ILC/GigaZ is taken from Ref.~\cite{Baak:2014ora}. There is no discussion of $\sigma_{A_\tau}$ at future lepton colliders in the literature; here we simply assume a value of $1/3$ of the current systematic error of $A_\tau$. We expect that future lepton collider measurements would reduce the statistical error to be negligible compared to the systematic error. 
Changing the phase of $\delta r_\tau$ does not appreciably affect the results.

To summarize: a new physics Yukawa coupling $|Y|>1.1$ implies that the Higgs effective potential goes unstable on a scale about three orders of magnitude (or less) above the vector-like lepton masses; $|Y|>2$ leads to instability about one order of magnitude above the vector-like lepton scale. Absence of signals in future $Z$-pole precision measurements would exclude the pure fermion model from generating $|\delta r_\tau| > 0.2$.

\subsubsection{Numerical verification for very light vector-like states}\label{sssec:num}
Vector-like leptons, especially if they only mix with the $\tau$ and not with the electron or muon, are experimentally allowed to be quite light (see App.~\ref{app:collider}). We therefore turn to examine numerically the validity of Eq.~(\ref{eq:drtauex1}) for very light vector-like states, with the goal of validating Fig.~\ref{fig:ZZHHtau}. To do this, we numerically scan over the parameter space of the model in Eq.~(\ref{eq:LDS}). To make the scan efficient, we reparameterise the model by writing the $3\times3$ lepton mass matrix $\mathcal{M}$ as
\be\label{eq:scan}-\mathcal{L}&=&\left(l^-\,L^-\,E^-\right)\mathcal{M}\left(\begin{array}{c}e^{c+}\\E^{c+}\\L^{c+} \end{array}\right),\;\;\;\;\mathcal{M}=VDU,\\
V&=&e^{i\theta_v\Lambda_7}e^{i\gamma_v\Lambda_3}e^{i\beta_v\Lambda_2}e^{i\phi\Lambda_3},\;\;\;\;\;U=e^{i\beta_u\Lambda_2}e^{i\gamma_u\Lambda_3}e^{i\theta_u\Lambda_7},\no
\ee
where $D={\rm diag}\left(m_\tau,M_1,M_2\right)$ is a positive-definite eigenvalue matrix (with $M_2\geq M_1$) and $V$ and $U$ are unitary diagonalization matrices that we express using a convenient basis of the Gell-Mann matrices $\Lambda$, with real parameters $\beta_{v,u},\theta_{v,u},\gamma_{v,u},\phi$. This parametrization is convenient and minimal, making use of the [SU(2)$\times$U(1)]$^2$ global symmetry of the gauge-kinetic terms in the action. It allows us to enter physical mass eigenvalues as input, and because the physical $Z\tau\tau$ couplings are given simply by $\delta g_L=\frac{1}{2}\left|V_{31}\right|^2$ and $\delta g_R=-\frac{1}{2}\left|U_{13}\right|^2$, allows us to skip phenomenologically unacceptable model  points from the outset. Moving back and forth between the basis we use for the numerical scan, Eq.~(\ref{eq:scan}), and the basis of Eq.~(\ref{eq:LDS}) is straightforward, and we do it to evaluate the size of the Yukawa couplings as defined in Eq.~(\ref{eq:LDS}) in order to make the connection with vacuum stability. In the scan we omit CP-violating phases, setting $\phi=\gamma_v=\gamma_u=0$. Besides from this simplification, we scan over the physical heavy mass eigenvalues $M_1$ and $M_2$ and over the angles $\theta_{v,u}$ and $\beta_{v,u}$. For each point in the scan, we evaluate $\delta r_\tau$ and $\delta g_{V\tau,A\tau}$ numerically, and, in addition, find the value of the maximal Yukawa coupling as given in the basis of Eq.~(\ref{eq:LDS}).

\begin{figure}[h]
\begin{center}
\includegraphics[width=0.45\textwidth]{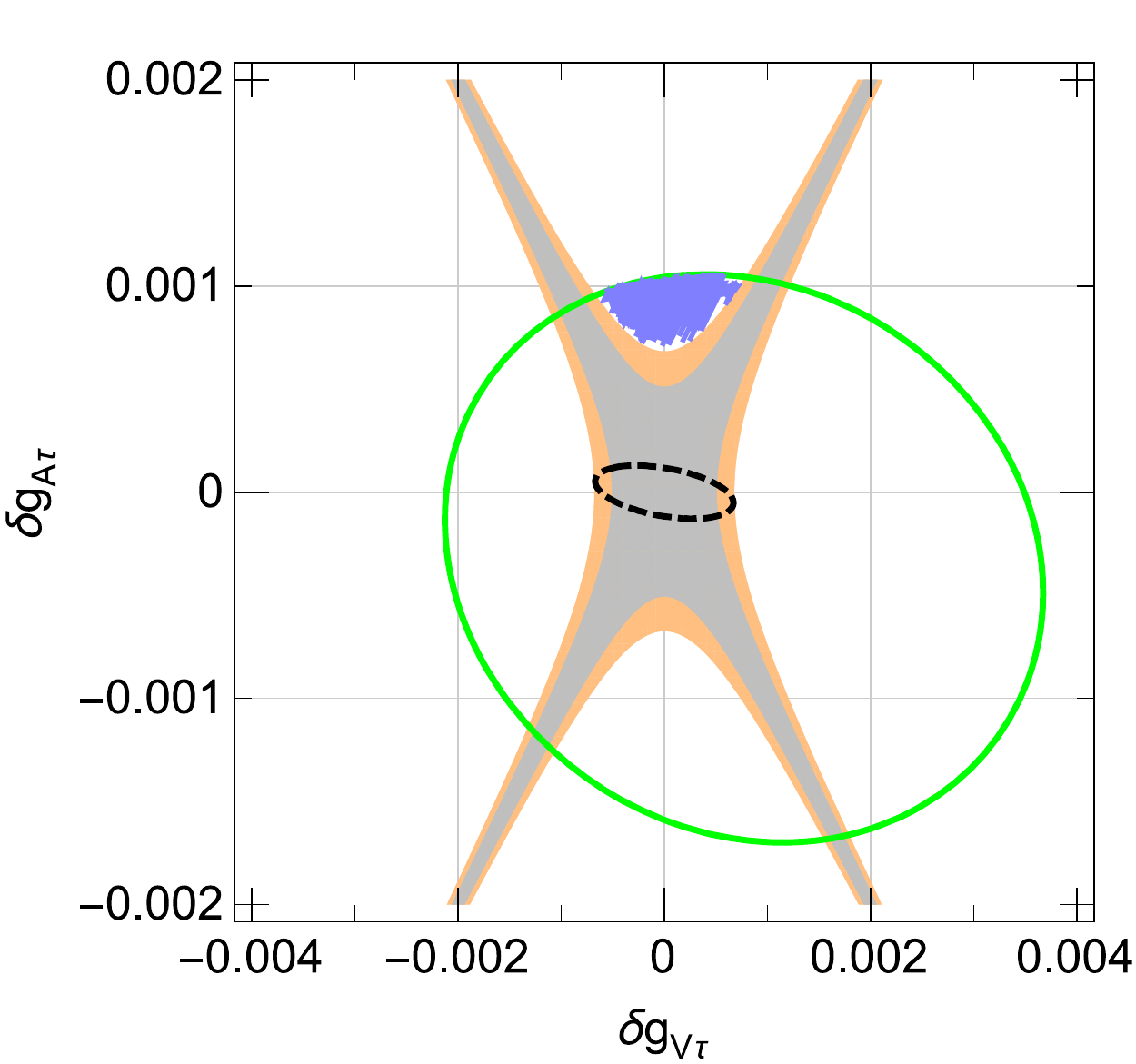}\quad
\caption{Same as in Fig.~\ref{fig:ZZHHtau}, but showing in blue points the results of the scan described in the text to validate the EFT analysis for low-mass vector-like states, where we impose $|Y|<$~1.5. We scan on the parameter space of the model, allowing the lightest vector-like fermion state to be as light as 200~GeV. 
}
\label{fig:ZZHHtaunum}
\end{center}
\end{figure}

The results of the scan are given in Fig.~\ref{fig:ZZHHtaunum}, where we repeat the calculation of Fig.~\ref{fig:ZZHHtau} and superimpose as blue points all of the scan points in which $\delta r_\tau>0.2$, the maximal Yukawa coupling is smaller than 1.5, and the $Z\tau\tau$ couplings are consistent with current data at the 95\%CL. Thus, the scan points in Fig.~\ref{fig:ZZHHtaunum} should complement the orange shaded region, within the 95\%CL contour. As can be seen from the plot, the results of the analytical study are confirmed by the numerical scan, and apply also for very light vector-like lepton states, here allowed to be as light as 200~GeV, with sizable Yukawa couplings. 

As final comments, we note that the occurrence of the scan points in the top region in Fig.~\ref{fig:ZZHHtaunum} is due to the exact relations $\delta g_V=\frac{1}{2}\left(\left|V_{31}\right|^2-\left|U_{13}\right|^2\right)$ and $\delta g_A=\frac{1}{2}\left(\left|V_{31}\right|^2+\left|U_{13}\right|^2\right)$. Other vector-like lepton representations that mix with the tau satisfy analogous expressions to Eq.~(\ref{eq:drtauex1}), that can be found in App.~\ref{app:Hqq}, Tab.~\ref{tab:lHZ}. Requesting a large $\delta r_\tau$ and imposing an upper bound to the Yukawa couplings, all of these representations are also restricted to specific regions in the $\delta g_{V\tau},\delta g_{A\tau}$ plane: rep' $LI$ and $LIII$ populate the top region shown by the blue points in Fig.~\ref{fig:ZZHHtaunum}; rep' $LII$ and $LIV$ populate the middle-right region, with $\delta g_{V\tau}\geq0$ and $\delta g_{A\tau}$ around zero; and rep' $LV$ populates the middle-left with $0\geq\delta g_{V\tau}$ and $\delta g_{A\tau}$ around zero. In all cases, the shaded orange or gray regions of Fig.~\ref{fig:ZZHHtau} indicate large Yukawa couplings and, therefore, vacuum instability. 

The numerical analysis that we report here applies with obvious changes to the $Hbb$ case as well, and demonstrates the validity of Eq.~(\ref{eq:drbex1}) on which we based the discussion in Sec.~\ref{ssec:Hqq}.

\subsection{$Htt$}
\label{ssec:Htt}

The top quark Yukawa coupling in the SM is $\mathcal{O}(1)$. Combined with the existing collider limits on fermion top partners, this means that large Yukawa couplings are necessary to give appreciable corrections to the $Htt$ vertex through mixing with vector-like fermions, implying a strong vacuum stability constraint. 

Consider the vector-like fermion representation  
\be Q(3,2)_{\frac{1}{6}},\;\;Q^c(\bar{3},2)_{-\frac{1}{6}},\;\;U(3,1)_{\frac{2}{3}},\;\;U^c(\bar{3},1)_{-\frac{2}{3}}\ee 
with the potential 
\be\label{eq:QUSexample}\mathcal{V}_{NP}=Y_{Qu^c}H^T\epsilon Qu^c+Y_{qU^c}H^T\epsilon qU^c+Y_{QU^c}H^T\epsilon QU^c+Y_{Q^cU}H^\dag Q^cU+M_QQ^T\epsilon Q^c+M_UUU^c+cc.\ee  This is rep' $UI$ in App.~\ref{app:Hqq}. Due to LHC constraints, the vector masses $M_Q,\,M_U$ must be larger than about 700~GeV. Integrating out the heavy fields, we find 
\be\label{eq:drtex}\delta r_t
&\approx&-v^2\left(\frac{|Y_{qU^c}|^2}{2|M_U|^2}+\frac{|Y_{Qu^c}|^2}{2|M_Q|^2}+\frac{Y_{Q^cU}Y_{Qu^c}Y_{qU^c}}{M_QM_U}\right),\ee
which implies
\be\label{eq:drtexlim}
|\delta r_t|&\lesssim&0.06\left(\left|\frac{Y_{qU^c}\,({\rm 700~GeV})}{M_U}\right|^2+\left|\frac{Y_{Qu^c}\,({\rm 700~GeV})}{M_Q}\right|^2+2|Y_{Q^cU}|\left|\frac{Y_{Qu^c}Y_{qU^c}\,({\rm 700~GeV})^2}{M_QM_U}\right|\right).\ee

We can make a conservative estimate of the maximum deviation in $Htt$, consistent with a pure fermion model, by letting all of the new Yukawa couplings be $\sim1$, tuning their phases to interfere constructively as in Eq.~(\ref{eq:drtexlim}), and at the same time allowing $M_Q,M_U\sim700$~GeV. This gives $|\delta r_t|<0.25$ and, judging from Fig.~\ref{fig:QDSLam2M1} that represents similar RGE, implies $\Lambda_{\rm UV}\lesssim100$~TeV. 

Note that the representation above gives a non-minimal example in terms of the field content. In contrast to the $Hbb$ and $H\tau\tau$ cases, where this possibility is precluded by the stringent $Z$-pole constraints, here we could omit one of the pairs of vector-like fermions (either the $Q,Q^c$ or the $U,U^c$ pair) and still produce a potentially sizable deviation in the $Htt$ vertex\footnote{Ref.~\cite{Brod:2014hsa} derived indirect constraints on the $Ztt$ couplings that read $|\delta g_{Vt}|,\,|\delta g_{At}|\lesssim0.1$. Thus a partial vector-like fermion model with either $Q,Q^c$ or $U,U^c$, but not both, could in principle induce $|\delta r_t|\sim0.2$, compatible with electroweak precision data, but at the cost of vacuum instability as shown in the text.}. The result would correspond to Eq.~(\ref{eq:drtexlim}) with one of $Y_{Qu^c}$ or $Y_{qU^c}$ set to zero. This does not change the conclusion. Finally, all of the vector-like fermion models listed in App.~\ref{app:Hqq} yield similar results to the example we analyzed here.

\section{Higgs couplings to massless gauge bosons}
\label{sec:HAA}
We now study the implications of vacuum stability constraints on $HGG$ and $H\gamma\gamma$ coupling modifications induced by vector-like fermions. In this section, we ignore the mixing of the new fermions with the SM fermions, which was analyzed in Sec.~\ref{sec:Hff}. Vacuum stability constraints in vector-like fermion models modifying $H\gamma\gamma$ and $HGG$ were considered in~\cite{ArkaniHamed:2012kq} and ~\cite{Reece:2012gi}. Our results here refine those analyses using the two-loop RG improved Coleman-Weinberg potential, as  described in App.~\ref{app:cw}.

The leading-log contribution of the new fermions to the $HGG$ and $H\gamma\gamma$ amplitudes can be derived from the Higgs low energy theorem~\cite{Ellis:1975ap, Shifman:1979eb}. For a fermion carrying electric charge $Q$, and transforming under SU(3)$_c$ in a representation of dimension $D$ with Dynkin index $T$, we have
\be
\delta r_\gamma&\approx&\frac{4\,Q^2\,D}{3\mathcal{A}_{SM}^\gamma}\left(\frac{\partial\log ||M||}{\partial\log v}\right),\label{eq:drgam}\\
\delta r_G&\approx&2T\left(\frac{\partial\log ||M||}{\partial\log v}\right),\label{eq:drG}\ee
where $||M||$ is the absolute value of the determinant of the fermion mass matrix $M$ and $\mathcal{A}_{SM}^\gamma\approx-6.5$ is the SM $H\gamma\gamma$ amplitude. Our convention is such that $T(3)=1/2$.

The minimal building block of interest for our vector-like fermion representation is~\cite{ArkaniHamed:2012kq}\footnote{For real color representations and $Q=0$ or $Q=\frac{1}{2}$, we could set $\chi=\chi^c$ or $\psi=\psi^c$.}\footnote{We could promote $\chi$ and $\chi^c$ to SU(2)$_W$ triplets, which would add more fields to the RGE and would not change our conclusions.}  
\be\psi(D,2)_{-Q+\frac{1}{2}},\;\psi^c(\bar D,2)_{Q-\frac{1}{2}},\;\chi(\bar D,1)_{Q},\;\chi^c(D,1)_{-Q}, \label{eq:reploop}\ee
with the mass matrix for charge $Q$ states
\be\label{eq:GNP}\mathcal{V}_{NP}=\left(\psi^{-Q}\;\chi^{c-Q}\right)\left(\begin{array}{cc}\frac{Yv}{\sqrt{2}}&-m_\psi\\m_\chi&-\frac{Y^cv}{\sqrt{2}}\end{array}\right)\left(\begin{array}{c}\chi^{+Q}\\\psi^{c+Q}\end{array}\right),\ee
corresponding to the potential given  in App.~\ref{app:cw}, Eq.~(\ref{eq:CWexample}). 
In what follows we ignore the physical complex phase $\arg(m_\psi^*m_\chi^* YY^c)$ and assume a basis where all of the parameters are real. We further choose $m_\psi,m_\chi$ to be positive. 
Denoting the mass eigenvalues as $M_1$ and $M_2$ with $0<M_1\leq M_2$, the log-determinant derivative is given by
\be\label{eq:ld}\left(\frac{\partial\log ||M||}{\partial\log v}\right)=-\frac{YY^cv^2}{M_1M_2}\,{\rm sign}\left(|M|\right).
\ee
Note that $|M|=m_\psi m_\chi-\frac{v^2}{2}YY^c$, so that a negative sign for $|M|$ in our basis requires large Yukawa and small vector-like fermion masses. For $m_\chi,m_\psi>174$~GeV, the product $YY^c$ must be larger than unity to have negative $|M|$, so that such configurations are automatically associated with low-scale vacuum instability.

Before proceeding to the numerical results, we comment on the effects of NLO corrections.
The first NLO effect involves two-loop diagrams that contribute to the Wilson coefficients of the effective dimension six $|H|^2GG$ and $|H|^2FF$ operators at the vector-like fermion mass threshold. These corrections may be sizable, of order tens of percent, especially for the $HGG$ case (see, e.g.~\cite{Djouadi:2005gi} for the analogous effect in the integration out of the SM top quark). However, note that as long as we restrict our discussion to fermions in the fundamental representation of color, the dominant effect is a multiplicative factor that acts on the SM top quark amplitude and new physics LO contribution alike, and so drops out in the relative correction to the $HGG$ vertex that we discuss here~\footnote{For an NNLO computation in a related model (partial vector-like quark representations), confirming these statements, see~\cite{Dawson:2012di}.}. Once we consider higher color representations (in the second part of Sec.~\ref{sec:HGG} below), the NLO K-factor ceases to be a common multiplicative effect for the SM and new physics contributions, and would alter our results to some extent.  
The second NLO effect pertains to the RGE running of the nonrenormalizable operators from the vector-like fermion mass scale down to $m_h$ in the SM EFT~\cite{Grojean:2013kd}. We have checked that the corresponding correction is limited to a few percent, and we omit it in what follows.

\subsection{$HGG$}\label{sec:HGG}

\begin{figure}[htbp]
\begin{center}
\includegraphics[width=0.5\textwidth]{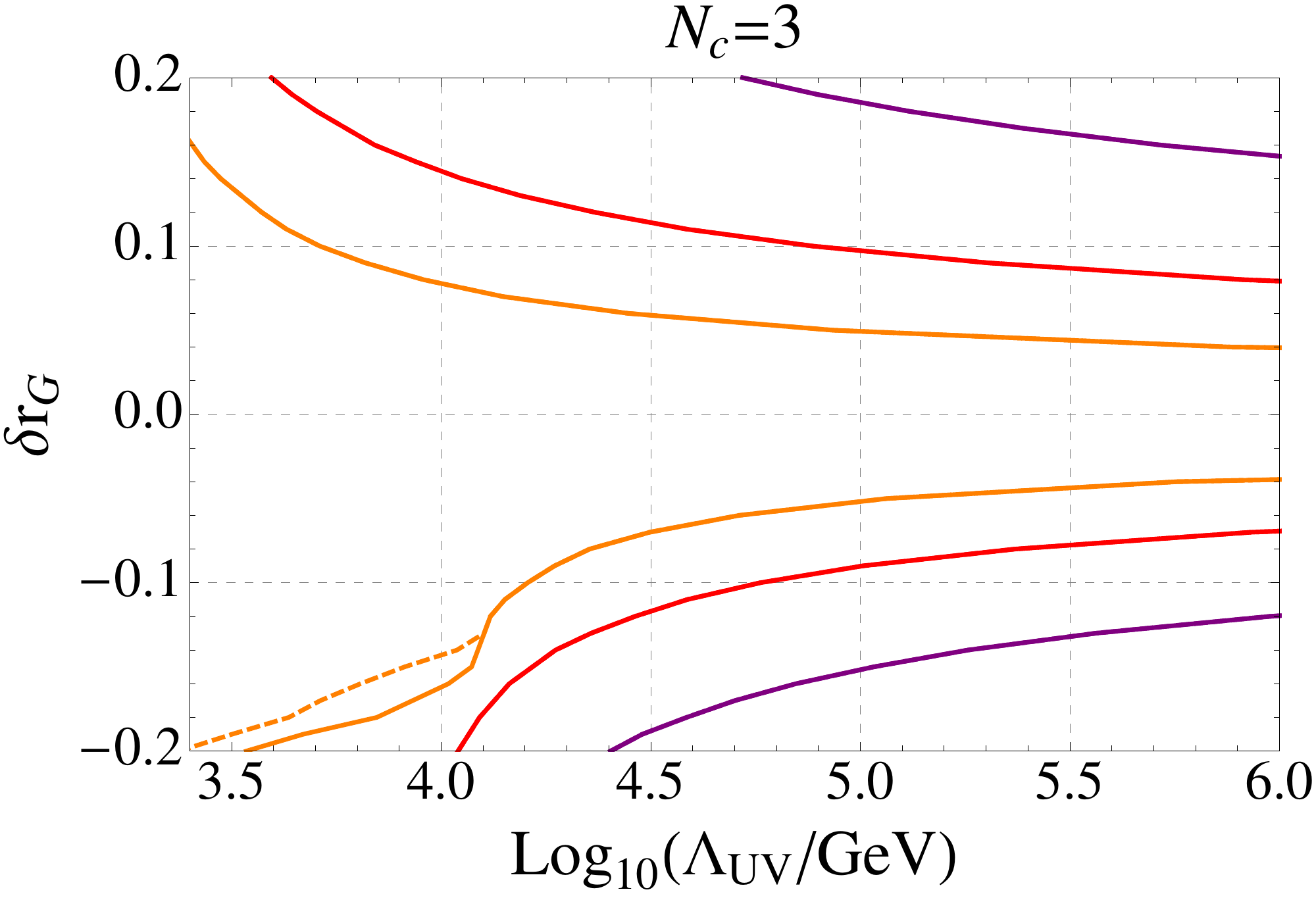}\quad
\caption{The maximum $HGG$ coupling deviation vs. the corresponding instability scale $\Lambda_{UV}$, obtained by adding vector-like fermions in the 3 representation of SU(3)$_c$. Purple, red, and orange lines correspond to fixing $M_1=0.5,0.7$ and 1~TeV, respectively. The dashed line, that is only visible for the $M_1=1$~TeV case at large negative $\delta r_G$, demonstrates the sensitivity of the result in this parameter region to our criterion for perturbativity; see text for details.}
\label{fig:drG3}
\end{center}
\end{figure}

In Fig.~\ref{fig:drG3} we show the maximal deviation $\delta r_G$ as a function of the vacuum instability scale $\Lambda_{UV}$, assuming a vector-like pair of fermions in the fundamental representation of SU(3)$_c$. The smooth purple, red, and orange curves correspond to setting the mass of the lighter colored fermion $M_1$ to 0.5, 0.7 and 1~TeV, respectively. 

In producing Fig.~\ref{fig:drG3} we use the following method. Typically, the most conservative vacuum stability constraint (largest $\Lambda_{\rm UV}$) is achieved when the new Yukawa couplings are as small as possible for fixed $\delta r_G$ and $M_1$. 
Considering $\delta r_G>0$, we see that Eqs.~(\ref{eq:drG}) and~(\ref{eq:ld}) require $YY^c<0$. The smallest consistent choice for $|Y|$ and $|Y^c|$, that we employ in Fig.~\ref{fig:drG3}, then corresponds to $Y\approx -Y^c$, with $M_1\approx M_2$ and as small as is allowed by direct collider searches for colored fermions. 
Considering $\delta r_G<0$, we need $YY^c>0$. The most conservative configuration for the vacuum stability analysis is different than that in the $\delta r_G>0$ case, due to the inequality
\be\label{eq:Gc}\left(M_2-M_1\right)^2\geq\frac{\left(Y+Y^c\right)^2v^2}{2}\ee
which can be derived from Eq.~(\ref{eq:GNP}). Due to Eq.~(\ref{eq:Gc}) we cannot tune $Y\approx Y^c$ and $M_1\approx M_2$ at the same time. Instead, the most conservative configuration for the stability analysis in this case, that we employ in Fig.~\ref{fig:drG3}, corresponds to saturating Eq.~(\ref{eq:Gc}) with $Y\approx Y^c$, $|Y|\approx\sqrt{-\frac{\delta r_G M_1M_2}{2T\,v^2}}$, and  
$M_2=M_1\left[1-\frac{\delta r_G}{2T}+\sqrt{-\frac{\delta r_G}{2T}\left(2-\frac{\delta r_G}{2T}\right)}\right]$. This explains the asymmetry of $\Lambda_{UV}$ for fixed in Fig.~\ref{fig:drG3} between positive and negative values of $\delta r_G$. 

We comment that the intuition by which larger $\Lambda_{\rm UV}$ corresponds to smaller Yukawa couplings, that we used to fix the model parameters in Fig.~\ref{fig:drG3}, holds well as long as the instability threshold occurs sufficiently far from the vector-like fermion mass threshold. When the instability occurs immediately above the vector-like fermion mass scale, we find in some cases that larger Yukawa couplings can lead to a slightly higher instability scale, due to threshold effects. We stress, however, that these effects only become relevant when the instability scale is very low in the first place, $\Lambda_{UV}<10$~TeV. We comment about these effects further in the next section.
%

In producing Fig.~\ref{fig:drG3}, we restrict our calculation to the region of parameter space where perturbation theory is under reasonable control by imposing $|Y|,|Y^c|<\frac{4\pi}{\sqrt{4D}}$. 
For fermions with $D=3$ we thus impose $|Y|,|Y^c|<3.6$. For most of the curves shown in Fig.~\ref{fig:drG3}, all of the couplings remain perturbative at all RGE scales up to the vacuum instability scale $\Lambda_{\rm UV}$. An exception occurs for the $M_1=1$~TeV example with large negative $\delta r_G<-0.1$. Here we find that $Y$ and $Y^c$ run large with increasing RGE scale, and cross the perturbativity threshold defined above before our nominal criterion for vacuum instability applies. When this happens, we redefine $\Lambda_{\rm UV}$ as the scale where the perturbativity threshold was crossed, which leads to the kink in the plot. Of course, this procedure is somewhat arbitrary; the main lesson is that for such large negative $\delta r_G$, the model runs strongly-coupled quickly, making our perturbative estimates less reliable. To highlight the sensitivity of our results in this parameter region to changing the perturbative prescription, we superimpose as a dashed line the result obtained when modifying the numerical perturbativity threshold by 20\%.

Higher color representations have a larger Dynkin index and thus reduce the size of the Yukawa couplings needed for a given $\delta r_G$, relaxing the vacuum stability constraint. At the same time, collider constraints on $M_1$ become stronger (see App.~\ref{app:collider}), balancing the effect to some extent. 

\begin{figure}[htbp]
\begin{center}
\includegraphics[width=0.45\textwidth]{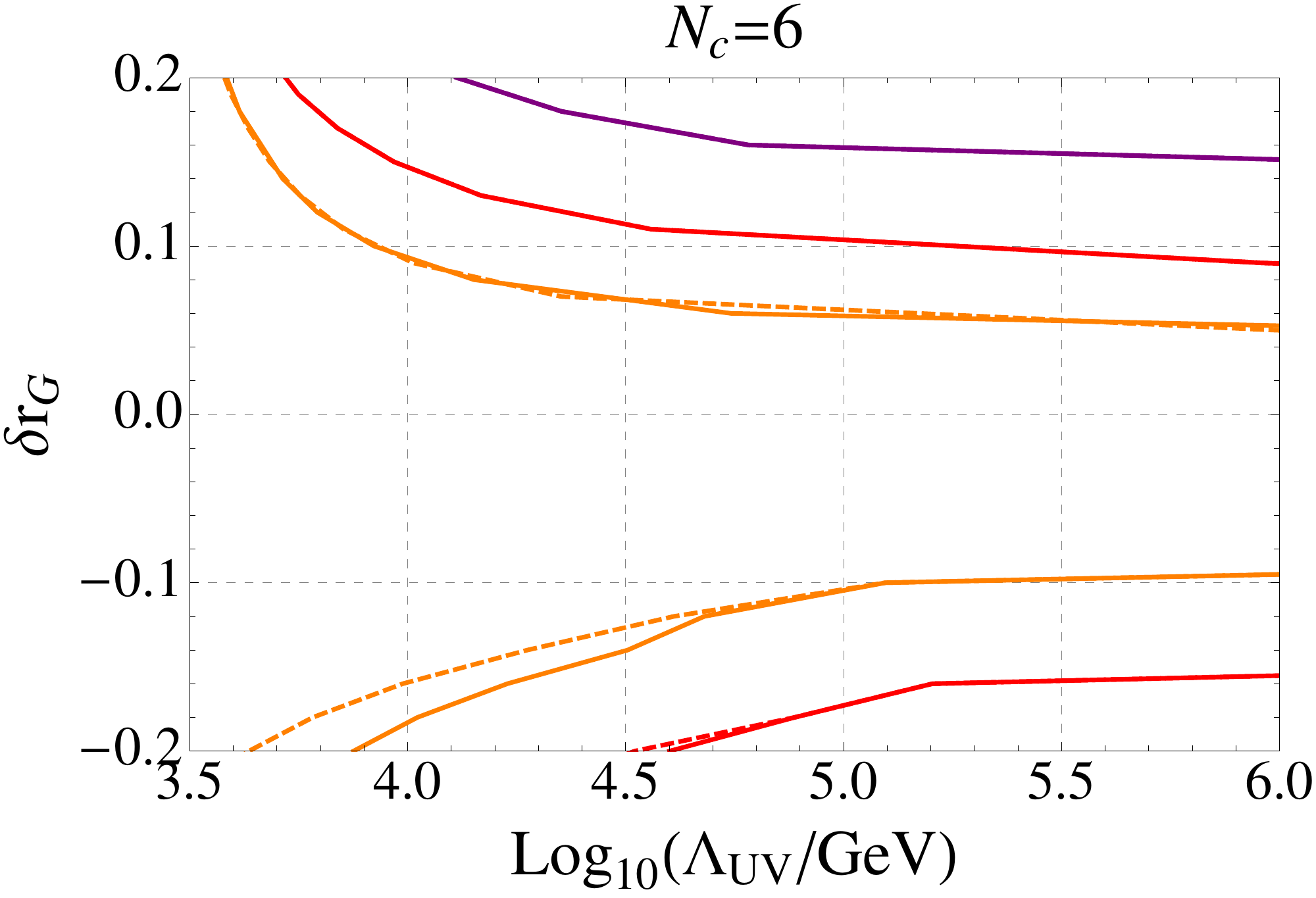}\quad
\includegraphics[width=0.45\textwidth]{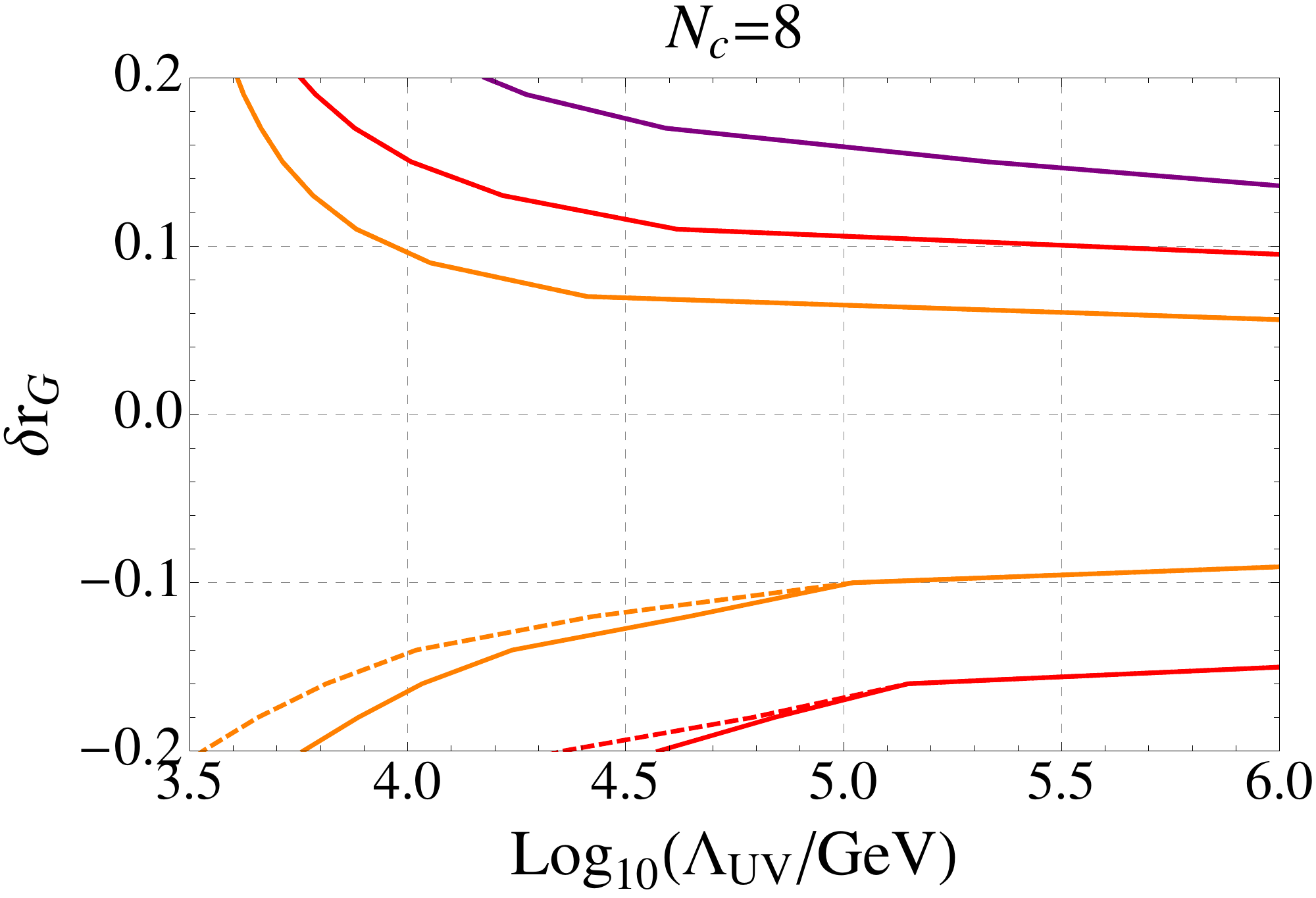}\quad
\caption{The maximum $HGG$ coupling deviation vs. the corresponding instability scale $\Lambda_{UV}$, obtained by adding vector-like fermions in the 6 representation of SU(3)$_c$ (left) and in the 8 representation of $SU(3)_c$ (right). Purple, red, and orange lines correspond to fixing $M_1=1.2,1.5$ and 2~TeV, respectively. The dashed lines correspond to reducing the perturbativity constraint by $20\%$. }
\label{fig:drG6}
\end{center}
\end{figure}

In Figure~\ref{fig:drG6} we show the vacuum stability bound for vector-like fermions in the $D=6$ ($T=5/2$, left panel) and $D=8$ ($T=3$, right panel) representations, using Eq.~(\ref{eq:reploop}) with $Q=1$, and taking $M_1=1.2, 1.5$ and $2$~TeV. We do not consider representations of $D>8$, as these cause the SU(3)$_c$ gauge coupling $g_3$ to run strong quickly and hit a Landau pole on scales very close to the vector-like fermion mass. 

Once again, in Fig.~\ref{fig:drG6} we restrict our calculation to the region of parameter space where $|Y|,|Y^c|<\frac{4\pi}{\sqrt{4D}}$. For $\delta r_G>0$, all couplings are perturbative according to this criterion up to the vacuum instability scale. For large negative $\delta r_G$, however, in both the $D=6$ and $D=8$ examples, the Yukawa couplings run strong and cross our perturbative reliability criterion at a scale lower than the vacuum instability. We indicate where this happens by showing, in dashed lines, the results obtained while modifying the perturbativity criterion by 20\%.

\subsection{$H\gamma\gamma$}\label{ssec:Hgg}

The largest effect in $H\gamma\gamma$ at a given stability cut-off $\Lambda_{UV}$ is obtained with color singlet fermions. This is because LHC searches put strong mass constraints on exotic colored fermions, whereas color singlets are still allowed to be quite light (see App.~\ref{app:collider}). For example, an exotic color octet vector-like representation would give $D=8$ in Eq.~(\ref{eq:drgam}), but this would come at the price of a large mass suppression, $M_1\gtrsim1$~TeV implying $\left|\left(\frac{\partial\log ||M||}{\partial\log v}\right)\right|\lesssim0.06$ even with large Yukawa couplings $Y, Y^c\sim1$. In contrast, vector-like leptons with $D=1$ are still allowed with $M_1\sim200$~GeV and $\left|\left(\frac{\partial\log ||M||}{\partial\log v}\right)\right|\sim1$ for the same $Y, Y^c\sim1$. The increase in $\left(\frac{\partial\log ||M||}{\partial\log v}\right)$ more than compensates for the decrease in $D$.
Therefore in what follows we focus on color singlet representations.

Given a value for $Q$ and having set $D=1$, we can calculate the maximal $\Lambda_{UV}$ for a given $\delta r_\gamma$ in analogy with the $\delta r_G$ case. 
Typically, the maximal instability scale for $\delta r_\gamma < 0$ is obtained for $Y\approx -Y^c$ and $M_1\approx M_2$, while for $\delta r_\gamma > 0$ we have it for $Y\approx Y^c$ and $M_2=M_1\left(1+x+\sqrt{x\left(2+x\right)}\right)$ with $x=\left|3 \mathcal{A}_{SM}^\gamma \delta r_\gamma/4 Q^2 D\right|$. We comment again that these relations are borne out in our calculation as long as the instability scale is sufficiently far from the vector-like fermion mass threshold, and imply simply that larger Yukawa couplings trigger an earlier vacuum instability. In the extreme case in which the instability scale occurs very close to the vector-like mass, we find that threshold effects can conspire to make a slightly higher instability scale pair with slightly larger Yukawa couplings. 

\begin{figure}[h]
\begin{center}
\includegraphics[width=0.45\textwidth]{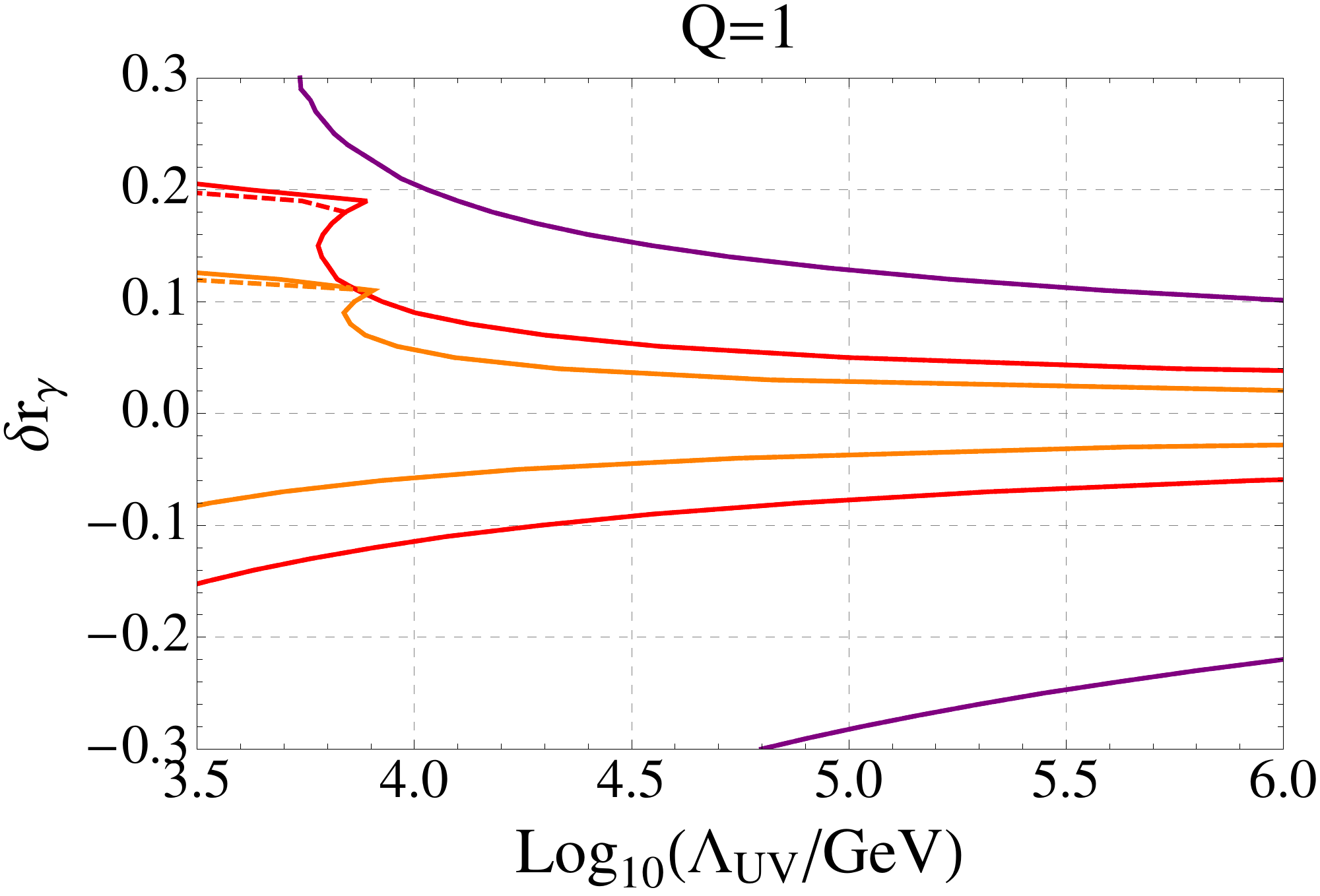}\quad \includegraphics[width=0.45\textwidth]{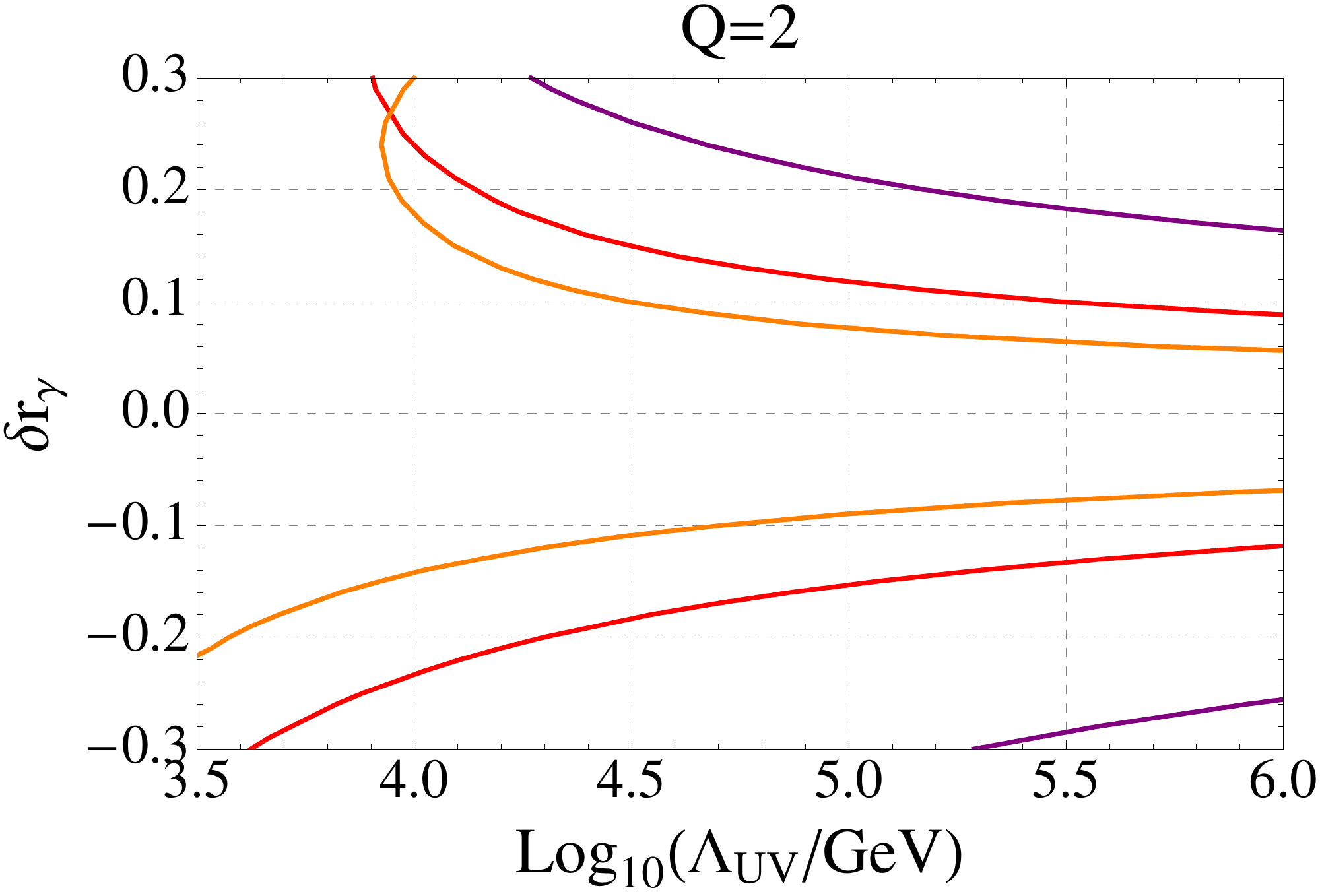}\quad
\caption{Left panel: the maximum $H\gamma\gamma$ coupling deviation vs. the corresponding instability scale $\Lambda_{UV}$, obtained by adding vector-like leptons with charged states carrying $Q=1$. Purple, red, and orange lines correspond to fixing $M_1=200, 400$ and 600~GeV, respectively. The dashed lines correspond to modifying the perturbativity constraint by $20\%$. Right panel: same as on the left, but for $Q=2$ and $M_1=400, 600$ and 800~GeV.}
\label{fig:drgamLDS}
\end{center}
\end{figure}

In the left panel of Fig.~\ref{fig:drgamLDS} we plot $\delta r_\gamma$ vs. $\Lambda_{UV}$ for a vector-like lepton representation, given by  Eq.~(\ref{eq:reploop}) with $D=1$ and $Q=1$. Purple, red, and orange lines correspond to fixing $M_1=200, 400$ and 600~GeV, respectively. In the right panel we set $Q=2$ and plot the results for $M_1=400, 600$ and 800~GeV. 
The general trend seen in Fig.~\ref{fig:drgamLDS} is that increasing $|\delta r_\gamma|$ implies a lower instability scale $\Lambda_{\rm UV}$. When the instability scale is very low, however -- lower than 10~TeV in these examples -- we note that some of the lines in Fig.~\ref{fig:drgamLDS} curve backwards and indicate a slightly larger $\Lambda_{\rm UV}$ for larger value of $\delta r_\gamma$. This is the threshold effect that was mentioned in the previous paragraph; again, this behavior is only apparent where $\Lambda_{\rm UV}$ is very low in the first place.

As already seen in the case of colored fermions, when the Yukawa couplings defined at the vector-like fermion threshold scale are sufficiently large, then their subsequent RGE leads to a break down of our perturbative calculation on a scale smaller than the nominal vacuum instability scale. When this happens, we repeat our procedure of the previous section and define $\Lambda_{\rm UV}$ as the scale at which either $|Y|$ or $|Y^c|$ reaches a magnitude $\frac{4\pi}{\sqrt{4D}}$. This is visible in the left panel of Fig.~\ref{fig:drgamLDS} for $M_1=400$ and 600 GeV and $\delta r_\gamma \gtrsim0.1$ and 0.2 respectively. The dashed lines show the results when modifying the perturbativity benchmark $4\pi/\sqrt{4D}$ by 20\%.  

\section{Conclusions}
\label{sec:sum}
In this paper we studied constraints on theories beyond the Standard Model, containing new physics fermions but no new scalar or vector-boson particles up to a very high scale. The basic constraint we discuss is due to the vacuum stability of the Higgs effective potential. In the presence of new fermions coupled to the SM through Yukawa interactions, the renormalization group evolution of the Higgs quartic coupling is negative definite, leading to vacuum instability. 

We surveyed Higgs couplings to the bottom and top quarks, to the tau lepton, and to gluons and photons, focusing on one particular coupling at a time. Our results imply that measuring a deviation at the level of 10\% or so in any of these couplings (combined with null detection of $Z$ coupling deviations in the $Hbb$ or $H\tau\tau$ case) would suggest a low cut-off scale $\Lambda_{\rm UV}$ for pure fermion models, below about 100~TeV, where new bosonic states must stabilize the Higgs potential. While we did not investigate here the question of the actual mass scale for these bosons, the decoupling theorem~\cite{Appelquist:1974tg} leads us to expect that these states should occur not far above $\Lambda_{\rm UV}$. For large deviations [still at the $\mathcal{O}(10\%)$ level], the vacuum instability occurs essentially immediately above the fermion mass threshold, implying that the pure fermion model is not self consistent even as an effective theory.  

Our main numerical results are summarized below. To derive these results, that are expressed as consistency relations between Higgs coupling deviations and the vacuum stability cut-off $\Lambda_{\rm UV}$ for pure fermion models, we have judiciously tuned the parameters of the models in order to obtain conservative estimates, namely, maximal $\Lambda_{\rm UV}$ per given Higgs coupling effect. For more generic parameter configurations we would expect the vacuum instability scale to occur earlier than these conservative estimates.
\begin{itemize}
\item $Hbb$ and $H\tau\tau$ 
\begin{itemize} 
\item
For pure fermion models, vacuum stability constraints imply that measuring an $Hbb$ deviation $|\delta r_b|\gtrsim0.2$ would require a corresponding deviation in the $Zbb$ couplings at the permille level, well within the expected sensitivity of future precision experiments with ILC-like capabilities. Finding a large $Hbb$ deviation without an accompanying deviation in $Zbb$ can therefore be used to rule out pure fermion models. The result for the $H\tau\tau$ coupling is similar. 
\item
Currently available $Z$-pole data from LEP imply that partial vector-like fermion representations -- models in which new fermions mix with either the SM left-handed quark, or the SM right-handed quark, but not both -- cannot induce $|\delta r_b|>0.1$. This result is independent of vacuum stability arguments. The result for the $H\tau\tau$ coupling is similar. 
\end{itemize}
\mbox{} \\
\item
$Htt$
\begin{itemize}
\item 
Vacuum stability constraints imply an upper bound on the possible $Htt$ coupling deviation in pure fermion models that we estimate, conservatively, as $|\delta r_t|\lesssim0.25$ for $\Lambda_{\rm UV}>100$~TeV.
\item 
In contrast to the $Hbb$ and $H\tau\tau$ examples, current precision electroweak data put only mild constraints on the effective $Ztt$ coupling, meaning that a partial vector-like fermion representation, mixing with either left- or right-handed SM quarks but not both, could still induce a sizable $Htt$ deviation. In such case, however, there should be a corresponding comparable deviation in $Ztt$.
\end{itemize}
\item $HGG$
\begin{itemize}
\item
Vacuum stability constraints on $HGG$ deviations induced by pure fermion models depend crucially on the experimental mass limits on the new states. Allowing for color triplet fermions as light as 500~GeV, we find that imposing that the vacuum instability scale $\Lambda_{\rm UV}>100$~TeV implies $|\delta r_G|\lesssim0.2$. If the new states are more massive than 1~TeV, imposing $\Lambda_{\rm UV}>100$~TeV implies $|\delta r_G|\lesssim0.05$.  
\item 
High color representations are also constrained. Allowing color octets or sextets as light as 1.2~TeV, we find that $\Lambda_{\rm UV}>100$~TeV implies $\delta r_G\lesssim0.2$, with somewhat weaker constraints on the magnitude of negative $\delta r_G$. If the new states are more massive than 2~TeV, the same $\Lambda_{\rm UV}$ bound implies $|\delta r_G|\lesssim0.1$.
\end{itemize}
\item $H\gamma\gamma$
\begin{itemize}
\item 
Considering new fermions with electromagnetic charge $|Q|\leq1$, and allowing for states as light as 200~GeV, we find that $\Lambda_{\rm UV}>100$~TeV implies $-0.3\lesssim\delta r_\gamma\lesssim0.15$. If the new states are more massive than 400~GeV, the same $\Lambda_{\rm UV}$ bound implies $-0.1\lesssim \delta r_\gamma\lesssim0.05$.
\item
Considering new fermions with exotic charge $|Q|=2$, and allowing for states as light as 400~GeV, we find that $\Lambda_{\rm UV}>100$~TeV implies $\delta r_\gamma\lesssim0.2$, with somewhat weaker constraints on the magnitude of negative $\delta r_\gamma$. If the new states are more massive than 800~GeV, the same $\Lambda_{\rm UV}$ bound implies $|\delta r_\gamma|\lesssim0.1$.
\end{itemize}
\end{itemize}

Finally, a few comments are in order. 
The experimentally accessible signal strength (production cross section times decay branching fraction) in various analysis channels depends on more than one underlying effective Higgs coupling. For example, the $HGG$ coupling affects the signal strength in $H\to\gamma\gamma$ measurements by modifying the gluon fusion production cross section, and the $Hbb$ coupling affects all other signal strength measurements through its effect on the total width and thus on the respective branching fractions. Clearly, moreover, a realistic new physics scenario may involve true deviations in a number of different Higgs couplings, adding to the complexity. In this paper we chose to ignore this complication in the interpretation of Higgs data, assuming simply that the various degeneracies can be resolved sufficiently well. Our theoretical vacuum stability constraints apply therefore to each underlying coupling individually, and should be cast into constraints on signal strengths once new data becomes accessible, during Run-II of the LHC or with future colliders.

Our calculation of the vacuum stability constraints relied on perturbation theory. In some models and some corners of the parameter space, the renormalization of the Yukawa couplings leads to a breakdown of the perturbative calculation on scales below the naively deduced vacuum instability scale. We indicated where this happens in the relevant sections of the $HGG$ and $H\gamma\gamma$ analyses. In practice, when this happens, the relevant scales are very low, in the ballpark of 10~TeV. 

We did not investigate the possibility of adding several different (non-minimal) vector-like fermion representations, and using them, e.g., to add a multiplicity factor in the $HGG$ or $H\gamma\gamma$ analyses or to cancel some of the correlations between Higgs and $Z$ observables in the $Hbb$ or $H\tau\tau$ analyses. Some considerations along these lines can be found in~\cite{ArkaniHamed:2012kq}. Since adding more fermions and more Yukawa couplings tends to make the instability problem more severe, we expect that, without significant accidental cancelations, such non-minimal models would result in vacuum stability constraints that are comparable to or stronger than those we derived here for minimal models.  

Our account of the experimental constraints on the various vector-like fermion models was partial, omitting potentially important constraints from, e.g., electroweak oblique parameters (see e.g.~\cite{ArkaniHamed:2012kq,Dawson:2012di,Batell:2012ca,Kearney:2012zi}) and precision flavor data. Including these additional constraints can only make our results stronger.

Lastly we comment on deviations to the $HWW$ and $HZZ$ couplings. 
In a pure fermion model, contributions to the effective $HWW$ and $HZZ$ couplings arise at the loop level. In contrast to the $HGG$ and $H\gamma\gamma$ cases, however, where we have analyzed the analogous loop contributions, here to estimate the Higgs coupling deviation one should compare a tree level coupling with a loop-suppressed effect. This means that the vacuum stability constraints on pure fermion models are strong, sufficiently strong to imply that observing modifications to the $HZZ$ or $HWW$ couplings with the precision expected at the LHC would most likely rule out any pure fermion model interpretation.

\section*{Acknowledgments}
We thank Nima Arkani-Hamed, Roni Harnik, Paul Langacker, Maurizio Pierini, Matt Reece and Nati Seiberg for useful discussions. KB was 
supported by the DOE grant DE-SC000998, the BSF, and the Bahcall Fellow Membership. RTD was supported by the NSF grant PHY-0907744.

\appendix

\section{The effective potential}\label{app:cw}

In this section we describe our calculation of the Higgs effective potential in the presence of the canonical vector-like fermion representation
\be \psi(D,2)_{-Q+\frac{1}{2}},\;\;\psi^c(\bar{D},2)_{Q-\frac{1}{2}},\;\;\chi(\bar{D},1)_{Q},\;\;\chi^c(D,1)_{-Q},\ee 
adding to the SM the potential  
\be\label{eq:CWexample}\mathcal{V}_{NP}&=&YH^\dag\psi\chi+Y^c H^T\epsilon\psi^c\chi^c+m_\psi\psi^T\epsilon \psi^c+m_\chi\chi\chi^c+cc.\ee
For simplicity, we do not include mixing of the heavy vector-like fermions with SM fermions. While this mixing is important for Higgs coupling deviations, the vacuum stability constraints that we are concerned with in this section are easier to derive without it, and provide constraints that are easily generalized to the mixing case in the phenomenologically interesting regimes. We also restrict our analysis to the case where all of the parameters $m_\psi,m_\chi,Y,$ and $Y^c$ are real.

We use the $\overline{\rm MS}$ one-loop Coleman-Weinberg (CW) potential in the Landau gauge improved by two loop RGE (see e.g.~\cite{Casas:1994us,Casas:1996aq}). Our problem involves two distinct physical scales -- the SM electroweak scale, that we fix here at $M_t\simeq173$~GeV, and the vector-like fermion mass scale, denoted by $M$, that we define as 
\be M\equiv\sqrt{(m_\psi^2+m_\chi^2)/2}.\ee
In most of our analysis, $M\gg M_t$. We thus need to implement explicitly the decoupling of heavy states in the calculation of the  effective potential, as we now explain.

The one-loop effective potential is a function of the classical Higgs field $h_c$ and the $\overline{\rm MS}$ renormalization scale $\mu$. We implement the decoupling of heavy states by defining two versions of the theory, one valid at $\mu<M$ and the other valid at $\mu>M$, and matching between these theories at $\mu=M$. In both of the regimes, $\mu<M$ and $\mu>M$, we define
\be V=V_0+V_1,\ee
where the tree level piece is given by  
\be\label{eq:V0}
V_0=\Omega(t)-\frac{m_h^2(t)}{2}h^2+\frac{\lambda(t)}{4}h^4.
\ee
For the one-loop contribution, on scales $\mu>M$ we use
\be\label{eq:V1UV} V_1=\sum_{i}\frac{n_i}{64\pi^2}m_i^4(h,t)\,\log\left(\frac{m_i^2(h,t)}{\mu^2e^{C_i}}\right),\ee
where the sum on $i$ includes $W,Z,t$, the Higgs $h$ and Goldstone bosons $G^{0,\pm}$, and the heavy vector-like fermions. The $\overline{MS}$ constants $C_i$ are equal to 5/6 for vector bosons and 3/2 for fermions and scalars. The effective number of degrees of freedom $n_i$ are $n_Z=3$, $n_W=6$, $n_{h,G^0}=1$, $n_{G^\pm}=2$, $n_{t}=-12$ and $n=-4D$ for each new fermion mass eigenstate transforming under SU(3)$_c$ in representation $D$. 
All couplings are running as function of the RGE parameter $t=\log\left(\mu/M_t\right)$ with two loop RGE that we derive using the package provided by~\cite{Lyonnet:2013dna}. 
The renormalized Higgs field is $h(t)=\zeta(t)h_c$ with $\zeta(t)=e^{-\int_0^tdt'\gamma(t')}$, where $\gamma$ is the Higgs anomalous dimension. 

On scales $\mu<M$, we replace Eq.~(\ref{eq:V1UV}) by
\be\label{eq:V1IR} V_1=\sum_{i\subset{\rm SM}}\frac{n_i}{64\pi^2}m_i^4(h,t)\,\log\left(\frac{m_i^2(h,t)}{\mu^2e^{C_i}}\right)+c_6h^6.\ee
Here, the sum on $i$ runs over the SM states, excluding the heavy vector-like fermions. The couplings run with beta functions and anomalous dimensions obtained with the SM RGE equations. The $c_6h^6$ term contains the leading non-renormalizable dimension six operator induced by integrating out the heavy states. To obtain it, we expand the contribution of the heavy fermions to Eq.~(\ref{eq:V1UV}) in powers of the Higgs field $h$, fixing $\mu=M$, and reading off the coefficient of $h^6$. The full expression for $c_6$ is not particularly illuminating; in the case $m_\chi=m_\psi=M$ and $Y^c=Y$, for example, it is given by $c_6=DY^6/(960\pi^2M^2)$. We have verified that adding higher order operators (i.e. $c_8h^8$) does not affect our results.

While we use the same symbols for the parameters $\lambda,m_h^2,$ and $\Omega$ in Eq.~(\ref{eq:V0}), above and below the matching scale $\mu=M$, their numerical values are different due to the one-loop threshold corrections. Writing $\lambda(\mu=M+)=\lambda(\mu=M-)+\Delta\lambda$, $m_h^2(\mu=M+)=m_h^2(\mu=M-)+\Delta m_h^2$, and $\Omega(\mu=M+)=\Omega(\mu=M-)+\Delta\Omega$, we have $\Delta\lambda=-4c_4$, $\Delta m_h^2=2c_2$, and $\Delta\Omega=-c_0$, where the corrections $c_n$ are given again by the coefficient of $h^n$ on expanding the vector-like fermion contribution to Eq.~(\ref{eq:V1UV}) in powers of $h$. In the case $m_\chi=m_\psi=M$ and $Y^c=Y$, for example, we have $c_0=3DM^4/(4\pi)^2$, $c_2=2DY^2M^2/(4\pi)^2$, and $c_4=-DY^4/(12\pi^2)$.

We fix the RGE initial conditions for the SM couplings at the scale $\mu=M_t$. While our calculation only requires one-loop matching, for convenience we use the two-loop values for the SM parameters at $\mu=M_t$ as given in Ref.~\cite{Buttazzo:2013uya}. For $m^2_h(t)$ and $\lambda(t)$ we include the non-renormalizable $c_6$ contribution through the correction
\be 
\lambda(0)&=&\lambda^{({\rm SM})}(0)+\frac{1}{2v^2}\left[\frac{1}{v}\left(\frac{\partial \Delta V_1}{\partial h}\right)_{h=v}-\left(\frac{\partial^2 \Delta V_1}{\partial h^2}\right)_{h=v}\right],\label{eq:matchingl}\\
m_h^2(0)&=&m_h^{2({\rm SM})}(0)+\frac{3}{2v}\left(\frac{\partial \Delta V_1}{\partial h}\right)_{h=v}-\frac{1}{2}\left(\frac{\partial^2 \Delta V_1}{\partial h^2}\right)_{h=v},\label{eq:matchingm}\ee
with $\Delta V_1=c_6h^6$ and where $\lambda^{({\rm SM})}(0)$ and $m_h^{2({\rm SM})}(0)$ are taken from Ref.~\cite{Buttazzo:2013uya}. Lastly, for the vacuum energy $\Omega$, we set $\Omega(0)=0$. This guarantees $\Omega(t)\approx0$ up to logarithmic corrections at all $t$.

The input values for the vector-like fermion parameters appearing in Eq.~(\ref{eq:CWexample}) are taken to be defined at the scale $\mu=M$. We note that, in principle, we should run the non-renormalizable $c_6$ contribution at $\mu<M$ using the SM anomalous dimension, as well as include insertions of $c_6$ in the RGE for the SM couplings~\cite{Elias-Miro:2013mua,Jenkins:2013zja,Jenkins:2013wua,Alonso:2013hga}. In practice, our interest in the effective potential mainly concerns scales $\mu>M$, where vacuum instability occurs, and where the non-renormalizable operators are replaced by the full one-loop contribution that we evolve properly using the two-loop RGE. The contribution of $c_6$ is thus mainly in setting the initial conditions for the running of $\lambda$. Here, however, the running of $c_6$ would be a two-loop correction to the one-loop matching that our approximation requires, and so we omit it. Concerning the insertion of $c_6$ in the running of $\lambda$, using the results of Ref.~\cite{Jenkins:2013zja,Jenkins:2013wua,Alonso:2013hga} we verified that including this effects leads to negligible corrections to our results.

Now that we have the effective potential $V(h_c,\mu)$ defined for all $\mu$, the final step in the calculation is the RGE improvement, that amounts to letting $\mu$ be a function of $h_c$ in order to control large logs far in field space. We set $\mu=h_c$. 

Finally we note that in some of the analyses in the body of the paper, in parameter regions allowing for very light fermions, the vector-like mass scale $M$ comes out lower than the SM top quark mass $M_t$ (in practice, this only happened for the $M_1=200$~GeV curve in the left panel of Fig.~\ref{fig:drgamLDS}, and only for $\delta r_\gamma<0$). In this case, instead of the matching procedure described above, we used the full theory effective potential and RGE starting from $M_t$ and setting the RGE initial conditions based on Eqs.~(\ref{eq:matchingl}-\ref{eq:matchingm}) with $\Delta V_1$ replaced by the full $m^4 \log m^2$ contribution of the new fermions.

\begin{table}[!b]
\caption{Vector-like fermion representations that can mix with SM quarks}\label{tab:qrkL}
\begin{center}
\begin{tabular}{|c|c|c|}
\hline
&representation&$\mathcal{V}_{NP}$\\
\hline\hline
DI&$Q(3,2)_{\frac{1}{6}},\;\;\;Q^c(\bar{3},2)_{-\frac{1}{6}}$&$Y_{Qd^c}H^\dag Qd^c+Y_{qD^c}H^\dag qD^c+Y_{QD^c}H^\dag QD^c+Y_{Q^cD}H^T\epsilon Q^cD$\\
&$D(3,1)_{-\frac{1}{3}},\;\;\;D^c(\bar{3},1)_{\frac{1}{3}}$&$+M_QQ^T\epsilon Q^c+M_DDD^c+cc$\\
\hline
DII&$Q'(3,2)_{-\frac{5}{6}},\;\;\;Q^{'c}(\bar{3},2)_{\frac{5}{6}}$&$Y_{Q'd^c}H^T\epsilon Q'd^c+Y_{qD^c}H^\dag qD^c+Y_{Q'D^c}H^T\epsilon Q' D^c+Y_{Q^{'c}D}H^\dag Q^{'c}D$\\
&$D(3,1)_{-\frac{1}{3}},\;\;\;D^c(\bar{3},1)_{\frac{1}{3}}$&$+M_{Q'}Q^{'T}\epsilon Q^{'c}+M_DDD^c+cc$\\
\hline
DIII&$Q(3,2)_{\frac{1}{6}},\;\;\;Q^c(\bar{3},2)_{-\frac{1}{6}}$&$Y_{Qd^c}H^\dag Qd^c+Y_{qD^{'c}}H^\dag \sigma q\cdot D^{'c}+Y_{QD^{'c}}H^\dag \sigma Q\cdot D^{'c}+Y_{Q^cD'}H^T\epsilon \sigma Q^c\cdot D'$\\
&$D'(3,3)_{-\frac{1}{3}},\;\;\;D^{'c}(\bar{3},3)_{\frac{1}{3}}$&$+M_QQ^T\epsilon Q^c+M_{D'}D'\cdot D^{'c}+cc$\\
\hline
DIV&$Q'(3,2)_{-\frac{5}{6}},\;\;\;Q^{'c}(\bar{3},2)_{\frac{5}{6}}$&$Y_{Q'd^c}H^T\epsilon Q'd^c+Y_{qD^{'c}}H^\dag \sigma q\cdot D^{'c}+Y_{Q'D^{'c}}H^T\epsilon \sigma Q'\cdot D^{'c}+Y_{Q^{'c}D'}H^\dag \sigma Q^{'c}\cdot D'$\\
&$D'(3,3)_{-\frac{1}{3}},\;\;\;D^{'c}(\bar{3},3)_{\frac{1}{3}}$&$+M_{Q'}Q^{'T}\epsilon Q^{'c}+M_{D'}D'\cdot D^{'c}+cc$\\
\hline\hline
UI&$Q(3,2)_{\frac{1}{6}},\;\;\;Q^c(\bar{3},2)_{-\frac{1}{6}}$&$Y_{Qu^c}H^T\epsilon Qu^c+Y_{qU^c}H^T\epsilon qU^c+Y_{QU^c}H^T\epsilon QU^c+Y_{Q^cU}H^\dag Q^cU$\\
&$U(3,1)_{\frac{2}{3}},\;\;\;U^c(\bar{3},1)_{-\frac{2}{3}}$&$+M_QQ^T\epsilon Q^c+M_UUU^c+cc$\\
\hline
UII&$Q^{''}(3,2)_{\frac{7}{6}},\;\;\;Q^{''c}(\bar{3},2)_{-\frac{7}{6}}$&$Y_{Q^{''}u^c}H^\dag Q^{''}u^c+Y_{qU^c}H^T \epsilon qU^c+Y_{Q^{''}U^c}H^\dag Q^{''}U^c+Y_{Q^{''c}U}H^T\epsilon Q^{''c}U$\\
&$U(3,1)_{\frac{2}{3}},\;\;\;U^c(\bar{3},1)_{-\frac{2}{3}}$&$+M_{Q''}Q^{''T}\epsilon Q^{''c}+M_UUU^c+cc$\\
\hline
UIII&$Q(3,2)_{\frac{1}{6}},\;\;\;Q^c(\bar{3},2)_{-\frac{1}{6}}$&$Y_{Qu^c}H^T\epsilon Qu^c+Y_{qU^{'c}}H^T\epsilon \sigma q\cdot U^{'c}+Y_{QU^{'c}}H^T\epsilon \sigma Q\cdot U^{'c}+Y_{Q^cU'}H^\dag \sigma Q^c\cdot U'$\\
&$U'(3,3)_{\frac{2}{3}},\;\;\;U^{'c}(\bar{3},3)_{-\frac{2}{3}}$&$+M_QQ^T\epsilon Q^c+M_{U'}U'\cdot U^{'c}+cc$\\
\hline
UIV&$Q^{''}(3,2)_{\frac{7}{6}},\;\;\;Q^{''c}(\bar{3},2)_{-\frac{7}{6}}$&$Y_{Q^{''}u^c}H^\dag Q^{''}u^c+Y_{qU^{'c}}H^T \epsilon \sigma q\cdot U^{'c}+Y_{Q^{''}U^{'c}}H^\dag \sigma Q^{''}\cdot U^{'c}+Y_{Q^{''c}U'}H^T\epsilon \sigma Q^{''c}\cdot U'$\\
&$U'(3,3)_{\frac{2}{3}},\;\;\;U^{'c}(\bar{3},3)_{-\frac{2}{3}}$&$+M_{Q''}Q^{''T}\epsilon Q^{''c}+M_{U'}U'\cdot U^{'c}+cc$\\
\hline\hline
DV&$Q'(3,2)_{-\frac{5}{6}},\;\;\;Q^{'c}(\bar{3},2)_{\frac{5}{6}}$&$Y_{Q'd^c}H^T\epsilon Q'd^c+Y_{Q'D^{''c}}H^\dag \sigma Q'\cdot D^{''c}+Y_{Q^{'c}D^{''}}H^T\epsilon \sigma Q^{'c}\cdot D^{''}$\\
&$D^{''}(3,3)_{-\frac{4}{3}},\;\;\;D^{''c}(\bar{3},3)_{\frac{4}{3}}$&$+M_{Q'}Q^{'T}\epsilon Q^{'c}+M_{D''}D^{''}\cdot D^{''c}+cc$\\
\hline
UV&$Q^{''}(3,2)_{\frac{7}{6}},\;\;\;Q^{''c}(\bar{3},2)_{-\frac{7}{6}}$&$Y_{Q^{''}u^c}H^\dag Q^{''}u^c+Y_{Q^{''}U^{''c}}H^T\epsilon \sigma Q^{''}\cdot U^{''c}+Y_{Q^{''c}U^{''}}H^\dag \sigma Q^{''c}\cdot U^{''}$\\
&$U^{''}(3,3)_{\frac{5}{3}},\;\;\;U^{''c}(\bar{3},3)_{-\frac{5}{3}}$&$+M_{Q''}Q^{''T}\epsilon Q^{''c}+M_{U''}U^{''}\cdot U^{''c}+cc$\\
\hline
\end{tabular}
\end{center}
\label{default}
\end{table}%
\begin{table}[!b]
\caption{Contributions to the non-renormalizable operators involving SM quarks, listed in Eq.~(\ref{eq:op})}\label{tab:qrkLeff2}
\begin{center}
\begin{tabular}{|c|c|}
\hline
&Effective operators\\
\hline\hline
&\\
DI&$c_{hd}=-\frac{|Y_{Qd^c}|^2}{2|M_Q|^2},\;\;\;\;c'_{hq}=c_{hq}=-\frac{|Y_{qD^c}|^2}{4|M_D|^2},\;\;\;\;\;c_{Yd}=\frac{y_d|Y_{qD^c}|^2}{2|M_D|^2}+\frac{y_d|Y_{Qd^c}|^2}{2|M_Q|^2}-\frac{Y_{Q^cD}Y_{Qd^c}Y_{qD^c}}{M_QM_D}$\\
&\\
\hline
&\\
DII&$c_{hd}=\frac{|Y_{Qd^c}|^2}{2|M_Q|^2},\;\;\;\;c'_{hq}=c_{hq}=-\frac{|Y_{qD^c}|^2}{4|M_D|^2},\;\;\;\;\;c_{Yd}=\frac{y_d|Y_{qD^c}|^2}{2|M_D|^2}+\frac{y_d|Y_{Qd^c}|^2}{2|M_Q|^2}+\frac{Y_{Q^cD}Y_{Qd^c}Y_{qD^c}}{M_QM_D}$\\
&\\
\hline
&\\
DIII&$c_{hd}=-\frac{|Y_{Qd^c}|^2}{2|M_Q|^2},\;\;\;\;3c'_{hq}=-c_{hq}=\frac{3|Y_{qD^c}|^2}{4|M_D|^2},\;\;\;\;\;c_{Yd}=\frac{y_d|Y_{qD^c}|^2}{2|M_D|^2}+\frac{y_d|Y_{Qd^c}|^2}{2|M_Q|^2}-\frac{Y_{Q^cD}Y_{Qd^c}Y_{qD^c}}{M_QM_D}$\\
&\\
\hline
&\\
DIV&$c_{hd}=\frac{|Y_{Qd^c}|^2}{2|M_Q|^2},\;\;\;\;3c'_{hq}=-c_{hq}=\frac{3|Y_{qD^c}|^2}{4|M_D|^2},\;\;\;\;\;c_{Yd}=\frac{y_d|Y_{qD^c}|^2}{2|M_D|^2}+\frac{y_d|Y_{Qd^c}|^2}{2|M_Q|^2}+\frac{Y_{Q^cD}Y_{Qd^c}Y_{qD^c}}{M_QM_D}$\\
&\\
\hline\hline
&\\
UI&$c_{hu}=\frac{|Y_{Qu^c}|^2}{2|M_Q|^2},\;\;\;\;c'_{hq}=-c_{hq}=-\frac{|Y_{qU^c}|^2}{4|M_U|^2},\;\;\;\;\;c_{Yu}=\frac{y_u|Y_{qU^c}|^2}{2|M_U|^2}+\frac{y_u|Y_{Qu^c}|^2}{2|M_Q|^2}+\frac{Y_{Q^cU}Y_{Ud^c}Y_{qU^c}}{M_QM_U}$\\
&\\
\hline
&\\
UII&$c_{hu}=-\frac{|Y_{Qu^c}|^2}{2|M_Q|^2},\;\;\;\;c'_{hq}=-c_{hq}=-\frac{|Y_{qU^c}|^2}{4|M_U|^2},\;\;\;\;\;c_{Yu}=\frac{y_u|Y_{qU^c}|^2}{2|M_U|^2}+\frac{y_u|Y_{Qu^c}|^2}{2|M_Q|^2}-\frac{Y_{Q^cU}Y_{Ud^c}Y_{qU^c}}{M_QM_U}$\\
&\\
\hline
&\\
UIII&$c_{hu}=\frac{|Y_{Qu^c}|^2}{2|M_Q|^2},\;\;\;\;3 c'_{hq}=c_{hq}=\frac{3|Y_{qU^c}|^2}{4|M_U|^2},\;\;\;\;\;c_{Yu}=\frac{y_u|Y_{Qu^c}|^2}{2|M_Q|^2}-\frac{y_u|Y_{qU^c}|^2}{2|M_U|^2}-\frac{Y_{Q^cU}Y_{Ud^c}Y_{qU^c}}{M_QM_U}$\\
&\\
\hline
&\\
UIV&$c_{hu}=-\frac{|Y_{Qu^c}|^2}{2|M_Q|^2},\;\;\;\;3 c'_{hq}=c_{hq}=\frac{3|Y_{qU^c}|^2}{4|M_U|^2},\;\;\;\;\;c_{Yu}=\frac{y_u|Y_{Qu^c}|^2}{2|M_Q|^2}-\frac{y_u|Y_{qU^c}|^2}{2|M_U|^2}+\frac{Y_{Q^cU}Y_{Ud^c}Y_{qU^c}}{M_QM_U}$\\
&\\
\hline
\end{tabular}
\end{center}
\label{default}
\end{table}%
\begin{table}[!b]
\caption{EFT contributions to Higgs and $Z$ couplings to SM quarks. The complex phase $\phi$ has a similar structure in all cases; for example, in model $DI$ it is given by $\phi={\rm arg}\left(Y_{Q^cD}Y_{qD^c}Y_{Qd^c}y_d^*M_Q^*M_D^*\right)$.}\label{tab:qHZ}
\begin{center}
\begin{tabular}{|c|c|}
\hline
&Coupling deviations\\
\hline\hline
&\\
DI&$\delta r_d\approx-2\delta g_{Ad}+\frac{2|Y_{Q^cD}|v}{\sqrt{2}m_d}\sqrt{\left|\delta g_{Vd}^2-\delta g_{Ad}^2\right|}e^{i\phi},\;\;\;\;\delta g_{Ad}=\frac{v^2}{2}\left(\frac{|Y_{qD^c}|^2}{2|M_D|^2}+\frac{|Y_{Qd^c}|^2}{2|M_Q|^2}\right),\;\;\;\delta g_{Vd}=\frac{v^2}{2}\left(\frac{|Y_{qD^c}|^2}{2|M_D|^2}-\frac{|Y_{Qd^c}|^2}{2|M_Q|^2}\right)$\\
&\\
\hline
&\\
DII&$\delta r_d\approx-2\delta g_{Vd}-\frac{2|Y_{Q^cD}|v}{\sqrt{2}m_d}\sqrt{\left|\delta g_{Vd}^2-\delta g_{Ad}^2\right|}e^{i\phi},\;\;\;\,\delta g_{Ad}=\frac{v^2}{2}\left(\frac{|Y_{qD^c}|^2}{2|M_D|^2}-\frac{|Y_{Qd^c}|^2}{2|M_Q|^2}\right),\;\;\;\delta g_{Vd}=\frac{v^2}{2}\left(\frac{|Y_{qD^c}|^2}{2|M_D|^2}+\frac{|Y_{Qd^c}|^2}{2|M_Q|^2}\right)$\\
&\\
\hline
&\\
DIII&$\delta r_d\approx-2\delta g_{Vd}-\frac{2|Y_{Q^cD}|v}{\sqrt{2}m_d}\sqrt{\left|\delta g_{Vd}^2-\delta g_{Ad}^2\right|}e^{i\phi},\;\;\;\;\delta g_{Ad}=\frac{v^2}{2}\left(\frac{|Y_{qD^c}|^2}{2|M_D|^2}+\frac{|Y_{Qd^c}|^2}{2|M_Q|^2}\right),\;\;\;\delta g_{Vd}=\frac{v^2}{2}\left(\frac{|Y_{qD^c}|^2}{2|M_D|^2}-\frac{|Y_{Qd^c}|^2}{2|M_Q|^2}\right)$\\&
\;\;\;\;\;\;\;\;\;\;\;\;\;\;\;\;\;\;\;\;\;\;\;$\delta g_{Au}=\delta g_{Vu}=\frac{v^2}{2}\frac{|Y_{qD^c}|^2}{|M_D|^2}$\\
&\\
\hline
&\\
DIV&$\delta r_d\approx-2\delta g_{Ad}+\frac{2|Y_{Q^cD}|v}{\sqrt{2}m_d}\sqrt{\left|\delta g_{Vd}^2-\delta g_{Ad}^2\right|}e^{i\phi},\;\;\;\,\delta g_{Ad}=\frac{v^2}{2}\left(\frac{|Y_{qD^c}|^2}{2|M_D|^2}-\frac{|Y_{Qd^c}|^2}{2|M_Q|^2}\right),\;\;\;\delta g_{Vd}=\frac{v^2}{2}\left(\frac{|Y_{qD^c}|^2}{2|M_D|^2}+\frac{|Y_{Qd^c}|^2}{2|M_Q|^2}\right)$\\&
\;\;\;\;\;\;\;\;\;\;\;\;\;\;\;\;\;\;\;\;\;\;\;$\delta g_{Au}=\delta g_{Vu}=\frac{v^2}{2}\frac{|Y_{qD^c}|^2}{|M_D|^2}$\\
&\\
\hline\hline
&\\
UI&$\delta r_u\approx2\delta g_{Au}-\frac{2|Y_{Q^cU}|v}{\sqrt{2}m_u}\sqrt{\left|\delta g_{Vu}^2-\delta g_{Au}^2\right|}e^{i\phi},\;\;\;\;\delta g_{Au}=-\frac{v^2}{2}\left(\frac{|Y_{qU^c}|^2}{2|M_U|^2}+\frac{|Y_{Qu^c}|^2}{2|M_Q|^2}\right),\;\;\;\delta g_{Vu}=-\frac{v^2}{2}\left(\frac{|Y_{qU^c}|^2}{2|M_U|^2}-\frac{|Y_{Qu^c}|^2}{2|M_Q|^2}\right)$\\
&\\
\hline
&\\
UII&$\delta r_u\approx2\delta g_{Vu}+\frac{2|Y_{Q^cU}|v}{\sqrt{2}m_u}\sqrt{\left|\delta g_{Vu}^2-\delta g_{Au}^2\right|}e^{i\phi},\;\;\;\,\delta g_{Au}=-\frac{v^2}{2}\left(\frac{|Y_{qU^c}|^2}{2|M_U|^2}-\frac{|Y_{Qu^c}|^2}{2|M_Q|^2}\right),\;\;\;\delta g_{Vu}=-\frac{v^2}{2}\left(\frac{|Y_{qU^c}|^2}{2|M_U|^2}+\frac{|Y_{Qu^c}|^2}{2|M_Q|^2}\right)$\\
&\\
\hline
&\\
UIII&$\delta r_u\approx-2\delta g_{Vu}+\frac{2|Y_{Q^cU}|v}{\sqrt{2}m_u}\sqrt{\left|\delta g_{Vu}^2-\delta g_{Au}^2\right|}e^{i\phi},\;\;\;\;\delta g_{Au}=-\frac{v^2}{2}\left(\frac{|Y_{qU^c}|^2}{2|M_U|^2}+\frac{|Y_{Qu^c}|^2}{2|M_Q|^2}\right),\;\;\;\delta g_{Vu}=-\frac{v^2}{2}\left(\frac{|Y_{qU^c}|^2}{2|M_U|^2}-\frac{|Y_{Qu^c}|^2}{2|M_Q|^2}\right)$\\&
\;\;\;\;\;\;\;\;\;\;\;\;\;\;\;\;\;\;\;\;\;\;\;$\delta g_{Ad}=\delta g_{Vd}=-\frac{v^2}{2}\frac{|Y_{qU^c}|^2}{|M_U|^2}$\\
&\\
\hline
&\\
UIV&$\delta r_u\approx-2\delta g_{Au}-\frac{2|Y_{Q^cU}|v}{\sqrt{2}m_u}\sqrt{\left|\delta g_{Vu}^2-\delta g_{Au}^2\right|}e^{i\phi},\;\;\;\,\delta g_{Au}=-\frac{v^2}{2}\left(\frac{|Y_{qU^c}|^2}{2|M_U|^2}-\frac{|Y_{Qu^c}|^2}{2|M_Q|^2}\right),\;\;\;\delta g_{Vu}=-\frac{v^2}{2}\left(\frac{|Y_{qU^c}|^2}{2|M_U|^2}+\frac{|Y_{Qu^c}|^2}{2|M_Q|^2}\right)$\\&
\;\;\;\;\;\;\;\;\;\;\;\;\;\;\;\;\;\;\;\;\;\;\;$\delta g_{Ad}=\delta g_{Vd}=-\frac{v^2}{2}\frac{|Y_{qU^c}|^2}{|M_U|^2}$\\
&\\
\hline
\end{tabular}
\end{center}
\label{default}
\end{table}%
\begin{table}[!t]
\caption{Vector-like fermion representations that can mix with the SM leptons}\label{tab:lepL}
\begin{center}
\begin{tabular}{|c|c|c|}
\hline
&representation&$\mathcal{V}_{NP}$\\
\hline\hline
LI&$L(1,2)_{-\frac{1}{2}},\;\;\;L^c(1,2)_{\frac{1}{2}}$&$Y_{Le^c}H^\dag Le^c+Y_{lE^c}H^\dag lE^c+Y_{LE^c}H^\dag LE^c+Y_{L^cE}H^T\epsilon L^cE$\\
&$E(1,1)_{-1},\;\;\;E^c(1,1)_{1}$&$+M_LL^T\epsilon L^c+M_EEE^c+cc$\\
\hline
LII&$L'(1,2)_{-\frac{3}{2}},\;\;\;L^{'c}(1,2)_{\frac{3}{2}}$&$Y_{L'e^c}H^T\epsilon L'e^c+Y_{lE^c}H^\dag lE^c+Y_{L'E^c}H^T\epsilon L' E^c+Y_{L^{'c}E}H^\dag L^{'c}E$\\
&$E(1,1)_{-1},\;\;\;E^c(1,1)_{1}$&$+M_{L'}L^{'T}\epsilon L^{'c}+M_EEE^c+cc$\\
\hline
LIII&$L(1,2)_{-\frac{1}{2}},\;\;\;L^c(1,2)_{\frac{1}{2}}$&$Y_{Le^c}H^\dag Le^c+Y_{lE^{'c}}H^\dag \sigma l\cdot E^{'c}+Y_{LE^{'c}}H^\dag \sigma L\cdot E^{'c}+Y_{L^cE'}H^T\epsilon \sigma L^c\cdot E'$\\
&$E'(1,3)_{-1},\;\;\;E^{'c}(1,3)_{1}$&$+M_LL^T\epsilon L^c+M_{E'}E'\cdot E^{'c}+cc$\\
\hline
LIV&$L'(1,2)_{-\frac{3}{2}},\;\;\;L^{'c}(1,2)_{\frac{3}{2}}$&$Y_{L'e^c}H^T\epsilon L'e^c+Y_{lE^{'c}}H^\dag \sigma l\cdot E^{'c}+Y_{L'E^{'c}}H^T\epsilon \sigma L'\cdot E^{'c}+Y_{L^{'c}E'}H^\dag \sigma L^{'c}\cdot E'$\\
&$E'(1,3)_{-1},\;\;\;E^{'c}(1,3)_{1}$&$+M_{L'}L^{'T}\epsilon L^{'c}+M_{E'}E'\cdot E^{'c}+cc$\\
\hline
LV&$L(1,2)_{-\frac{1}{2}},\;\;\;L^{c}(1,2)_{\frac{1}{2}}$&$Y_{Le^c}H^\dag Le^c+Y_{lE^{'c}}H^T \epsilon \sigma l\cdot E^{''c}+Y_{L^cE^{''}}H^\dag \sigma L^c\cdot E^{''}+Y_{LE^{''c}}H^T\epsilon \sigma L\cdot E^{''c}$\\
&$E^{''}(1,3)_{0},\;\;\;E^{''c}(1,3)_{0}$&$+M_{L}L^{T}\epsilon L^{c}+M_{E''}E^{''}\cdot E^{''c}+cc$\\
\hline
\end{tabular}
\end{center}
\label{default}
\end{table}%
\begin{table}[!h]
\caption{Contributions to the non-renormalizable operators involving SM leptons, listed in Eq.~(\ref{eq:op})}\label{tab:lepLeff2}
\begin{center}
\begin{tabular}{|c|c|}
\hline
&Effective operators\\
\hline\hline
&\\
LI&$c_{he}=-\frac{|Y_{Le^c}|^2}{2|M_L|^2},\;\;\;\;c'_{hl}=c_{hl}=-\frac{|Y_{lE^c}|^2}{4|M_E|^2},\;\;\;\;\;c_{Ye}=\frac{Y_{le^c}|Y_{lE^c}|^2}{2|M_E|^2}+\frac{Y_{le^c}|Y_{Le^c}|^2}{2|M_L|^2}-\frac{Y_{L^cE}Y_{Le^c}Y_{lE^c}}{M_LM_E}$\\
&\\
\hline
&\\
LII&$c_{he}=\frac{|Y_{Le^c}|^2}{2|M_L|^2},\;\;\;\;c'_{hl}=c_{hl}=-\frac{|Y_{lE^c}|^2}{4|M_E|^2},\;\;\;\;\;c_{Ye}=\frac{Y_{le^c}|Y_{lE^c}|^2}{2|M_E|^2}+\frac{Y_{le^c}|Y_{Le^c}|^2}{2|M_L|^2}+\frac{Y_{L^cE}Y_{Le^c}Y_{lE^c}}{M_LM_E}$\\
&\\
\hline
&\\
LIII&$c_{he}=-\frac{|Y_{Le^c}|^2}{2|M_L|^2},\;\;\;\;3c'_{hl}=-c_{hl}=\frac{3|Y_{lE^c}|^2}{4|M_E|^2},\;\;\;\;\;c_{Ye}=\frac{Y_{le^c}|Y_{lE^c}|^2}{2|M_E|^2}-\frac{Y_{L^cE}Y_{Le^c}Y_{lE^c}}{M_LM_E}+\frac{Y_{le^c}|Y_{Le^c}|^2}{2|M_L|^2}$\\
&\\
\hline
&\\
LIV&$c_{he}=\frac{|Y_{Le^c}|^2}{2|M_L|^2},\;\;\;\;3c'_{hl}=-c_{hl}=\frac{3|Y_{lE^c}|^2}{4|M_E|^2},\;\;\;\;\;c_{Ye}=\frac{Y_{le^c}|Y_{lE^c}|^2}{2|M_E|^2}+\frac{Y_{L^cE}Y_{Le^c}Y_{lE^c}}{M_LM_E}+\frac{Y_{le^c}|Y_{Le^c}|^2}{2|M_L|^2}$\\
&\\
\hline
&\\
LV&$c_{he}=-\frac{|Y_{Le^c}|^2}{2|M_L|^2},\;\;\;\;3 c'_{hl}=c_{hl}=\frac{3|Y_{lE^c}|^2}{4|M_E|^2},\;\;\;\;\;c_{Ye}=\frac{Y_{le^c}|Y_{Le^c}|^2}{2|M_L|^2}+\frac{Y_{le^c}|Y_{lE^c}|^2}{|M_E|^2}-2\frac{Y_{L^cE}Y_{Le^c}Y_{lE^c}}{M_LM_E}$\\
&\\
\hline
\end{tabular}
\end{center}
\label{default}
\end{table}%
\begin{table}[!t]
\caption{EFT contributions to Higgs and $Z$ couplings to SM leptons. The complex phase $\phi$ has a similar structure in all cases; for example, in model $LI$ it is given by $\phi={\rm arg}\left(Y_{L^cE}Y_{lE^c}Y_{Le^c}y_e^*M_L^*M_E^*\right)$.}\label{tab:lHZ}
\begin{center}
\begin{tabular}{|c|c|}
\hline
&Coupling deviations\\
\hline\hline
&\\
LI&$\delta r_e\approx-2\delta g_{Ae}+\frac{2|Y_{L^cE}|v}{\sqrt{2}m_e}\sqrt{\left|\delta g_{Ve}^2-\delta g_{Ae}^2\right|}e^{i\phi},\;\;\;\;\delta g_{Ae}=\frac{v^2}{2}\left(\frac{|Y_{lE^c}|^2}{2|M_E|^2}+\frac{|Y_{Le^c}|^2}{2|M_L|^2}\right),\;\;\;\delta g_{Ve}=\frac{v^2}{2}\left(\frac{|Y_{lE^c}|^2}{2|M_E|^2}-\frac{|Y_{Le^c}|^2}{2|M_L|^2}\right)$\\
&\\
\hline
&\\
LII&$\delta r_e\approx-2\delta g_{Ve}-\frac{2|Y_{L^cE}|v}{\sqrt{2}m_e}\sqrt{\left|\delta g_{Ve}^2-\delta g_{Ae}^2\right|}e^{i\phi},\;\;\;\,\delta g_{Ae}=\frac{v^2}{2}\left(\frac{|Y_{lE^c}|^2}{2|M_E|^2}-\frac{|Y_{Le^c}|^2}{2|M_L|^2}\right),\;\;\;\delta g_{Ve}=\frac{v^2}{2}\left(\frac{|Y_{lE^c}|^2}{2|M_E|^2}+\frac{|Y_{Le^c}|^2}{2|M_L|^2}\right)$\\
&\\
\hline
&\\
LIII&$\delta r_e\approx-2\delta g_{Ve}-\frac{2|Y_{L^cE}|v}{\sqrt{2}m_e}\sqrt{\left|\delta g_{Ve}^2-\delta g_{Ae}^2\right|}e^{i\phi},\;\;\;\;\delta g_{Ae}=\frac{v^2}{2}\left(\frac{|Y_{lE^c}|^2}{2|M_E|^2}+\frac{|Y_{Le^c}|^2}{2|M_L|^2}\right),\;\;\;\delta g_{Ve}=\frac{v^2}{2}\left(\frac{|Y_{lE^c}|^2}{2|M_E|^2}-\frac{|Y_{Le^c}|^2}{2|M_L|^2}\right)$\\&
\;\;\;\;\;\;\;\;\;\;\;\;\;\;\;\;\;\;\;\;\;\;\;$\delta g_{A\nu}=\delta g_{V\nu}=\frac{v^2}{2}\frac{|Y_{lE^c}|^2}{|M_E|^2}$\\
&\\
\hline
&\\
LIV&$\delta r_e\approx-2\delta g_{Ae}+\frac{2|Y_{L^cE}|v}{\sqrt{2}m_e}\sqrt{\left|\delta g_{Ve}^2-\delta g_{Ae}^2\right|}e^{i\phi},\;\;\;\,\delta g_{Ae}=\frac{v^2}{2}\left(\frac{|Y_{lE^c}|^2}{2|M_E|^2}-\frac{|Y_{Le^c}|^2}{2|M_L|^2}\right),\;\;\;\delta g_{Ve}=\frac{v^2}{2}\left(\frac{|Y_{lE^c}|^2}{2|M_E|^2}+\frac{|Y_{Le^c}|^2}{2|M_L|^2}\right)$\\&
\;\;\;\;\;\;\;\;\;\;\;\;\;\;\;\;\;\;\;\;\;\;\;$\delta g_{A\nu}=\delta g_{V\nu}=\frac{v^2}{2}\frac{|Y_{lE^c}|^2}{|M_E|^2}$\\
&\\
\hline
&\\
LV&$\delta r_e\approx2\delta g_{Ve}+\frac{2|Y_{L^cE}|v}{\sqrt{2}m_e}\sqrt{\left|\delta g_{Ve}^2-\delta g_{Ae}^2\right|}e^{i\phi},\;\;\;\;\delta g_{Ae}=\frac{v^2}{2}\left(-\frac{|Y_{lE^c}|^2}{|M_E|^2}+\frac{|Y_{Le^c}|^2}{2|M_L|^2}\right),\;\;\;\delta g_{Ve}=\frac{v^2}{2}\left(-\frac{|Y_{lE^c}|^2}{|M_E|^2}-\frac{|Y_{Le^c}|^2}{2|M_L|^2}\right)$\\&
\;\;\;\;\;\;\;\;\;\;\;\;\;\;\;\;\;\;\;\;\;\;\;$\delta g_{A\nu}=\delta g_{V\nu}=-\frac{v^2}{4}\frac{|Y_{lE^c}|^2}{|M_E|^2}$\\
&\\
\hline
\end{tabular}
\end{center}
\label{default}
\end{table}%

\section{EFT analysis of deviations in $Z$ and Higgs couplings to SM fermions}\label{app:Hqq}

We examine the vector-like fermion representations that can modify the effective Higgs Yukawa couplings at tree level (for related analyses see, e.g.~\cite{delAguila:2000rc, delAguila:2000aa,Batell:2012ca}). Before we introduce the new fermion fields, we first describe our conventions for the SM effective theory below the vector-like fermion scale.

Our notation for the SM Higgs and fermion representations is  
$H(1,2)_{\frac{1}{2}}$, $q(3,2)_{\frac{1}{6}}$, $d^c(\bar{3},1)_{\frac{1}{3}}$, $u^c(\bar{3},1)_{-\frac{2}{3}}$, $l(1,2)_{\frac{1}{2}}$, $e^c(1,1)_1$. The first and second numbers in parenthesis denote the $SU(3)_c$ and $SU(2)_W$ representation, respectively, and the subscript denotes the hypercharge. The fermion sector of the SM Lagrangian is 
\be\label{eq:LSM}\mathcal{L}_{SM}&=&i\bar{q}\bar{\sigma}_\mu\mathcal{D}^\mu q+i\overline{d^c}\bar{\sigma}_\mu\mathcal{D}^\mu d^c+i\overline{u^c}\bar{\sigma}_\mu\mathcal{D}^\mu u^c+i\bar{l}\bar{\sigma}_\mu\mathcal{D}^\mu l+i\overline{e^c}\bar{\sigma}_\mu\mathcal{D}^\mu e^c\no\\
&&-\left\{y_dH^\dag qd^c+y_uH^T\epsilon qu^c+y_eH^\dag le^c+cc\right\}\ee
with $\epsilon_{12}=-\epsilon_{21}=1$. In unitary gauge $H=(0\;\;h)^T/\sqrt{2}$, with $\langle h\rangle=v\simeq246.22$~GeV. 

Integrating out heavy fermions leads to effective $Hff$ and $Zff$ couplings that are modified compared to their SM values. The main effect is captured by considering non-renormalizable operators~\cite{Buchmuller:1985jz,Grzadkowski:2010es}, 
\be\label{eq:effop}\Delta\mathcal{L}_{eff}&=&c_i\mathcal{O}_i+cc.\ee
Following Ref.~\cite{Barbieri:1999tm} and adding operators that affect Yukawa couplings, we list the operators of interest as follows\footnote{Note that our definitions for the operators $\mathcal{O}_{hd}$, $\mathcal{O}_{hu}$, and $\mathcal{O}_{he}$ differs by a sign compared to the corresponding operators in Ref.~\cite{Barbieri:1999tm}. In addition, we define $v=246$~GeV while Ref.~\cite{Barbieri:1999tm} works with $v=174$~GeV.},
\be
&\mathcal{O}_{hl}=iH^\dag\mathcal{D}^\mu H\,\bar{l}\bar{\sigma}_\mu l,\;\;\;\;\;
\mathcal{O}'_{hl}=i\left(H^\dag\mathcal{D}^\mu \sigma H\right)\cdot\left(\bar{l}\bar{\sigma}_\mu \sigma l\right),\;\;\;\;\;\mathcal{O}_{he}=iH^\dag\mathcal{D}^\mu H\,\overline{e^c}\bar{\sigma}_\mu e^c\no\\
&\mathcal{O}_{hq}=iH^\dag\mathcal{D}^\mu H\,\bar{q}\bar{\sigma}_\mu q,\;\;\;\;\;\mathcal{O}'_{hq}=i\left(H^\dag\mathcal{D}^\mu \sigma H\right)\cdot\left(\bar{q}\bar{\sigma}_\mu \sigma q\right)\no\\
&\mathcal{O}_{hd}=iH^\dag\mathcal{D}^\mu H\,\overline{d^c}\bar{\sigma}_\mu d^c,\;\;\;\;\;\mathcal{O}_{hu}=iH^\dag\mathcal{D}^\mu H\,\overline{u^c}\bar{\sigma}_\mu u^c\no\\
&\mathcal{O}_{Yd}=|H|^2H^\dag qd^c,\;\;\;\;\;\mathcal{O}_{Yu}=|H|^2H^T\epsilon qu^c,\;\;\;\;\;\mathcal{O}_{Ye}=|H|^2H^\dag le^c.
\label{eq:op}
\ee

For fermion $f$, we define the modification to the Higgs coupling as $\delta r_f\equiv \left(g_{hff}-g_{hff}^{\rm SM}\right)/g_{hff}^{\rm SM}$. We find (see also~\cite{Ciuchini:2013pca})
\be\label{eq:eftd}
&\delta r_d=-\frac{v^3}{\sqrt{2}m_d}c_{Yd},\;\;\;\;\;
\delta g_{Vd}=-\frac{v^2}{2}\left(c_{hq}+c'_{hq}-c_{hd}\right),\;\;\;\;\;
\delta g_{Ad}=-\frac{v^2}{2}\left(c_{hq}+c'_{hq}+c_{hd}\right),\no\\
&\delta r_u=-\frac{v^3}{\sqrt{2}m_u}c_{Yu},\;\;\;\;\;
\delta g_{Vu}=-\frac{v^2}{2}\left(c_{hq}-c'_{hq}-c_{hu}\right),\;\;\;\;\;
\delta g_{Au}=-\frac{v^2}{2}\left(c_{hq}-c'_{hq}+c_{hu}\right),\no\\
&\delta r_e=-\frac{v^3}{\sqrt{2}m_e}c_{Ye},\;\;\;\;\;
\delta g_{Ve}=-\frac{v^2}{2}\left(c_{hl}+c'_{hl}-c_{he}\right),\;\;\;\;\;
\delta g_{Ae}=-\frac{v^2}{2}\left(c_{hl}+c'_{hl}+c_{he}\right),\no\\
&\delta g_{V\nu}=-\frac{v^2}{2}\left(c_{hl}-c'_{hl}\right),\;\;\;\;\;
\delta g_{A\nu}=-\frac{v^2}{2}\left(c_{hl}-c'_{hl}\right).
\ee
In addition, for any $f$, the coupling $y_f$ of Eq.~(\ref{eq:LSM}) is constrained by
\be\label{eq:mf}y_f=\frac{\sqrt{2}m_f}{v}\left(1-\frac{\delta r_f}{2}\right).\ee
Eqs.~(\ref{eq:eftd}-\ref{eq:mf}) are valid to $\mathcal{O}\left(Y^2v^2/M^2\right)$, where $M$ represents the heavy vector-like fermion mass scale and $Y$ is a Yukawa coupling.

Vector-like fermion representations that can mix with SM quarks are given in Tab.~\ref{tab:qrkL}. Integrating out the heavy fields we collect contributions to effective operators of interest in Tab.~\ref{tab:qrkLeff2}. This is further summarized in terms of relations between the $Hff$ and $Zff$ effective couplings in Tab.~\ref{tab:qHZ}. Note that the expressions in Tab.~\ref{tab:qHZ} include an expansion in $|\delta g_A|$ and $|\delta g_V|$, and are valid to leading order in the $\delta g$'s. 

Vector-like fermion representations that can mix with SM leptons are given in Tab.~\ref{tab:lepL}. Integrating out the heavy fields, we further summarize the contributions of the effective operators in Tab.~\ref{tab:lepLeff2} and the modifications to the $Hff$ and $Zff$ effective couplings obtained via the EFT analysis in Tab.~\ref{tab:lHZ}, again valid to leading order in $|\delta g_{A,V}|$. These results are used in Sec.~\ref{sec:Hff}, where we also comment on cases in which some of the expressions derived here using EFT remain valid even for $Y^2v^2/M^2=\mathcal{O}(1)$. 

We comment that the derivative operators in Eq.~(\ref{eq:op}), for which we have highlighted the effect on $Zff$ couplings, also enter into Higgs $t\bar t$ associated production and $h\to bb,\tau\tau$ decays. However, in contrast to the modified Yukawa couplings [parametrized by the $\delta r_f$ terms in Eq.~(\ref{eq:eftd})], the contribution due to the derivative operators does not interfere with the leading SM contribution to the Higgs decay or production matrix element. Thus, for the $Htt$, $Hbb$, and $H\tau\tau$ couplings of interest to us in this paper, their contribution to the respective signal strength is suppressed in comparison to the $\delta r_f$ terms, and we neglect it here and in the main text.

\section{Collider constraints}\label{app:collider}
We discuss here the collider bounds on vector-like quarks and leptons. We also include limits on states with exotic color or electromagnetic charge assignments that can be relevant for the $H\gamma\gamma$ and $HGG$ analyses. 

Unless specified otherwise, we list constraints on a single Dirac fermion at 95\% C.L., taking into account the color multiplicity, but ignoring additional constraints due to e.g. isospin multiplicity. For the vector-like fermion models at hand, these constraints are conservative since they ignore the contributions due to some of the states. Thus our bounds are weaker, for example, than those derived in~\cite{Falkowski:2013jya, Altmannshofer:2013zba, Dermisek:2014qca} for vector-like leptons. 

We discuss the constraints on states that are either stable or decay promptly on collider time scales, separating the discussion for colored and color singlet fermions. We do not discuss the constraints on states that decay non-promptly but yet within the tracker volume. We comment however that for $c\tau\gtrsim0.1$~cm the constraints on such non-prompt decays are typically comparable to or stronger than the constraints obtained in either the stable or prompt cases (see, e.g.~\cite{CMS:2013oea, CMS:2014mca}). 

We stress that the broad summary of constraints presented here, that applies to numerous different vector-like fermion models, is only intended to provide a rough estimate for the mass scales of such fermions still allowed after Run-I of the LHC. For the purpose of the current paper we find these rough estimates sufficient; a more careful analysis will be motivated in the case that Higgs coupling deviations end up being discovered at the LHC. 

\subsection{Colored particles}
As discussed in Section~\ref{sec:HGG} we are interested only in states transforming as 3, 6 or 8 of $SU(3)_c$. For the vector-like models we deal with here, higher color representations induce a Landau pole in $\alpha_s$ on scales close to the vector-like fermion masses. To obtain the mass bounds quoted in the following, for the color 3 we use the $t^\prime$ cross section computed in~\cite{Chatrchyan:2013uxa} using {\tt HATHOR}~\cite{Aliev:2010zk} and for the 8 we compute it at NLO+NLL using {\tt NLLfast}~\cite{Beenakker:1996ch, Kulesza:2008jb,Kulesza:2009kq, Beenakker:2009ha, Beenakker:2011fu}. The mass bounds for the 6 and 8 representations are similar~\cite{Ilisie:2012cc}.

\subsubsection{Stable states}
If we are interested only in $HGG$ or $H\gamma\gamma$ deviations we can imagine that the new fermions possess a conserved quantum number that prevents them from decaying into SM states. In this case we are looking for stable colored particles that travel inside the detector and can form charged bound states. The strongest bounds on heavy stable charged particles are set by the CMS search that exploits mainly their long time of flight and their anomalous energy loss per unit length~\cite{Chatrchyan:2013oca}. The behavior of heavy colored particles inside the detector is subject to large uncertainties, so we quote a range for the mass exclusion given by the different hadronization models considered by the collaboration~\cite{Mackeprang:2009ad,Kraan:2004tz, Mackeprang:2006gx}. The bounds we quote here refer to a combination of the CMS tracker and time of flight (TOF) analyses. Similar bounds are obtained from the tracker analysis alone.

The CMS collaboration considered colored particles with $Q=0$ and $Q=2/3$. 
If we use the cross section limit obtained by the collaboration on gluino pair production, we find the mass bounds $m_3 >1050\pm 30$~GeV and $m_8 >1295\pm35$~GeV for a single Dirac fermion and $m_3 >1165\pm35$~GeV and $m_8 >1390\pm35$~GeV for three Dirac fermions (corresponding to a vector-like doublet and a singlet with the same mass). The bounds are the same for $Q=2/3$ and $Q=0$ states. We expect the limit on the 6 to be similar to that on the 8. This procedure is partially justified by the fact that the cross section limit for stops (i.e. $N_c=3$ and $Q=2/3$) converges to the one for gluinos ($Q=0$ and $N_c=8$) at high masses. However it can overestimate the bound for particles with $Q=0$ and $N_c=3,6$ and a more thorough collider study is needed to go beyond this very rough estimate. We expect similar bounds for other electromagnetic charges, including e.g. $Q=1/3$.

\subsubsection{States decaying promptly to first and second generation quarks}
\paragraph{Wq, Zq and hq final states} Vector-like quarks with mass mixing with the first or second generation SM quarks can decay  to $W, Z$ or $h$ plus one jet. This situation could be relevant for the $HGG$ and $H\gamma\gamma$ analyses, where mixing with third generation quarks is not guaranteed. Bounds on pair produced heavy quarks for these final states were examined in~\cite{D'Agnolo:2012ie}. The corresponding experimental analyses where not updated since the 7 TeV run, and the leading explicit constraint we find is $m_{q^\prime} > 350$~GeV from ATLAS~\cite{Aad:2012bt}, where the search was performed on $WW$+2 jets final states and the bound assumes ${\rm BR}(q^\prime \to W q)=1$. For three Dirac copies of $q^\prime$, the ATLAS search would exclude $m_{q^\prime} > 420$~GeV. ATLAS also reported a multijet search for RPV gluinos~\cite{TheATLAScollaboration:2013xia} that is a counting experiment,   does not exploit the shape of the jet invariant masses and can be reinterpreted in the context of our models. A possible caveat is that in most cases, ATLAS finds that requiring a seventh jet improves the bounds. Our signal, that includes $WWjj$, $ZZjj$ and $hhjj$ final states, would has a different probability of radiating an extra jet with respect to a pair of gluinos, but we ignore this correction for the purpose of the current rough estimate. In the case ${\rm BR}(q^\prime \to W q)=1$, no $b$-quarks are produced and the RPV gluino search yields a weaker bound than the one we obtain from~\cite{Aad:2012bt} discussed above. Instead for ${\rm BR}(q^\prime \to hq)=1$ we get for three Dirac fermions $m_{q^\prime} >700$~GeV while for ${\rm BR}(q^\prime \to Zq)=1$, $m_{q^\prime} > 450$~GeV. The $ZZjj$ and $hhjj$ final states could also be constrained by the 8 TeV CMS multilepton search performed with an integrated luminosity of 19.5 fb$^{-1}$~\cite{Chatrchyan:2014aea}. For ${\rm BR}(q^\prime \to hq)=1$ we find that the mass bounds for three Dirac fermions is $m_{q^\prime} >540$~GeV, which is weaker than the multijet bound quoted above, while for ${\rm BR}(q^\prime \to Zq)=1$, $m_{q^\prime} > 650$~GeV, stronger than the multijet bound.

\paragraph{Dijet decays} 
Both the 3 and the 6 could couple to an SM quark and a gluon via nonrenormalizable magnetic interactions, allowing dijet decays. However we find somewhat unlikely for the dijet mode to be the dominant one, as producing the nonrenormalizable magnetic interaction requires flavor changing scalar or boson loop diagrams. If the operator is formed with a $W$ boson loop (relevant for the 3), then we expect the irreducible $q^\prime \to W q$ to dominate and lead to the bound discussed in the previous paragraph. For the 6,  new bosons with masses larger than $\Lambda_{UV}$ must be involved and we expect a tree level three-jet decay (discussed below) with branching fraction comparable or larger than the dijet mode. However from the the CMS search for pairs of dijet resonances~\cite{Khachatryan:2014lpa} we can make a rough estimate $m_3>500$~GeV, valid for a single Dirac fermion, with a somewhat stronger limit on the 6.

\paragraph{Three jet, and higher jet multiplicity decays} Limits on this final state were set by the ATLAS multijet search discussed above~\cite{TheATLAScollaboration:2013xia}, with somewhat weaker bounds reported by CMS~\cite{Chatrchyan:2013gia}. For a single Dirac 8 we find $m_8 > 1.1$~TeV, while for three copies of equal mass we obtain a limit stronger than $m_8 > 1.2$~TeV. The latter bound is conservative, since what prevented us from quoting a higher number is simply the fact that the ATLAS exclusion plot extends only up to 1.2~TeV. The constraints on the 6 representation are the same. For the 3 representation, we find essentially no bound from~\cite{TheATLAScollaboration:2013xia}. However, our new fermions necessarily include electromagnetically charged states, leading to irreducible three jet plus weak gauge boson (or Higgs) decays which translate with $\mathcal{O}(1)$ branching fraction into five jet modes. From the 10 jet bin of the general purpose analysis in~\cite{ATLAScrazy}, we find $m_3>850$~GeV or so. In addition to the five jet channel, leptonic modes [with an $\mathcal{O}(10\%)$ branching fraction] suggest even stronger bounds of order a TeV.

\subsubsection{States decaying promptly to third generation quarks}

The discussion here is aimed to address vector-like quarks mixing with the third generation SM quarks, relevant to the $Hbb$ and $Htt$ coupling analyses. The constraints in this case are generally stronger than those discussed above for decays to $W,Z,h+q$, involving only first or second generation quarks. 
We focus on the CMS searches for pair production.
CMS results are presented as a function of branching ratios and can be directly read off the tables in~\cite{Chatrchyan:2013uxa, CMS-PAS-B2G-12-019, CMS:2013una}. ATLAS bounds are comparable~\cite{ATLAS-CONF-2014-036, ATLAS-CONF-2013-051, The ATLAScollaboration:2013sha, ATLAS:2013ima}. 

In the case of a $t^\prime$, the weakest bound is $m_{t^\prime} > 700$~GeV, obtained for ${\rm BR}(t^\prime \to W b) \approx 1$~\cite{Chatrchyan:2013uxa}. Adding a finite ${\rm BR}(t^\prime \to Z t)$ or ${\rm BR}(t^\prime \to h t)$ makes the bound a few tens of GeV stronger. 
For a $b^\prime$ the situation is reversed, since top rich final states are easier to distinguish from the background. The most constraining case corresponds to ${\rm BR}(b^\prime \to W t) \approx 1$ and $m_{b^\prime} > 730$~GeV, while the weakest bound reads $m_{b^\prime} > 600$~GeV, obtained for ${\rm BR}(b^\prime \to W t) =0$ and ${\rm BR}(b^\prime \to h b) \approx {\rm BR}(b^\prime \to Z b)$~\cite{CMS-PAS-B2G-12-019, CMS:2013una}. 

Even if the lightest state has an exotic charge, the bounds do not change dramatically. For a quark with charge $5/3$, we have ${\rm BR}(t^{5/3} \to W t) = 1$ and $m_{t^{5/3}} \gtrsim 800$~GeV~\cite{Chatrchyan:2013wfa}. A charge $4/3$ quark will decay to $Wb$, and the limit is the same as in the $t^\prime$ case discussed above. 

\subsection{Leptons}
We focus on pair production $pp \to L_1L_1$ where $L_1$ stands for the lightest new lepton, that in our minimal models is always electromagnetically charged if Higgs coupling deviations are to be induced. Our limits on this single state are less stringent than those set on complete models by~\cite{Falkowski:2013jya, Altmannshofer:2013zba, Dermisek:2014qca}, and should be thought of as conservative estimates. We consider separately the cases in which $L_1$ is mostly an SU(2)$_W$ singlet, doublet or triplet.

We compute the relevant cross sections at QCD NLO with {\tt Pospino}~\cite{Beenakker:1996ed} decoupling all supersymmetric partners except for Winos (Higgsinos) when considering SU(2)$_W$ triplets (doublets). The singlet Drell-Yan production cross section was computed at LO in~\cite{Chatrchyan:2013oca} using {\tt PYTHIA} v6.426~\cite{Sjostrand:2006za} and CTEQ6L1 PDFs~\cite{Pumplin:2002vw}. For doubly charged leptons we generate the models in UFO format~\cite{Degrande:2011ua} using FeynRules~\cite{Alloul:2013bka,Christensen:2008py} and compute the cross section at LO with {\tt Madgraph5}~\cite{Alwall:2011uj}.

We first discuss the case in which the lightest charged lepton is stable, relevant to the $H\gamma\gamma$ analysis, and then consider the case in which it decays to the SM states. For the latter possibility, the decay channels of $L_1$ are $W\nu, Z \tau(\ell), h \tau(\ell)$ with branching fractions depending on its electroweak quantum numbers. For instance, if $L_1$ is mostly an SU(2)$_W$ singlet or triplet and $m_{L_1} \gg m_h$, it will decay to $W\nu, Z \tau(\ell), h \tau(\ell)$ with ratio $2:1:1$. If $L_1$ is mostly an SU(2)$_W$ doublet, it will decay to $Z \tau(\ell), h \tau(\ell)$ with equal probability, but not to $W\nu$, again for $m_{L_1} \gg m_h$. In the following we use these ratios of decay probabilities to set our bounds. Note however that near kinematical thresholds there can be important differences. In particular, if $W\nu$ is the only allowed decay mode, LHC constraints essentially vanish, as discussed in Section~\ref{sec:WN}.

While quoting mass limits on $Q=1$ fermions we distinguish between singlets, doublets and triplets, referring to $Y=1$ singlets, $Y=1/2$ doublets and $Y=0$ triplets. Bounds on $Y=3/2$ doublets can be inferred from the numbers presented here by increasing the $Y=1/2$ doublets cross section by $\approx 50\%$. $Q=1$ fermions from $Y=1$ triplets instead have the same production cross section as $Q=1$ singlets.

In general we expect multiple leptons in the final state and we find that the strongest constraints at the moment come from the 8 TeV CMS multilepton search performed with an integrated luminosity of 19.5 fb$^{-1}$~\cite{Chatrchyan:2014aea}. We compute the asymptotic CLs defined in the appendix of~\cite{ArkaniHamed:2012kq} in each individual signal region of the search to constrain $\sigma \times \epsilon$ where $\sigma$ is the production cross section and $\epsilon$ includes acceptance, trigger and identification efficiencies, efficiencies of the kinematical cuts and the branching ratio of the vector-like lepton. In evaluating $\epsilon$, we computed the branching ratio for each signal region and assumed a flat 70\% efficiency times acceptance for electrons and muons, and an hadronic tau identification efficiency of 35\%, based on~\cite{2012JInst.7.1001C, PFT-10-004, PFT-08-001, thomas}. We also assume a flat 70\% efficiency for identifying a $b$-quark~\cite{Chatrchyan:2014aea}. We always quote the bound from the most constraining search region. Clearly, stronger bounds could be derived in a more refined analysis by combining the results in different regions.

\subsubsection{Stable states}\label{sec:stablept}
The same CMS search relevant to stable colored states also sets the strongest bound on stable leptons~\cite{Chatrchyan:2013oca}. For $Q=1$, the limits read $m_{L_1} > 574$~GeV for singlets, $m_{L_1} > 670$~GeV for doublets and $m_{L_1} > 800$~GeV for triplets. For $Q=2$ we find $m_{L^{++}} > 705$~GeV for a singlet or doublet, and $m_{L^{++}} > 790$~GeV for a triplet. 

\subsubsection{Prompt decays to charged first and second generation leptons}
Considering $Q=1$ states, a singlet $L_1$ that only mixes with the first two generation leptons must have $m_{L_1} > 140$ GeV, from the 4 lepton search region with missing $E_T < 50$ GeV, $H_T < 200$ GeV, one pair of leptons from a $Z$ decay, no hadronic $\tau$ and no $b$ jets. When $L_1$ is mostly a doublet the bound is $m_{L_1} > 260$ GeV,  from the 4 lepton search region with missing $E_T < 50$ GeV, $H_T < 200$ GeV, two pairs of opposite-sign same-flavor leptons where at least one pair originates from $Z$ decays, no hadronic $\tau$ and at least one $b$ jet. For a triplet we find $m_{L_1} > 230$ GeV from the same search region. For the case of doublet $Q=2$ states decaying to $W \ell$, Ref.~\cite{Altmannshofer:2013zba} obtains $m_{L^{++}}\gtrsim 460$~GeV. We expect the bound on singlets and triplets to be comparable.

\subsubsection{Prompt decays to charged third generation leptons}
Considering $Q=1$ states, in the singlet case, if $L_1$ only mixes with the $\tau$, the multilepton search bound on $m_{L_1}$ is below 100 GeV, weaker than the LEP bound. If $L_1$ is mostly a doublet we have $m_{L_1} > 170$ GeV from the 4 lepton search region with missing $E_T < 50$ GeV, $H_T < 200$ GeV, one pair of leptons from a $Z$ decay, no hadronic $\tau$ and no $b$ jets. If the lightest charged lepton is dominantly an SU(2)$_W$ triplet the mass bound is $m_{L_1} > 185$ GeV from the same search region. For the case of doublet $Q=2$ states decaying to $W \ell$, Ref.~\cite{Altmannshofer:2013zba} obtains $m_{L^{++}}\gtrsim 320$~GeV. We expect the bound on singlets and triplets to be comparable.

\subsubsection{Prompt decays to a W and missing energy}\label{sec:WN}
It is possible to add to our minimal models a SM singlet $N$ (a sterile neutrino or a bino) mixing with the doublets $L, L^c$ via Yukawa interactions. This can improve the agreement with electroweak precision tests~\cite{ArkaniHamed:2012kq}, though the additional Yukawa couplings imply somewhat more stringent vacuum stability constraints than would be obtained from the minimal models. 
In this setting the lightest charged $L_1$ lepton can decay predominantly to $W N$ and $N$ can be stable. If the decay is prompt, LEP still sets the most stringent constraint for small mass splittings between the neutral and charged fermion, giving $m_{L_1}\gtrsim 100$~GeV~\cite{Agashe:2014kda} for $m_{L_1}-m_N > 3$~GeV. The ATLAS search for charginos is also relevant~\cite{Aad:2014vma}, excluding masses up to $180$~GeV for triplets, provided that $m_N < 20$~GeV. If $N$ is stable it could be searched for in mono-jet and mono-boson final states but current LHC bounds are not strong enough to constrain electroweak production.

\subsubsection{Prompt cascade decays of $Q=2$ fermions}
In case of small mixing with the SM and largish mass difference between the new leptons, $L^{++}\to L_1^+ W^+$ can be the dominant decay mode. Assuming ${\rm BR}(L^{++} \to L_1^+W^+) = 1$ and that $L_1$ only decays to the third generation SM leptons, we used the CMS multilepton search~\cite{Chatrchyan:2014aea} to obtain $m_{L^{++}} > (140 - 250)$~GeV for a singlet and $m_{L^{++}} > (180 - 300)$~GeV for a triplet. The lower end of the range is achieved when ${\rm BR} (L_1^+ \to W\nu) = 1$ while the upper end of the range is achieved when ${\rm BR} (L_1^+ \to Z \tau) = 1$ or ${\rm BR} (L_1^+ \to h \tau) = 1$. If $L_1$ only decays to the first two generation SM leptons, we have $m_{L^{++}} > (140 - 420)$~GeV for a singlet and $m_{L^{++}} > (180 - 490)$~GeV for a triplet, where the lower end of the range is achieved when ${\rm BR} (L_1^+ \to W\nu) = 1$ while the upper end of the range is achieved when ${\rm BR} (L_1^+ \to h \ell) = 1$.

\bibliography{ref}

\providecommand{\href}[2]{#2}\begingroup\raggedright\begin{thebibliography}{100}

\bibitem{Chatrchyan:2014vua}
{ CMS} Collaboration, S.~Chatrchyan {\em et~al.}, ``{Evidence for the direct
  decay of the 125 GeV Higgs boson to fermions},''
  \href{http://dx.doi.org/10.1038/nphys3005}{{\em Nature Phys.} {\bfseries 10}
  (2014) },
\href{http://arxiv.org/abs/1401.6527}{{\ttfamily arXiv:1401.6527 [hep-ex]}}.

\bibitem{Khachatryan:2014ira}
{ CMS} Collaboration, V.~Khachatryan {\em et~al.}, ``{Observation of the
  diphoton decay of the Higgs boson and measurement of its properties},''
\href{http://arxiv.org/abs/1407.0558}{{\ttfamily arXiv:1407.0558 [hep-ex]}}.

\bibitem{CMScom}
{ CMS} Collaboration, ``Precise determination of the mass of the higgs boson
  and studies of the compatibility of its couplings with the standard model,''
  CMS Physics Analysis Summary, CMS-PAS-HIG-14-009, 2014.
\newblock \url{http://cds.cern.ch/record/1728249}.

\bibitem{Aad:2014xzb}
{ ATLAS} Collaboration, G.~Aad {\em et~al.}, ``{Search for the $b\bar{b}$ decay
  of the Standard Model Higgs boson in associated $(W/Z)H$ production with the
  ATLAS detector},''
\href{http://arxiv.org/abs/1409.6212}{{\ttfamily arXiv:1409.6212 [hep-ex]}}.

\bibitem{Aad:2014eha}
{ ATLAS} Collaboration, G.~Aad {\em et~al.}, ``{Measurement of Higgs boson
  production in the diphoton decay channel in $pp$ collisions at center-of-mass
  energies of 7 and 8 TeV with the ATLAS detector},''
\href{http://arxiv.org/abs/1408.7084}{{\ttfamily arXiv:1408.7084 [hep-ex]}}.

\bibitem{Aad:2014eva}
{ ATLAS} Collaboration, G.~Aad {\em et~al.}, ``{Measurements of Higgs boson
  production and couplings in the four-lepton channel in pp collisions at
  center-of-mass energies of 7 and 8 TeV with the ATLAS detector},''
\href{http://arxiv.org/abs/1408.5191}{{\ttfamily arXiv:1408.5191 [hep-ex]}}.

\bibitem{Dimopoulos:1981zb}
S.~Dimopoulos and H.~Georgi, ``{Softly Broken Supersymmetry and SU(5)},''
\href{http://dx.doi.org/10.1016/0550-3213(81)90522-8}{{\em Nucl.Phys.}
  {\bfseries B193} (1981) 150}.

\bibitem{Kaplan:1983fs}
D.~B. Kaplan and H.~Georgi, ``{SU(2) x U(1) Breaking by Vacuum Misalignment},''
\href{http://dx.doi.org/10.1016/0370-2693(84)91177-8}{{\em Phys.Lett.}
  {\bfseries B136} (1984) 183}.

\bibitem{Kaplan:1983sm}
D.~B. Kaplan, H.~Georgi, and S.~Dimopoulos, ``{Composite Higgs Scalars},''
\href{http://dx.doi.org/10.1016/0370-2693(84)91178-X}{{\em Phys.Lett.}
  {\bfseries B136} (1984) 187}.

\bibitem{Zhang:2000zy}
B.~Zhang and H.-q. Zheng, ``{Top quark, heavy fermions and the composite Higgs
  boson},'' \href{http://dx.doi.org/10.1088/0253-6102/35/2/162}{{\em
  Commun.Theor.Phys.} {\bfseries 35} (2001) 162--166},
\href{http://arxiv.org/abs/hep-ph/0003065}{{\ttfamily arXiv:hep-ph/0003065
  [hep-ph]}}.

\bibitem{Sher:1988mj}
M.~Sher, ``{Electroweak Higgs Potentials and Vacuum Stability},''
\href{http://dx.doi.org/10.1016/0370-1573(89)90061-6}{{\em Phys.Rept.}
  {\bfseries 179} (1989) 273--418}.

\bibitem{Altarelli:1994rb}
G.~Altarelli and G.~Isidori, ``{Lower limit on the Higgs mass in the standard
  model: An Update},''
\href{http://dx.doi.org/10.1016/0370-2693(94)91458-3}{{\em Phys.Lett.}
  {\bfseries B337} (1994) 141--144}.

\bibitem{Casas:1994qy}
J.~Casas, J.~Espinosa, and M.~Quiros, ``{Improved Higgs mass stability bound in
  the standard model and implications for supersymmetry},''
  \href{http://dx.doi.org/10.1016/0370-2693(94)01404-Z}{{\em Phys.Lett.}
  {\bfseries B342} (1995) 171--179},
\href{http://arxiv.org/abs/hep-ph/9409458}{{\ttfamily arXiv:hep-ph/9409458
  [hep-ph]}}.

\bibitem{Casas:1994us}
J.~Casas, J.~Espinosa, M.~Quiros, and A.~Riotto, ``{The Lightest Higgs boson
  mass in the minimal supersymmetric standard model},''
  \href{http://dx.doi.org/10.1016/0550-3213(94)00508-C}{{\em Nucl.Phys.}
  {\bfseries B436} (1995) 3--29},
\href{http://arxiv.org/abs/hep-ph/9407389}{{\ttfamily arXiv:hep-ph/9407389
  [hep-ph]}}.

\bibitem{Espinosa:1995se}
J.~Espinosa and M.~Quiros, ``{Improved metastability bounds on the standard
  model Higgs mass},''
  \href{http://dx.doi.org/10.1016/0370-2693(95)00572-3}{{\em Phys.Lett.}
  {\bfseries B353} (1995) 257--266},
\href{http://arxiv.org/abs/hep-ph/9504241}{{\ttfamily arXiv:hep-ph/9504241
  [hep-ph]}}.

\bibitem{Hung:1995in}
P.~Hung and M.~Sher, ``{Implications of a Higgs discovery at LEP},''
  \href{http://dx.doi.org/10.1016/0370-2693(96)00123-2}{{\em Phys.Lett.}
  {\bfseries B374} (1996) 138--144},
\href{http://arxiv.org/abs/hep-ph/9512313}{{\ttfamily arXiv:hep-ph/9512313
  [hep-ph]}}.

\bibitem{Casas:1996aq}
J.~Casas, J.~Espinosa, and M.~Quiros, ``{Standard model stability bounds for
  new physics within LHC reach},''
  \href{http://dx.doi.org/10.1016/0370-2693(96)00682-X}{{\em Phys.Lett.}
  {\bfseries B382} (1996) 374--382},
\href{http://arxiv.org/abs/hep-ph/9603227}{{\ttfamily arXiv:hep-ph/9603227
  [hep-ph]}}.

\bibitem{Degrassi:2012ry}
G.~Degrassi, S.~Di~Vita, J.~Elias-Miro, J.~R. Espinosa, G.~F. Giudice, {\em
  et~al.}, ``{Higgs mass and vacuum stability in the Standard Model at NNLO},''
  \href{http://dx.doi.org/10.1007/JHEP08(2012)098}{{\em JHEP} {\bfseries 1208}
  (2012) 098},
\href{http://arxiv.org/abs/1205.6497}{{\ttfamily arXiv:1205.6497 [hep-ph]}}.

\bibitem{Andreassen:2014gha}
A.~Andreassen, W.~Frost, and M.~D. Schwartz, ``{Consistent Use of the Standard
  Model Effective Potential},''
\href{http://arxiv.org/abs/1408.0292}{{\ttfamily arXiv:1408.0292 [hep-ph]}}.

\bibitem{Branchina:2014rva}
V.~Branchina, E.~Messina, and M.~Sher, ``{The lifetime of the electroweak
  vacuum and sensitivity to Planck scale physics},''
\href{http://arxiv.org/abs/1408.5302}{{\ttfamily arXiv:1408.5302 [hep-ph]}}.

\bibitem{Peskin:2012we}
M.~E. Peskin, ``{Comparison of LHC and ILC Capabilities for Higgs Boson
  Coupling Measurements},''
\href{http://arxiv.org/abs/1207.2516}{{\ttfamily arXiv:1207.2516 [hep-ph]}}.

\bibitem{ArkaniHamed:2004fb}
N.~Arkani-Hamed and S.~Dimopoulos, ``{Supersymmetric unification without low
  energy supersymmetry and signatures for fine-tuning at the LHC},''
  \href{http://dx.doi.org/10.1088/1126-6708/2005/06/073}{{\em JHEP} {\bfseries
  0506} (2005) 073},
\href{http://arxiv.org/abs/hep-th/0405159}{{\ttfamily arXiv:hep-th/0405159
  [hep-th]}}.

\bibitem{Giudice:2004tc}
G.~Giudice and A.~Romanino, ``{Split supersymmetry},''
  \href{http://dx.doi.org/10.1016/j.nuclphysb.2004.11.048}{{\em Nucl.Phys.}
  {\bfseries B699} (2004) 65--89},
\href{http://arxiv.org/abs/hep-ph/0406088}{{\ttfamily arXiv:hep-ph/0406088
  [hep-ph]}}.

\bibitem{ArkaniHamed:2004yi}
N.~Arkani-Hamed, S.~Dimopoulos, G.~Giudice, and A.~Romanino, ``{Aspects of
  split supersymmetry},''
  \href{http://dx.doi.org/10.1016/j.nuclphysb.2004.12.026}{{\em Nucl.Phys.}
  {\bfseries B709} (2005) 3--46},
\href{http://arxiv.org/abs/hep-ph/0409232}{{\ttfamily arXiv:hep-ph/0409232
  [hep-ph]}}.

\bibitem{Wells:2004di}
J.~D. Wells, ``{PeV-scale supersymmetry},''
  \href{http://dx.doi.org/10.1103/PhysRevD.71.015013}{{\em Phys.Rev.}
  {\bfseries D71} (2005) 015013},
\href{http://arxiv.org/abs/hep-ph/0411041}{{\ttfamily arXiv:hep-ph/0411041
  [hep-ph]}}.

\bibitem{Asner:2013psa}
D.~Asner, T.~Barklow, C.~Calancha, K.~Fujii, N.~Graf, {\em et~al.}, ``{ILC
  Higgs White Paper},''
\href{http://arxiv.org/abs/1310.0763}{{\ttfamily arXiv:1310.0763 [hep-ph]}}.

\bibitem{Gomez-Ceballos:2013zzn}
{ TLEP Design Study Working Group} Collaboration, M.~Bicer {\em et~al.},
  ``{First Look at the Physics Case of TLEP},''
  \href{http://dx.doi.org/10.1007/JHEP01(2014)164}{{\em JHEP} {\bfseries 1401}
  (2014) 164},
\href{http://arxiv.org/abs/1308.6176}{{\ttfamily arXiv:1308.6176 [hep-ex]}}.

\bibitem{Dawson:2013bba}
S.~Dawson, A.~Gritsan, H.~Logan, J.~Qian, C.~Tully, {\em et~al.}, ``{Working
  Group Report: Higgs Boson},''
\href{http://arxiv.org/abs/1310.8361}{{\ttfamily arXiv:1310.8361 [hep-ex]}}.

\bibitem{ZhuTalk}
S.~Zhu, ``{Report from Theory: CEPC/CPPC Pre-cdr Status},''.
  \url{http://indico.ihep.ac.cn/getFile.py/access?http://indico.ihep.ac.cn/getFile.py/access?contribId=10&sessionId=6&resId=0&materialId=slides&confId=4338}.

\bibitem{Fan:2014vta}
J.~Fan, M.~Reece, and L.-T. Wang, ``{Possible Futures of Electroweak Precision:
  ILC, FCC-ee, and CEPC},''
\href{http://arxiv.org/abs/1411.1054}{{\ttfamily arXiv:1411.1054 [hep-ph]}}.

\bibitem{Joglekar:2012vc}
A.~Joglekar, P.~Schwaller, and C.~E. Wagner, ``{Dark Matter and Enhanced Higgs
  to Di-photon Rate from Vector-like Leptons},''
  \href{http://dx.doi.org/10.1007/JHEP12(2012)064}{{\em JHEP} {\bfseries 1212}
  (2012) 064},
\href{http://arxiv.org/abs/1207.4235}{{\ttfamily arXiv:1207.4235 [hep-ph]}}.

\bibitem{Kearney:2012zi}
J.~Kearney, A.~Pierce, and N.~Weiner, ``{Vectorlike Fermions and Higgs
  Couplings},'' \href{http://dx.doi.org/10.1103/PhysRevD.86.113005}{{\em
  Phys.Rev.} {\bfseries D86} (2012) 113005},
\href{http://arxiv.org/abs/1207.7062}{{\ttfamily arXiv:1207.7062 [hep-ph]}}.

\bibitem{Batell:2012ca}
B.~Batell, S.~Gori, and L.-T. Wang, ``{Higgs Couplings and Precision
  Electroweak Data},'' \href{http://dx.doi.org/10.1007/JHEP01(2013)139}{{\em
  JHEP} {\bfseries 1301} (2013) 139},
\href{http://arxiv.org/abs/1209.6382}{{\ttfamily arXiv:1209.6382 [hep-ph]}}.

\bibitem{Altmannshofer:2013zba}
W.~Altmannshofer, M.~Bauer, and M.~Carena, ``{Exotic Leptons: Higgs, Flavor and
  Collider Phenomenology},''
  \href{http://dx.doi.org/10.1007/JHEP01(2014)060}{{\em JHEP} {\bfseries 1401}
  (2014) 060},
\href{http://arxiv.org/abs/1308.1987}{{\ttfamily arXiv:1308.1987 [hep-ph]}}.

\bibitem{Ellis:2014dza}
S.~A. Ellis, R.~M. Godbole, S.~Gopalakrishna, and J.~D. Wells, ``{Survey of
  vector-like fermion extensions of the Standard Model and their
  phenomenological implications},''
\href{http://arxiv.org/abs/1404.4398}{{\ttfamily arXiv:1404.4398 [hep-ph]}}.

\bibitem{Fairbairn:2013xaa}
M.~Fairbairn and P.~Grothaus, ``{Baryogenesis and Dark Matter with Vector-like
  Fermions},'' \href{http://dx.doi.org/10.1007/JHEP10(2013)176}{{\em JHEP}
  {\bfseries 1310} (2013) 176},
\href{http://arxiv.org/abs/1307.8011}{{\ttfamily arXiv:1307.8011 [hep-ph]}}.

\bibitem{Maru:2013ooa}
N.~Maru and N.~Okada, ``{Diphoton decay excess and 125 GeV Higgs boson in
  gauge-Higgs unification},''
  \href{http://dx.doi.org/10.1103/PhysRevD.87.095019}{{\em Phys.Rev.}
  {\bfseries D87} no.~9, (2013) 095019},
\href{http://arxiv.org/abs/1303.5810}{{\ttfamily arXiv:1303.5810 [hep-ph]}}.

\bibitem{Feng:2013mea}
W.-Z. Feng and P.~Nath, ``{Higgs diphoton rate and mass enhancement with
  vectorlike leptons and the scale of supersymmetry},''
  \href{http://dx.doi.org/10.1103/PhysRevD.87.075018}{{\em Phys.Rev.}
  {\bfseries D87} no.~7, (2013) 075018},
\href{http://arxiv.org/abs/1303.0289}{{\ttfamily arXiv:1303.0289 [hep-ph]}}.

\bibitem{Huo:2012tw}
R.~Huo, G.~Lee, A.~M. Thalapillil, and C.~E. Wagner, ``{SU(2)SU(2) gauge
  extensions of the MSSM revisited},''
  \href{http://dx.doi.org/10.1103/PhysRevD.87.055011}{{\em Phys.Rev.}
  {\bfseries D87} no.~5, (2013) 055011},
\href{http://arxiv.org/abs/1212.0560}{{\ttfamily arXiv:1212.0560 [hep-ph]}}.

\bibitem{Carena:2012mw}
M.~Carena, S.~Gori, I.~Low, N.~R. Shah, and C.~E. Wagner, ``{Vacuum Stability
  and Higgs Diphoton Decays in the MSSM},''
  \href{http://dx.doi.org/10.1007/JHEP02(2013)114}{{\em JHEP} {\bfseries 1302}
  (2013) 114},
\href{http://arxiv.org/abs/1211.6136}{{\ttfamily arXiv:1211.6136 [hep-ph]}}.

\bibitem{Davoudiasl:2012tu}
H.~Davoudiasl, I.~Lewis, and E.~Ponton, ``{Electroweak Phase Transition, Higgs
  Diphoton Rate, and New Heavy Fermions},''
  \href{http://dx.doi.org/10.1103/PhysRevD.87.093001}{{\em Phys.Rev.}
  {\bfseries D87} (2013) 093001},
\href{http://arxiv.org/abs/1211.3449}{{\ttfamily arXiv:1211.3449 [hep-ph]}}.

\bibitem{Batell:2012zw}
B.~Batell, S.~Jung, and H.~M. Lee, ``{Singlet Assisted Vacuum Stability and the
  Higgs to Diphoton Rate},''
  \href{http://dx.doi.org/10.1007/JHEP01(2013)135}{{\em JHEP} {\bfseries 1301}
  (2013) 135},
\href{http://arxiv.org/abs/1211.2449}{{\ttfamily arXiv:1211.2449 [hep-ph]}}.

\bibitem{Lee:2012wz}
H.~M. Lee, M.~Park, and W.-I. Park, ``{Axion-mediated dark matter and Higgs
  diphoton signal},'' \href{http://dx.doi.org/10.1007/JHEP12(2012)037}{{\em
  JHEP} {\bfseries 1212} (2012) 037},
\href{http://arxiv.org/abs/1209.1955}{{\ttfamily arXiv:1209.1955 [hep-ph]}}.

\bibitem{Chun:2012jw}
E.~J. Chun, H.~M. Lee, and P.~Sharma, ``{Vacuum Stability, Perturbativity, EWPD
  and Higgs-to-diphoton rate in Type II Seesaw Models},''
  \href{http://dx.doi.org/10.1007/JHEP11(2012)106}{{\em JHEP} {\bfseries 1211}
  (2012) 106},
\href{http://arxiv.org/abs/1209.1303}{{\ttfamily arXiv:1209.1303 [hep-ph]}}.

\bibitem{Kobakhidze:2012wb}
A.~Kobakhidze, ``{Standard Model with a distorted Higgs sector and the enhanced
  Higgs diphoton decay rate},''
\href{http://arxiv.org/abs/1208.5180}{{\ttfamily arXiv:1208.5180 [hep-ph]}}.

\bibitem{Kitahara:2012pb}
T.~Kitahara, ``{Vacuum Stability Constraints on the Enhancement of the h to
  gamma gamma rate in the MSSm},''
  \href{http://dx.doi.org/10.1007/JHEP11(2012)021}{{\em JHEP} {\bfseries 1211}
  (2012) 021},
\href{http://arxiv.org/abs/1208.4792}{{\ttfamily arXiv:1208.4792 [hep-ph]}}.

\bibitem{Giudice:2012pf}
G.~F. Giudice, P.~Paradisi, A.~Strumia, and A.~Strumia, ``{Correlation between
  the Higgs Decay Rate to Two Photons and the Muon g - 2},''
  \href{http://dx.doi.org/10.1007/JHEP10(2012)186}{{\em JHEP} {\bfseries 1210}
  (2012) 186},
\href{http://arxiv.org/abs/1207.6393}{{\ttfamily arXiv:1207.6393 [hep-ph]}}.

\bibitem{ArkaniHamed:2012kq}
N.~Arkani-Hamed, K.~Blum, R.~T. D'Agnolo, and J.~Fan, ``{2:1 for Naturalness at
  the LHC?},'' \href{http://dx.doi.org/10.1007/JHEP01(2013)149}{{\em JHEP}
  {\bfseries 1301} (2013) 149},
\href{http://arxiv.org/abs/1207.4482}{{\ttfamily arXiv:1207.4482 [hep-ph]}}.

\bibitem{Reece:2012gi}
M.~Reece, ``{Vacuum Instabilities with a Wrong-Sign Higgs-Gluon-Gluon
  Amplitude},'' \href{http://dx.doi.org/10.1088/1367-2630/15/4/043003}{{\em New
  J.Phys.} {\bfseries 15} (2013) 043003},
\href{http://arxiv.org/abs/1208.1765}{{\ttfamily arXiv:1208.1765 [hep-ph]}}.

\bibitem{Choudhury:2001hs}
D.~Choudhury, T.~M. Tait, and C.~Wagner, ``{Beautiful mirrors and precision
  electroweak data},'' \href{http://dx.doi.org/10.1103/PhysRevD.65.053002}{{\em
  Phys.Rev.} {\bfseries D65} (2002) 053002},
\href{http://arxiv.org/abs/hep-ph/0109097}{{\ttfamily arXiv:hep-ph/0109097
  [hep-ph]}}.

\bibitem{Morrissey:2003sc}
D.~Morrissey and C.~Wagner, ``{Beautiful mirrors, unification of couplings and
  collider phenomenology},''
  \href{http://dx.doi.org/10.1103/PhysRevD.69.053001}{{\em Phys.Rev.}
  {\bfseries D69} (2004) 053001},
\href{http://arxiv.org/abs/hep-ph/0308001}{{\ttfamily arXiv:hep-ph/0308001
  [hep-ph]}}.

\bibitem{Dawson:2012di}
S.~Dawson and E.~Furlan, ``{A Higgs Conundrum with Vector Fermions},''
  \href{http://dx.doi.org/10.1103/PhysRevD.86.015021}{{\em Phys.Rev.}
  {\bfseries D86} (2012) 015021},
\href{http://arxiv.org/abs/1205.4733}{{\ttfamily arXiv:1205.4733 [hep-ph]}}.

\bibitem{Isidori:2001bm}
G.~Isidori, G.~Ridolfi, and A.~Strumia, ``{On the metastability of the standard
  model vacuum},'' \href{http://dx.doi.org/10.1016/S0550-3213(01)00302-9}{{\em
  Nucl.Phys.} {\bfseries B609} (2001) 387--409},
\href{http://arxiv.org/abs/hep-ph/0104016}{{\ttfamily arXiv:hep-ph/0104016
  [hep-ph]}}.

\bibitem{Ciuchini:2013pca}
M.~Ciuchini, E.~Franco, S.~Mishima, and L.~Silvestrini, ``{Electroweak
  Precision Observables, New Physics and the Nature of a 126 GeV Higgs
  Boson},'' \href{http://dx.doi.org/10.1007/JHEP08(2013)106}{{\em JHEP}
  {\bfseries 1308} (2013) 106},
\href{http://arxiv.org/abs/1306.4644}{{\ttfamily arXiv:1306.4644 [hep-ph]}}.

\bibitem{Agashe:2014kda}
{ Particle Data Group} Collaboration, K.~Olive {\em et~al.}, ``{Review of
  Particle Physics},''
\href{http://dx.doi.org/10.1088/1674-1137/38/9/090001}{{\em Chin.Phys.}
  {\bfseries C38} (2014) 090001}.

\bibitem{Baer:2013cma}
H.~Baer, T.~Barklow, K.~Fujii, Y.~Gao, A.~Hoang, {\em et~al.}, ``{The
  International Linear Collider Technical Design Report - Volume 2: Physics},''
\href{http://arxiv.org/abs/1306.6352}{{\ttfamily arXiv:1306.6352 [hep-ph]}}.

\bibitem{Elias-Miro:2013mua}
J.~Elias-Miro, J.~Espinosa, E.~Masso, and A.~Pomarol, ``{Higgs windows to new
  physics through d=6 operators: constraints and one-loop anomalous
  dimensions},'' \href{http://dx.doi.org/10.1007/JHEP11(2013)066}{{\em JHEP}
  {\bfseries 1311} (2013) 066},
\href{http://arxiv.org/abs/1308.1879}{{\ttfamily arXiv:1308.1879 [hep-ph]}}.

\bibitem{Jenkins:2013zja}
E.~E. Jenkins, A.~V. Manohar, and M.~Trott, ``{Renormalization Group Evolution
  of the Standard Model Dimension Six Operators I: Formalism and lambda
  Dependence},'' \href{http://dx.doi.org/10.1007/JHEP10(2013)087}{{\em JHEP}
  {\bfseries 1310} (2013) 087},
\href{http://arxiv.org/abs/1308.2627}{{\ttfamily arXiv:1308.2627 [hep-ph]}}.

\bibitem{Jenkins:2013wua}
E.~E. Jenkins, A.~V. Manohar, and M.~Trott, ``{Renormalization Group Evolution
  of the Standard Model Dimension Six Operators II: Yukawa Dependence},''
  \href{http://dx.doi.org/10.1007/JHEP01(2014)035}{{\em JHEP} {\bfseries 1401}
  (2014) 035},
\href{http://arxiv.org/abs/1310.4838}{{\ttfamily arXiv:1310.4838 [hep-ph]}}.

\bibitem{Alonso:2013hga}
R.~Alonso, E.~E. Jenkins, A.~V. Manohar, and M.~Trott, ``{Renormalization Group
  Evolution of the Standard Model Dimension Six Operators III: Gauge Coupling
  Dependence and Phenomenology},''
  \href{http://dx.doi.org/10.1007/JHEP04(2014)159}{{\em JHEP} {\bfseries 1404}
  (2014) 159},
\href{http://arxiv.org/abs/1312.2014}{{\ttfamily arXiv:1312.2014 [hep-ph]}}.

\bibitem{Baak:2014ora}
{ Gfitter Group} Collaboration, M.~Baak {\em et~al.}, ``{The global electroweak
  fit at NNLO and prospects for the LHC and ILC},''
  \href{http://dx.doi.org/10.1140/epjc/s10052-014-3046-5}{{\em Eur.Phys.J.}
  {\bfseries C74} no.~9, (2014) 3046},
\href{http://arxiv.org/abs/1407.3792}{{\ttfamily arXiv:1407.3792 [hep-ph]}}.

\bibitem{Brod:2014hsa}
J.~Brod, A.~Greljo, E.~Stamou, and P.~Uttayarat, ``{Probing anomalous $t\bar t
  Z$ interactions with rare meson decays},''
\href{http://arxiv.org/abs/1408.0792}{{\ttfamily arXiv:1408.0792 [hep-ph]}}.

\bibitem{Ellis:1975ap}
J.~R. Ellis, M.~K. Gaillard, and D.~V. Nanopoulos, ``{A Phenomenological
  Profile of the Higgs Boson},''
\href{http://dx.doi.org/10.1016/0550-3213(76)90382-5}{{\em Nucl.Phys.}
  {\bfseries B106} (1976) 292}.

\bibitem{Shifman:1979eb}
M.~A. Shifman, A.~Vainshtein, M.~Voloshin, and V.~I. Zakharov, ``{Low-Energy
  Theorems for Higgs Boson Couplings to Photons},''
{\em Sov.J.Nucl.Phys.} {\bfseries 30} (1979) 711--716.

\bibitem{Djouadi:2005gi}
A.~Djouadi, ``{The Anatomy of electro-weak symmetry breaking. I: The Higgs
  boson in the standard model},''
  \href{http://dx.doi.org/10.1016/j.physrep.2007.10.004}{{\em Phys.Rept.}
  {\bfseries 457} (2008) 1--216},
\href{http://arxiv.org/abs/hep-ph/0503172}{{\ttfamily arXiv:hep-ph/0503172
  [hep-ph]}}.

\bibitem{Grojean:2013kd}
C.~Grojean, E.~E. Jenkins, A.~V. Manohar, and M.~Trott, ``{Renormalization
  Group Scaling of Higgs Operators and $h \to\gamma \gamma$ decay},''
  \href{http://dx.doi.org/10.1007/JHEP04(2013)016}{{\em JHEP} {\bfseries 1304}
  (2013) 016},
\href{http://arxiv.org/abs/1301.2588}{{\ttfamily arXiv:1301.2588 [hep-ph]}}.

\bibitem{Appelquist:1974tg}
T.~Appelquist and J.~Carazzone, ``{Infrared Singularities and Massive
  Fields},''
\href{http://dx.doi.org/10.1103/PhysRevD.11.2856}{{\em Phys.Rev.} {\bfseries
  D11} (1975) 2856}.

\bibitem{Lyonnet:2013dna}
F.~Lyonnet, I.~Schienbein, F.~Staub, and A.~Wingerter, ``{PyR@TE:
  Renormalization Group Equations for General Gauge Theories},''
  \href{http://dx.doi.org/10.1016/j.cpc.2013.12.002}{{\em Comput.Phys.Commun.}
  {\bfseries 185} (2014) 1130--1152},
\href{http://arxiv.org/abs/1309.7030}{{\ttfamily arXiv:1309.7030 [hep-ph]}}.

\bibitem{Buttazzo:2013uya}
D.~Buttazzo, G.~Degrassi, P.~P. Giardino, G.~F. Giudice, F.~Sala, {\em et~al.},
  ``{Investigating the near-criticality of the Higgs boson},''
  \href{http://dx.doi.org/10.1007/JHEP12(2013)089}{{\em JHEP} {\bfseries 1312}
  (2013) 089},
\href{http://arxiv.org/abs/1307.3536}{{\ttfamily arXiv:1307.3536 [hep-ph]}}.

\bibitem{delAguila:2000rc}
F.~del Aguila, M.~Perez-Victoria, and J.~Santiago, ``{Observable contributions
  of new exotic quarks to quark mixing},''
  \href{http://dx.doi.org/10.1088/1126-6708/2000/09/011}{{\em JHEP} {\bfseries
  0009} (2000) 011},
\href{http://arxiv.org/abs/hep-ph/0007316}{{\ttfamily arXiv:hep-ph/0007316
  [hep-ph]}}.

\bibitem{delAguila:2000aa}
F.~del Aguila, M.~Perez-Victoria, and J.~Santiago, ``{Effective description of
  quark mixing},'' \href{http://dx.doi.org/10.1016/S0370-2693(00)01071-6}{{\em
  Phys.Lett.} {\bfseries B492} (2000) 98--106},
\href{http://arxiv.org/abs/hep-ph/0007160}{{\ttfamily arXiv:hep-ph/0007160
  [hep-ph]}}.

\bibitem{Buchmuller:1985jz}
W.~Buchmuller and D.~Wyler, ``{Effective Lagrangian Analysis of New
  Interactions and Flavor Conservation},''
\href{http://dx.doi.org/10.1016/0550-3213(86)90262-2}{{\em Nucl.Phys.}
  {\bfseries B268} (1986) 621--653}.

\bibitem{Grzadkowski:2010es}
B.~Grzadkowski, M.~Iskrzynski, M.~Misiak, and J.~Rosiek, ``{Dimension-Six Terms
  in the Standard Model Lagrangian},''
  \href{http://dx.doi.org/10.1007/JHEP10(2010)085}{{\em JHEP} {\bfseries 1010}
  (2010) 085},
\href{http://arxiv.org/abs/1008.4884}{{\ttfamily arXiv:1008.4884 [hep-ph]}}.

\bibitem{Barbieri:1999tm}
R.~Barbieri and A.~Strumia, ``{What is the limit on the Higgs mass?},''
  \href{http://dx.doi.org/10.1016/S0370-2693(99)00882-5}{{\em Phys.Lett.}
  {\bfseries B462} (1999) 144--149},
\href{http://arxiv.org/abs/hep-ph/9905281}{{\ttfamily arXiv:hep-ph/9905281
  [hep-ph]}}.

\bibitem{Falkowski:2013jya}
A.~Falkowski, D.~M. Straub, and A.~Vicente, ``{Vector-like leptons: Higgs
  decays and collider phenomenology},''
  \href{http://dx.doi.org/10.1007/JHEP05(2014)092}{{\em JHEP} {\bfseries 1405}
  (2014) 092},
\href{http://arxiv.org/abs/1312.5329}{{\ttfamily arXiv:1312.5329 [hep-ph]}}.

\bibitem{Dermisek:2014qca}
R.~Dermíšek, J.~P. Hall, E.~Lunghi, and S.~Shin, ``{Limits on Vectorlike
  Leptons from Searches for Anomalous Production of Multi-Lepton Events},''
\href{http://arxiv.org/abs/1408.3123}{{\ttfamily arXiv:1408.3123 [hep-ph]}}.

\bibitem{CMS:2013oea}
{ CMS} Collaboration, ``{Search for long-lived neutral particles decaying to
  dijets},'' CMS Physics Analysis Summary, CMS-PAS-EXO-12-038, 2013.
\newblock
  \url{https://twiki.cern.ch/twiki/bin/view/CMSPublic/PhysicsResultsEXO12038}.

\bibitem{CMS:2014mca}
{ CMS} Collaboration, ``{Search for long-lived particles decaying to final
  states that include dileptons},'' CMS Physics Analysis Summary,
  CMS-PAS-EXO-12-037, 2014.
\newblock \url{https://cds.cern.ch/record/1669814?ln=en}.

\bibitem{Chatrchyan:2013uxa}
{ CMS} Collaboration, S.~Chatrchyan {\em et~al.}, ``{Inclusive search for a
  vector-like T quark with charge 2/3 in pp collisions at sqrt(s) = 8 TeV},''
  \href{http://dx.doi.org/10.1016/j.physletb.2014.01.006}{{\em Phys.Lett.}
  {\bfseries B729} (2014) 149--171},
\href{http://arxiv.org/abs/1311.7667}{{\ttfamily arXiv:1311.7667 [hep-ex]}}.

\bibitem{Aliev:2010zk}
M.~Aliev, H.~Lacker, U.~Langenfeld, S.~Moch, P.~Uwer, {\em et~al.}, ``{HATHOR:
  HAdronic Top and Heavy quarks crOss section calculatoR},''
  \href{http://dx.doi.org/10.1016/j.cpc.2010.12.040}{{\em Comput.Phys.Commun.}
  {\bfseries 182} (2011) 1034--1046},
\href{http://arxiv.org/abs/1007.1327}{{\ttfamily arXiv:1007.1327 [hep-ph]}}.

\bibitem{Beenakker:1996ch}
W.~Beenakker, R.~Hopker, M.~Spira, and P.~Zerwas, ``{Squark and gluino
  production at hadron colliders},''
  \href{http://dx.doi.org/10.1016/S0550-3213(97)80027-2}{{\em Nucl.Phys.}
  {\bfseries B492} (1997) 51--103},
\href{http://arxiv.org/abs/hep-ph/9610490}{{\ttfamily arXiv:hep-ph/9610490
  [hep-ph]}}.

\bibitem{Kulesza:2008jb}
A.~Kulesza and L.~Motyka, ``{Threshold resummation for squark-antisquark and
  gluino-pair production at the LHC},''
  \href{http://dx.doi.org/10.1103/PhysRevLett.102.111802}{{\em Phys.Rev.Lett.}
  {\bfseries 102} (2009) 111802},
\href{http://arxiv.org/abs/0807.2405}{{\ttfamily arXiv:0807.2405 [hep-ph]}}.

\bibitem{Kulesza:2009kq}
A.~Kulesza and L.~Motyka, ``{Soft gluon resummation for the production of
  gluino-gluino and squark-antisquark pairs at the LHC},''
  \href{http://dx.doi.org/10.1103/PhysRevD.80.095004}{{\em Phys.Rev.}
  {\bfseries D80} (2009) 095004},
\href{http://arxiv.org/abs/0905.4749}{{\ttfamily arXiv:0905.4749 [hep-ph]}}.

\bibitem{Beenakker:2009ha}
W.~Beenakker, S.~Brensing, M.~Kramer, A.~Kulesza, E.~Laenen, {\em et~al.},
  ``{Soft-gluon resummation for squark and gluino hadroproduction},''
  \href{http://dx.doi.org/10.1088/1126-6708/2009/12/041}{{\em JHEP} {\bfseries
  0912} (2009) 041},
\href{http://arxiv.org/abs/0909.4418}{{\ttfamily arXiv:0909.4418 [hep-ph]}}.

\bibitem{Beenakker:2011fu}
W.~Beenakker, S.~Brensing, M.~Kramer, A.~Kulesza, E.~Laenen, {\em et~al.},
  ``{Squark and Gluino Hadroproduction},''
  \href{http://dx.doi.org/10.1142/S0217751X11053560}{{\em Int.J.Mod.Phys.}
  {\bfseries A26} (2011) 2637--2664},
\href{http://arxiv.org/abs/1105.1110}{{\ttfamily arXiv:1105.1110 [hep-ph]}}.

\bibitem{Ilisie:2012cc}
V.~Ilisie and A.~Pich, ``{QCD exotics versus a Standard Model Higgs},''
  \href{http://dx.doi.org/10.1103/PhysRevD.86.033001}{{\em Phys.Rev.}
  {\bfseries D86} (2012) 033001},
\href{http://arxiv.org/abs/1202.3420}{{\ttfamily arXiv:1202.3420 [hep-ph]}}.

\bibitem{Chatrchyan:2013oca}
{ CMS} Collaboration, S.~Chatrchyan {\em et~al.}, ``{Searches for long-lived
  charged particles in pp collisions at $\sqrt{s}$=7 and 8 TeV},''
  \href{http://dx.doi.org/10.1007/JHEP07(2013)122}{{\em JHEP} {\bfseries 1307}
  (2013) 122},
\href{http://arxiv.org/abs/1305.0491}{{\ttfamily arXiv:1305.0491 [hep-ex]}}.

\bibitem{Mackeprang:2009ad}
R.~Mackeprang and D.~Milstead, ``{An Updated Description of Heavy-Hadron
  Interactions in GEANT-4},''
  \href{http://dx.doi.org/10.1140/epjc/s10052-010-1262-1}{{\em Eur.Phys.J.}
  {\bfseries C66} (2010) 493--501},
\href{http://arxiv.org/abs/0908.1868}{{\ttfamily arXiv:0908.1868 [hep-ph]}}.

\bibitem{Kraan:2004tz}
A.~C. Kraan, ``{Interactions of heavy stable hadronizing particles},''
  \href{http://dx.doi.org/10.1140/epjc/s2004-01997-7}{{\em Eur.Phys.J.}
  {\bfseries C37} (2004) 91--104},
\href{http://arxiv.org/abs/hep-ex/0404001}{{\ttfamily arXiv:hep-ex/0404001
  [hep-ex]}}.

\bibitem{Mackeprang:2006gx}
R.~Mackeprang and A.~Rizzi, ``{Interactions of Coloured Heavy Stable Particles
  in Matter},'' \href{http://dx.doi.org/10.1140/epjc/s10052-007-0252-4}{{\em
  Eur.Phys.J.} {\bfseries C50} (2007) 353--362},
\href{http://arxiv.org/abs/hep-ph/0612161}{{\ttfamily arXiv:hep-ph/0612161
  [hep-ph]}}.

\bibitem{D'Agnolo:2012ie}
R.~T. D'Agnolo and D.~M. Straub, ``{Gauged flavour symmetry for the light
  generations},'' \href{http://dx.doi.org/10.1007/JHEP05(2012)034}{{\em JHEP}
  {\bfseries 1205} (2012) 034},
\href{http://arxiv.org/abs/1202.4759}{{\ttfamily arXiv:1202.4759 [hep-ph]}}.

\bibitem{Aad:2012bt}
{ ATLAS} Collaboration, G.~Aad {\em et~al.}, ``{Search for pair-produced heavy
  quarks decaying to Wq in the two-lepton channel at $\sqrt{s}=7$ TeV with the
  ATLAS detector},'' \href{http://dx.doi.org/10.1103/PhysRevD.86.012007}{{\em
  Phys.Rev.} {\bfseries D86} (2012) 012007},
\href{http://arxiv.org/abs/1202.3389}{{\ttfamily arXiv:1202.3389 [hep-ex]}}.

\bibitem{TheATLAScollaboration:2013xia}
{ ATLAS} Collaboration, ``Search for massive particles in multijet signatures
  with the atlas detector in $\sqrt{s} = 8$ tev pp collisions at the lhc,''
  ATLAS Conference Note, ATLAS-CONF-2013-091, 2013.
\newblock \url{https://cds.cern.ch/record/1595753}.

\bibitem{Chatrchyan:2014aea}
{ CMS} Collaboration, S.~Chatrchyan {\em et~al.}, ``{Search for anomalous
  production of events with three or more leptons in pp collisions at
  $\sqrt{s}$=8 TeV},''
\href{http://arxiv.org/abs/1404.5801}{{\ttfamily arXiv:1404.5801 [hep-ex]}}.

\bibitem{Khachatryan:2014lpa}
{ CMS Collaboration} Collaboration, V.~Khachatryan {\em et~al.}, ``{Search for
  pair-produced resonances decaying to jet pairs in proton-proton collisions at
  $\sqrt{s}$ = 8 TeV},''
\href{http://arxiv.org/abs/1412.7706}{{\ttfamily arXiv:1412.7706 [hep-ex]}}.

\bibitem{Chatrchyan:2013gia}
{ CMS Collaboration} Collaboration, S.~Chatrchyan {\em et~al.}, ``{Searches for
  light- and heavy-flavour three-jet resonances in pp collisions at $\sqrt{s} =
  8$ TeV},'' \href{http://dx.doi.org/10.1016/j.physletb.2014.01.049}{{\em
  Phys.Lett.} {\bfseries B730} (2014) 193--214},
\href{http://arxiv.org/abs/1311.1799}{{\ttfamily arXiv:1311.1799 [hep-ex]}}.

\bibitem{ATLAScrazy}
T.~A. collaboration, ``{A general search for new phenomena with the ATLAS
  detector in pp collisions at $\sqrt{s}=8$ TeV},''.
\url{https://cds.cern.ch/record/1666536/files/ATLAS-CONF-2014-006.pdf}.

\bibitem{CMS-PAS-B2G-12-019}
{ CMS} Collaboration, ``{Search for pair-produced vector-like quarks of charge
  $-1/3$ in leptos$+$jets final state in pp collisions at $\sqrt{s}=8$ TeV},''
  CMS Physics Analysis Summary, CMS-PAS-B2G-12-019, 2012.
\newblock \url{http://cds.cern.ch/record/1599719/}.

\bibitem{CMS:2013una}
{ CMS} Collaboration, ``Search for vector-like b' pair production with
  multilepton final states in pp collisions at sqrt(s) = 8 tev,'' CMS Physics
  Analysis Summary, CMS-PAS-B2G-13-003, 2013.
\newblock \url{https://cds.cern.ch/record/1629574?ln=en}.

\bibitem{ATLAS-CONF-2014-036}
{ ATLAS} Collaboration, G.~Aad {\em et~al.}, ``{Search for pair and single
  production of new heavy quarks that decay to a $Z$ boson and a
  third-generation quark in $pp$ collisions at $\sqrt{s}=8$ TeV with the ATLAS
  detector},''
\href{http://arxiv.org/abs/1409.5500}{{\ttfamily arXiv:1409.5500 [hep-ex]}}.

\bibitem{ATLAS-CONF-2013-051}
{ ATLAS} Collaboration, ``{Search for anomalous production of events with
  same-sign dileptons and b jets in 14.3/fb of pp collisions at sqrt(s) = 8 TeV
  with the ATLAS detector},'' ATLAS conference note, ATLAS-CONF-2013-051, 2013.
\newblock \url{http://cds.cern.ch/record/1547567}.

\bibitem{TheATLAScollaboration:2013sha}
{ ATLAS} Collaboration, ``{Search for pair production of heavy top-like quarks
  decaying to a high-$p_{\rm T}$ $W$ boson and a $b$ quark in the lepton plus
  jets final state in $pp$ collisions at $\sqrt{s}=8$ TeV with the ATLAS
  detector},'' ATALS Conference Note, ATLAS-CONF-2013-060, 2013.
\newblock \url{http://cds.cern.ch/record/1557777}.

\bibitem{ATLAS:2013ima}
{ ATLAS} Collaboration, ``{Search for heavy top-like quarks decaying to a Higgs
  boson and a top quark in the lepton plus jets final state in pp collisions at
  sqrt(s)=8 TeV with the ATLAS detector},'' ATLAS conference note,
  ATLAS-CONF-2013-018, 2013.
\newblock \url{http://cds.cern.ch/record/1525525}.

\bibitem{Chatrchyan:2013wfa}
{ CMS} Collaboration, S.~Chatrchyan {\em et~al.}, ``{Search for top-quark
  partners with charge 5/3 in the same-sign dilepton final state},''
  \href{http://dx.doi.org/10.1103/PhysRevLett.112.171801}{{\em Phys.Rev.Lett.}
  {\bfseries 112} (2014) 171801},
\href{http://arxiv.org/abs/1312.2391}{{\ttfamily arXiv:1312.2391 [hep-ex]}}.

\bibitem{Beenakker:1996ed}
W.~Beenakker, R.~Hopker, and M.~Spira, ``{PROSPINO: A Program for the
  production of supersymmetric particles in next-to-leading order QCD},''
\href{http://arxiv.org/abs/hep-ph/9611232}{{\ttfamily arXiv:hep-ph/9611232
  [hep-ph]}}.

\bibitem{Sjostrand:2006za}
T.~Sjostrand, S.~Mrenna, and P.~Z. Skands, ``{PYTHIA 6.4 Physics and Manual},''
  \href{http://dx.doi.org/10.1088/1126-6708/2006/05/026}{{\em JHEP} {\bfseries
  0605} (2006) 026},
\href{http://arxiv.org/abs/hep-ph/0603175}{{\ttfamily arXiv:hep-ph/0603175
  [hep-ph]}}.

\bibitem{Pumplin:2002vw}
J.~Pumplin, D.~Stump, J.~Huston, H.~Lai, P.~M. Nadolsky, {\em et~al.}, ``{New
  generation of parton distributions with uncertainties from global QCD
  analysis},'' \href{http://dx.doi.org/10.1088/1126-6708/2002/07/012}{{\em
  JHEP} {\bfseries 0207} (2002) 012},
\href{http://arxiv.org/abs/hep-ph/0201195}{{\ttfamily arXiv:hep-ph/0201195
  [hep-ph]}}.

\bibitem{Degrande:2011ua}
C.~Degrande, C.~Duhr, B.~Fuks, D.~Grellscheid, O.~Mattelaer, {\em et~al.},
  ``{UFO - The Universal FeynRules Output},''
  \href{http://dx.doi.org/10.1016/j.cpc.2012.01.022}{{\em Comput.Phys.Commun.}
  {\bfseries 183} (2012) 1201--1214},
\href{http://arxiv.org/abs/1108.2040}{{\ttfamily arXiv:1108.2040 [hep-ph]}}.

\bibitem{Alloul:2013bka}
A.~Alloul, N.~D. Christensen, C.~Degrande, C.~Duhr, and B.~Fuks, ``{FeynRules
  2.0 - A complete toolbox for tree-level phenomenology},''
  \href{http://dx.doi.org/10.1016/j.cpc.2014.04.012}{{\em Comput.Phys.Commun.}
  {\bfseries 185} (2014) 2250--2300},
\href{http://arxiv.org/abs/1310.1921}{{\ttfamily arXiv:1310.1921 [hep-ph]}}.

\bibitem{Christensen:2008py}
N.~D. Christensen and C.~Duhr, ``{FeynRules - Feynman rules made easy},''
  \href{http://dx.doi.org/10.1016/j.cpc.2009.02.018}{{\em Comput.Phys.Commun.}
  {\bfseries 180} (2009) 1614--1641},
\href{http://arxiv.org/abs/0806.4194}{{\ttfamily arXiv:0806.4194 [hep-ph]}}.

\bibitem{Alwall:2011uj}
J.~Alwall, M.~Herquet, F.~Maltoni, O.~Mattelaer, and T.~Stelzer, ``{MadGraph 5
  : Going Beyond},'' \href{http://dx.doi.org/10.1007/JHEP06(2011)128}{{\em
  JHEP} {\bfseries 1106} (2011) 128},
\href{http://arxiv.org/abs/1106.0522}{{\ttfamily arXiv:1106.0522 [hep-ph]}}.

\bibitem{2012JInst.7.1001C}
{CMS Collaboration}, ``{Performance of {$\tau$}-lepton reconstruction and
  identification in CMS},''
  \href{http://dx.doi.org/10.1088/1748-0221/7/01/P01001}{{\em J. Instrum.}
  {\bfseries 7} (2012) 1001}, \href{http://arxiv.org/abs/1109.6034}{{\ttfamily
  arXiv:1109.6034 [physics.ins-det]}}.

\bibitem{PFT-10-004}
{ CMS} Collaboration, ``Study of tau reconstruction algorithms using pppp
  collisions data collected at $\sqrt{s} = 7\,${TeV},'' CMS Physics Analysis
  Summary, CMS-PAS-PFT-10-004, 2010.
\newblock \url{http://cdsweb.cern.ch/record/1279358}.

\bibitem{PFT-08-001}
{ CMS} Collaboration, ``Cms strategies for tau reconstruction and
  identification using particle-flow techniques,'' CMS Physics Analysis
  Summary, CMS-PAS-PFT-08-001, 2009.
\newblock \url{http://cdsweb.cern.ch/record/1198228}.

\bibitem{thomas}
``Private communication with scott thomas and sunil somalwar,''.

\bibitem{Aad:2014vma}
{ ATLAS} Collaboration, G.~Aad {\em et~al.}, ``{Search for direct production of
  charginos, neutralinos and sleptons in final states with two leptons and
  missing transverse momentum in $pp$ collisions at $\sqrt{s} =$ 8 TeV with the
  ATLAS detector},'' \href{http://dx.doi.org/10.1007/JHEP05(2014)071}{{\em
  JHEP} {\bfseries 1405} (2014) 071},
\href{http://arxiv.org/abs/1403.5294}{{\ttfamily arXiv:1403.5294 [hep-ex]}}.

\end{thebibliography}\endgroup
\bibliographystyle{utphys}

\end{document}